\documentclass[aps,prd,10pt,twocolumn,
preprintnumbers,nofootinbib,longbibliography]{revtex4-2}

\usepackage{graphicx}

\usepackage{mathtools}
\usepackage{amssymb,bm}

\usepackage{comment}
\usepackage{mathrsfs}
\usepackage{upgreek}

\usepackage[svgnames,table]{xcolor}
\usepackage[super]{nth}

\newcommand{\ri}{\mathrm{i}}
\newcommand{\rd}{\mathrm{d}}

\DeclareMathOperator{\Tr}{Tr}

\newcommand{\ra}{\rangle}
\newcommand{\la}{\langle}
\newcommand{\pd}{\partial} 

\newcommand{\ssf}[1]{\mathsf{#1}}
\newcommand{\mb}[1]{\mathbf{#1}}
\newcommand{\mf}[1]{\mathfrak{#1}}
\newcommand{\cl}[1]{\mathcal{#1}}
\newcommand{\tx}[1]{\mathrm{#1}}
\newcommand{\scr}[1]{\mathscr{#1}}

\newcommand{\fixket}[1]{|#1\rangle}
\newcommand{\fixbra}[1]{\langle#1|}
\newcommand{\fixbraket}[2]{\langle#1|#2\rangle}
\newcommand{\projector}[1]{|#1\rangle \! \langle#1|}
\newcommand{\dblket}[1]{|#1\rangle \! \rangle}
\newcommand{\dblbra}[1]{\langle \! \langle#1|}
\newcommand{\dblbraket}[2]{\langle \! \langle#1|#2\rangle \! \rangle}
\newcommand{\bkk}[2]{\langle#1|#2\rangle \! \rangle}
\newcommand{\bbk}[2]{\langle \! \langle#2|#1\rangle }

\newcommand{\weakeq}[1]{\mathrel{\overset{\scriptscriptstyle #1}{\approx}}}

\newcommand{\pmtx}[1]{\begin{pmatrix}#1\end{pmatrix}}

\newcommand{\eqn}[1]{\begin{equation}\begin{split}#1\end{split}\end{equation}}

\usepackage{silence}
\usepackage[colorlinks=true, allcolors=DarkSlateGray
]{hyperref}

\begin{document}

\preprint{YITP-25-198}
\preprint{FQSP-2026-2}

\author{Kengo Shimada}
\email{kengo.shimada@yukawa.kyoto-u.ac.jp}
\affiliation{
Center for Gravitational Physics and Quantum Information, Yukawa Institute for Theoretical Physics, \\
Kyoto University, Kitashirakawa Oiwakecho, Sakyo-Ku, Kyoto 606-8502, Japan
}
\affiliation{
Fundamental Quantum Science Program, TRIP Headquarters, RIKEN, Wako 351-0198, Japan
}

\title{
Unitarity with respect to the Clock of a Dynamical Local Observer \texorpdfstring{\\}{ }
and the Einstein Equivalence Principle
}

\begin{abstract}
We investigate the unitarity of quantum evolution with respect to an internal time defined by a local observer's clock. The observer is modeled as a relativistic particle carrying both a clock and a matter-field detector, analyzed first on a fixed curved background and subsequently within a fully diffeomorphism-invariant theory of dynamical gravity.
In the former case, we find that evolution relative to the internal clock is generally nonunitary, implying a violation of the Einstein equivalence principle at the quantum level. In contrast, in the latter case, diffeomorphism invariance allows us to adopt observer-centric coordinates. On the resulting partially reduced phase space, one of the diffeomorphism generators becomes linear in the clock Hamiltonian, generating a relational evolution that is consistent with the remaining diffeomorphism constraints.
Assuming the existence of an effective quantum theory that consistently preserves diffeomorphism invariance,
these features imply unitary evolution relative to the internal clock, thereby preserving the equivalence principle even in the quantum regime. Our results highlight the fundamental role of diffeomorphism invariance in shaping relational unitary evolution from the perspective of a local observer.
\end{abstract}
\maketitle

\section{Introduction \label{sec:introduction}}
Quantum theory is widely regarded as a fundamental and universal framework for describing nature.
In its standard formulation, time is assumed to be given \textit{a priori} as an external parameter, and the time evolution of a closed system is unitary, governed by the Schr\"odinger equation.
By contrast, incorporating gravity into the quantum framework has remained a longstanding challenge.
The classical theory of gravity is based on diffeomorphism invariance that is believed to be a fundamental requirement for any theory describing nature.  
While the nonrenormalizability of gravitational interactions is often emphasized as a technical obstacle, quantum gravity also faces profound conceptual difficulties.
 
\ 

The problem of time is known as one such issue.
In diffeomorphism-invariant theories such as general relativity, evolution with respect to the ``external'' time as one of the spacetime coordinates loses its physical meaning, and the Hamiltonian is constrained to vanish.
In canonical quantum gravity, this problem manifests itself as the Wheeler-DeWitt (WDW) equation \cite{DeWitt:1967yk}, which is the condition for physical states $\dblket{\Psi_\tx{ph}}$ to satisfy,
\eqn{
\hat{\ssf{C}}^\tx{diff} \dblket{\Psi_\tx{ph}} = 0 ~, \label{eq:WDW_eq}
}
where $\hat{\ssf{ C}}^\tx{diff}$ represents the diffeomorphism generators whose linear combination forms the Hamiltonian.   
Then, the standard notion of unitary time evolution familiar from quantum theory is no longer straightforwardly applicable to quantum gravity.

\

A widely explored approach to this problem is to describe dynamics relationally, by identifying one of the dynamical degrees of freedom (d.o.f.) as a ``clock'' and expressing the evolution of the remaining d.o.f. with respect to it. 
Let $\hat{A}$ denote the operator corresponding to such a clock variable, acting on the clock Hilbert space $\cl{H}_\tx{C}$, which is a tensor factor of the ``kinematical'' Hilbert space $\cl{H}_\tx{kin}  = \cl{H}_\tx{C} \otimes \cl{H}_{\overline{\tx{C}}}$ based on which the physical space $\cl{H}_\tx{ph}$ is constructed under the diffeomorphism constraints (\ref{eq:WDW_eq}).
Following Page and Wootters (PW) \cite{Page:1983uc,Wootters:1984wfv}, the probability of obtaining a value $B$ for another observable $\hat{B}$ acting on $\cl{H}_{\overline{\tx{C}}}$, given that the clock reads $A$, is defined as
\eqn{
P ( B | A ) &= \frac{\dblbra{\Psi_\tx{ph}} \hat{\Pi}(A) \otimes \hat{\Pi}(B)  \dblket{\Psi_\tx{ph}} }{ \dblbra{\Psi_\tx{ph}} \hat{\Pi} (A) \otimes \hat{I}_{\overline{\tx{C}}}  \dblket{\Psi_\tx{ph}} } \\
&= \frac{\fixbra{\Psi_\tx{ph} (A)}  \hat{\Pi}(B)  \fixket{\Psi_\tx{ph} (A)} }{ \fixbraket{\Psi_\tx{ph} (A)}{\Psi_\tx{ph} (A)} } ~, \label{eq:PW_conditional-probability}}
where $\hat{\Pi}(A)$ is the projector onto $\hat{A}$'s eigenstate $\fixket{A}$ with the eigenvalue $A$, and $\hat{\Pi}(B)$ is defined likewise.
It is normalized so that it integrates, or sums, to $1$, with the denominator in which $\fixket{\Psi_\tx{ph} (A)} := \bkk{A}{\Psi_\tx{ph}}$ and $\hat{I}_{\overline{\tx{C}}}$ is the identity operator on $\cl{H}_{\overline{\tx{C}}}$.
If the square norm  $\fixbraket{\Psi_\tx{ph} (A)}{\Psi_\tx{ph} (A)}$ is independent of the clock reading $A$, the above conditional-probability interpretation reduces to the standard probability interpretation in quantum theory.
Furthermore, if $\fixket{\Psi_\tx{ph} (A)} = \hat{\ssf{U}}(A) \fixket{\Psi_\tx{0}}$ with a unitary operator $\hat{\ssf{U}}(A)$ for any element of $\cl{H}_\tx{ph}$, then the dynamics can be described in a framework equivalent to quantum theory.
In what follows, this form of evolution is said to be ``unitary'' with respect to the clock variable $A$; yet this is not the case in general \cite{Rovelli:1989jn,Rovelli:1990jm,Kiefer:1990pt,Gambini:2006yj}.

\

The PW formalism has faced criticisms, notably those raised by Kucha\v{r} \cite{Kuchar:1991qf}.
Two of these objections are directly apparent in Eq.(\ref{eq:PW_conditional-probability}): the projectors do not commute with the constraint operator, and applying the conditional-probability formula to a two-time case appears to yield wrong propagators.
These issues have been addressed in subsequent works by appropriately formulating relational probabilities in terms of Dirac observables \cite{Gambini:2004pe,Gambini:2008ke,Hoehn:2019fsy}.

\

Owing to diffeomorphism invariance, gravity couples universally to all d.o.f.
As a consequence, anything that plays the role of an observer is necessarily part of the Universe itself and subject to the constraints (\ref{eq:WDW_eq}).
From an operational and everyday perspective, it is natural to consider an observer that is spatially localized.
For such a local observer, time is most naturally understood as a local quantity intrinsic to the observer, giving rise to the notion of an ``internal'' time, and relative to which local observables evolve \cite{Rovelli:1990ph,Rovelli:1990pi}.
This immediately leads to a central question of this work:
\begin{center}
\emph{Is the evolution with respect to such an internal observer-dependent time unitary?}
\end{center}
If so, we can further ask whether, in the presence of multiple observers, the corresponding internal-time unitarities are mutually compatible.

\ 

In gravitational theories, a local observer is given a special status, particularly when in a state of free fall.
The Einstein equivalence principle (EEP) states that the outcomes of local experiments performed within a free-falling laboratory are independent of the background gravitational field, at least at the classical level.
The inertial nature of such a laboratory implies the absence of an external reference; in other words, the description of local experiments must rely solely on internal d.o.f., whose dynamics are fundamentally relational. 
As we extend our focus to quantum theory, another question that we can address is the following:
\begin{center}
\emph{Does the EEP remain valid at the quantum level and\\ how is it related to the unitarity of the evolution relative to the internal clock?}
\end{center}
While there is a proposal of the quantum EEP as a guiding principle for constructing quantum gravity \cite{Giacomini:2017zju,Giacomini:2020ahk} based on the concept of the quantum reference frame (QRF, see Ref.\cite{Hoehn:2019fsy} and references therein), here we take a bottom-up approach based on the conventional formulation of canonical quantum theory to examine whether the EEP naturally emerges from the underlying dynamics.

\ 

As discussed above, the notion of an internal clock emerges from diffeomorphism invariance.
By contrast, the classical EEP is traditionally formulated on a fixed background, a framework that does not accommodate diffeomorphism invariance.
To isolate the role played by this invariance in the context of unitarity and EEP at the quantum level, we start with a theory on a nondynamical curved background.
This provides a useful baseline for examining how the physics changes when diffeomorphism invariance is fully realized within the framework of dynamical gravity. 

\

We adopt units in which $c = \hbar = 1$.
With $d$ being the spatial dimensionality, Greek indices $\mu, \nu,...$ run over spacetime coordinates ranging from $0$ to $d$, while Latin indices $i,j,...$ run over spatial coordinates from $1$ to $d$.
We employ the Einstein summation convention unless otherwise stated.
By separating the external coordinate time $t := x^0$ from the spatial coordinates $\tx{x}^i$, we write the spacetime coordinates as $x^\mu = (t , \mb{x})$, where $\mb{x}$ collectively denotes the spatial components $\tx{x}^i$.
We use the same notation for spacetime vectors and spatial vectors.
When convenient, fields $f(x)$ defined on spacetime are written as $f(t, \mb{x})$.
In quantum mechanical discussions, we drop the identity operator acting on a component of the tensor product space, such as $\hat{I}_{\overline{\tx{C}}}$ in Eq.(\ref{eq:PW_conditional-probability}), when no confusion arises.

\section{Summary / overview \label{sec:summary/overview}}
\begin{figure}[t]
\begin{center}
\includegraphics[width=8cm]{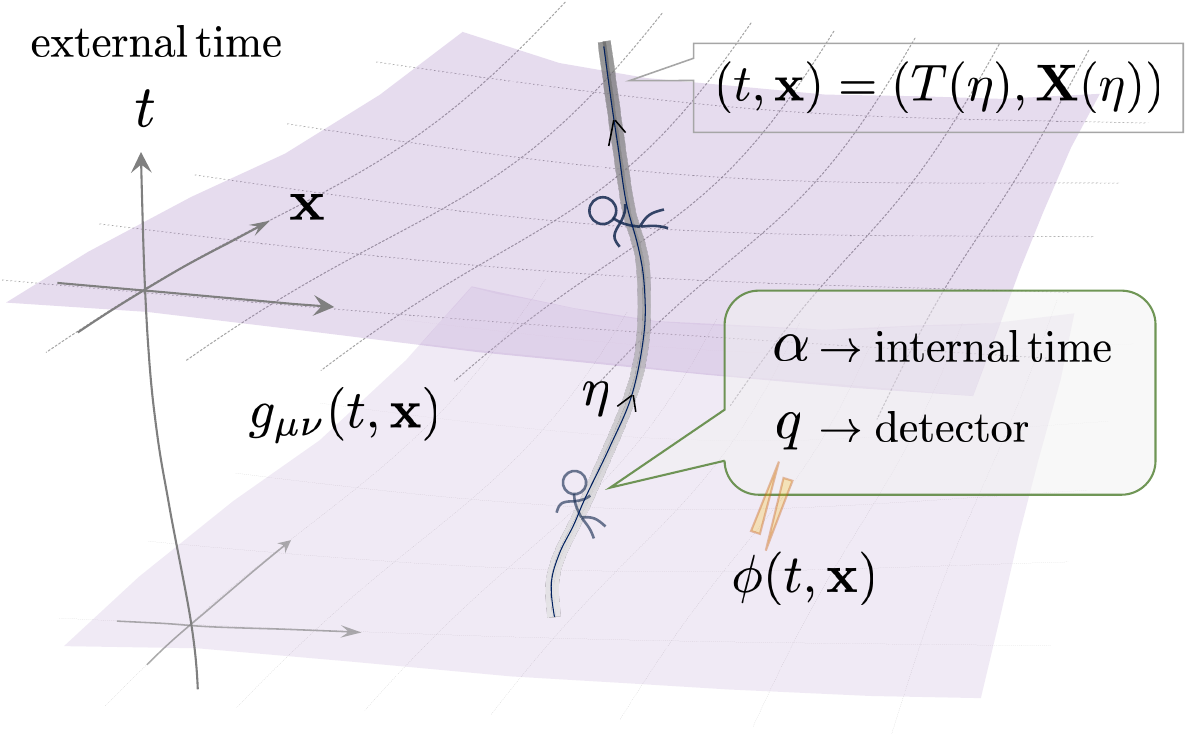}
\caption{The observer's trajectory $X^\mu (\eta)$ is parametrized by $\eta$. It is assumed that the observer minimally couples to the spacetime metric field $g_{\mu \nu}$, and then the observer's internal d.o.f. simply contribute to its effective mass. One of the internal d.o.f. $\alpha$ with no direct coupling to the matter field serves as the internal clock, whereas the other one $q$, which directly couples to the matter field $\phi$, is referred to as the detector. When the coupling is turned off, $q$ can represent all d.o.f. involved in a local experiment performed in a freely falling laboratory.
}
\label{fig:observer}
\end{center}
\end{figure}

In this paper, we model the local observer as a relativistic particle with internal d.o.f., one of which is assumed to have no direct coupling to a matter field and identified as the clock variable, denoted by $\alpha$.
On the other hand, the internal variable $q$ directly coupled to the matter field is referred to as the ``matter-field detector''; see Fig.\ref{fig:observer}.
Other variables, such as the fields and also the observer's location in spacetime, are referred to as the external d.o.f.

\

We discuss how the rest of the system, excluding the clock $\alpha$, evolves relative to the shift of $\alpha$ generated by the so-called clock Hamiltonian $H_\tx{C}$.
To clarify the importance of diffeomorphism invariance, we start with a model on a fixed curved background, which naturally admits the standard formulation of quantum theory from the external perspective defined by the unitary evolution with respect to the external coordinate time.
Then, we discuss a diffeomorphism-invariant model on dynamical spacetime, where physical evolution is necessarily described relative to the clock $\alpha$.

\

When the matter-field detector is turned off, the observer is in free fall, and then $q$ represents all variables that are involved in a local experiment on the worldline.
With respect to the local time defined by $\alpha$, the evolution of $q$ should be independent of external information according to the EEP.
In this case, our main findings can be summarized as follows: 
\begin{itemize}
\item On a general fixed background:
\begin{itemize}
\item  The standard formulation based on the external perspective leads to the nonunitary evolution relative to the internal clock $\alpha$, described by the Schr\"odinger-type equation (\ref{summary:Schrodinger_eq_with_nonHermitian_Hamiltonian}) with a non-Hermitian operator.
\item Consequently, the EEP is found to be violated in the sense that the probability of finding the internal variable $q$ in a certain state depends on the state of the external d.o.f., as seen in Eq.(\ref{summary:internal-time_derivative_of_conditional_probability}).
\item This follows from the fact that the constraint function, $C^\tx{ex}_+$ given in Eq.(\ref{summary:constraint_function_for_external_perspective}) and defining the physical state by the condition (\ref{summary:physical_state_condition_for_external_perspective}), is nonlinear in the Hamiltonian of the internal d.o.f., especially $H_\tx{C}$, and couples the internal d.o.f. to the external d.o.f., albeit indirectly.
\end{itemize}
\end{itemize}
We note that the nonunitarity of relational evolution based on physical clocks has already been discussed in previous works \cite{Gambini:2004pe,Gambini:2006yj,Gambini:2006ph,Paiva:2021wbv,Paiva:2022mxf}.
Our focus here, however, is on its implications for the EEP.
This nonunitarity-induced violation of the EEP motivates us to investigate whether this tension can be resolved by treating spacetime as dynamical.
\begin{itemize}
\item On dynamical spacetime:
\begin{itemize}
\item Diffeomorphism invariance allows us to work on a partially reduced phase space $\cl Z$ by adopting a coordinate system in which the observer appears to be at rest.
From diffeomorphism constraints, a specific element, $\ssf{C}$ given by Eq.(\ref{summary:singled_out_from_C^diff}), can be isolated as the only one that involves $H_\tx{C}$ and depends on it linearly. 
\item
With the generator $\ssf{C}$, the local quantities (\ref{summary:classical_observable_relative_to_internal_time}) constructed by group-averaging evolve relative to $\alpha$ consistently with the rest of the diffeomorphisms, ensured by Eqs.(\ref{summary:closed_sub-algebra}) and (\ref{summary:invariance_under_sub-diffeo}), and therefore are physical observables. 
\item These suggest that, if the theory admits a consistent quantization, the corresponding evolution is described by the Schr\"odinger equation (\ref{summary:Schrodinger_eq_with_gravity}) and is unitary.
In addition, the decoupling between the internal and external d.o.f. is manifest on $\cl Z$, thereby recovering the EEP.
\end{itemize}
\end{itemize}

When the matter-field detector is turned on, the observer is no longer in free fall, and thus the EEP is not addressed.
In this case, we find the following:
\begin{itemize}
\item The unitarity with respect to the internal time does not hold in the formulation on a fixed background, even on Minkowski spacetime for a noninertial observer.
Nevertheless, for the same reason presented above, the unitarity is expected to be recovered on dynamical spacetime due to diffeomorphism invariance.
\end{itemize}

In the remainder of this section, we will elucidate the above statements without delving into technical details, providing an overview of the paper's structure so that readers can easily locate the corresponding details in Secs.\ref{sec:relativistic_particle_with_clock_and_detector}, \ref{sec:violation_of_EEP}, and \ref{sec:observers_with_dynamical_gravity}.
A brief discussion on the relations with other approaches and implications for related fields will be given in Sec.\ref{sec:conclusion_and_outlook}.

\subsection{On fixed background spacetime}
In Sec.\ref{sec:relativistic_particle_with_clock_and_detector}, we discuss a model on a fixed background spacetime, mainly focusing on the standard formulation from the external perspective; namely, the evolution of the system is described with respect to the external coordinate time.
For simplicity, in this subsection, we assume a unit lapse function and a vanishing shift vector.

\subsubsection{Classical theory \label{summary-subsubsec:classical_theory}}
The classical description of the dynamical local observer is presented in Sec.\ref{subsec:classical_theory}, starting from an action analogous to the familiar one of the relativistic particle.
Let us first assume that the matter fields are also nondynamical; in other words, the only external d.o.f. are the observer's spacetime location $X^\mu = (T,\mb{X})$ whose conjugate momentum is denoted by $P_\mu = ( -E , \mb{P})$ and satisfies $\{ X^\mu , P_\nu \}_\tx{P} = \delta^\mu_\nu$ under the Poisson bracket $\{ \cdot , \cdot \}_\tx{P}$.
The reparametrization invariance on the worldline gives rise to a constraint that fully governs the system consisting of these external d.o.f. as well as the internal d.o.f.
Consistency with the physical requirement that the observer's energy $E$ is positive recasts this constraint into the condition that the constraint function
\eqn{
C^\tx{ex}_+ := - E + \omega ~
\label{summary:constraint_function_for_external_perspective}}
vanishes.
As we shall see, $\omega := ( |\mb{P}|_{h}^2 +  M_\tx{eff}^2  )^{1/2}$ turns out to be the Hamiltonian generating the evolution with respect to the external time, where $|\mb{P}|_{h}^2$ is the squared norm of the observer's spatial momentum with respect to the metric $h_{ij}$ induced on a reference time slice specified by a value of $T$ and $M_\tx{eff} := M + H_\tx{C} + H_\tx{D}$ is the observer's effective mass with its constant part $M > 0$ and the contributions from the internal d.o.f., which are the clock Hamiltonian $H_\tx{C} \geq 0$ defined to be the generator of $\alpha$'s shift satisfying $\{ \alpha , H_\tx{C} \}_\tx{P} = 1$ and the detector Hamiltonian $H_\tx{D} \geq 0$ describing the local interaction of the detector variable $q$ with the matter fields at the observer's location $x^\mu = X^\mu$.

\subsubsection{Quantum theory and internal-time nonunitarity}
The quantum theory of this model is formulated from the external perspective based on the constraint $C_+^\tx{ex} = 0$ in Sec.\ref{subsec:quantization}.
The kinematical Hilbert space $\cl{ H}_\tx{kin} = \cl{H}_\tx{IN} \otimes \cl{H}_\tx{EX}$ is a tensor product of those for the internal and external d.o.f., each of which is further factorized as $\cl{H}_\tx{IN} = \cl{ H}_\tx{C} \otimes \cl{ H}_\tx{D}$ and $\cl{H}_\tx{EX} = \cl{ H}_\tx{T} \otimes \cl{ H}_\tx{X}$, where $\cl{ H}_\tx{C}$, $\cl{ H}_\tx{D}$, $\cl{ H}_\tx{T}$ and $\cl{ H}_\tx{X}$ are the Hilbert spaces for the clock, the detector, the observer's time and position coordinates, respectively.
Then, the physical Hilbert space $\cl{H}_\tx{ph}^\tx{ex}$ is identified as a projection of $\cl{H}_\tx{kin}$ spanned by the states that are annihilated by the operator corresponding to the constraint function (\ref{summary:constraint_function_for_external_perspective}) as
\eqn{ 
\hat{C}_+^\tx{ex} \dblket{\psi^\tx{ex}_\tx{ph}} = 0 ~.
\label{summary:physical_state_condition_for_external_perspective}}
Then, $\cl{H}_\tx{ph}^\tx{ex}$ is no longer factorizable.
Projecting this to the eigenstate $\fixket{T}$ of the external time operator $\hat{T}$, we get the Schr\"odinger equation for $\fixket{\psi^\tx{ex}_\tx{ph} (T)} := \bkk{T}{\psi^\tx{ex}_\tx{ph}}$ as 
\eqn{
(\ri \pd_T  - \hat{\omega}(T)  ) \fixket{\psi^\tx{ex}_\tx{ph} (T)} =  0  ~
\label{summary:Schrodinger_eq_in_T-rep}}
with the generally $T$-dependent Hamiltonian $\hat{\omega}(T)$, which is the representation of the operator corresponding to $\omega$ with respect to the basis $\fixket{T}$.
This makes it explicit that the evolution with respect to the external reference time $T$ is unitary. 
More specifically, the external-time unitarity follows from the fact that the constraint function ${C}^\tx{ex}_+$ is linear in $E$, the canonical conjugate to $T$.

\

In Sec.\ref{subsec:evolution_relative_to_the_internal_clock}, the evolution relative to the internal clock is investigated within this theory.
In the spirit of the PW formulation discussed in relation with Eq.(\ref{eq:PW_conditional-probability}), we consider $\fixket{ \psi_\tx{ph}^\tx{ex} (\alpha)} := \bkk{\alpha}{\psi_\tx{ph}^\tx{ex}}$, an element of $\cl{ H}_{\overline{\tx{C}}} := \cl{ H}_\tx{D} \otimes \cl{H}_\tx{EX}$, where $\fixket{\alpha}$ is the so-called clock state defined to satisfy $e^{- \ri \alpha' \hat{H}_\tx{C}} \fixket{\alpha} = \fixket{\alpha + \alpha'}$.
Then, the evolution with respect to the internal reference time $\alpha$ can be read off from the projection of the physical state condition (\ref{summary:physical_state_condition_for_external_perspective}) to $\fixket{\alpha}$, i.e.,
\eqn{
\fixbra{\alpha} \hat{C}_+^\tx{ex} \dblket{\psi_\tx{ph}^\tx{ex}} = 0 ~.
\label{summary:alpha-representation_of_physical_state_condition_for_external_perspective}}
In this ``$\alpha$ representation,'' the clock Hamiltonian operator $\hat{H}_\tx{C}$ turns to $- \ri \pd_\alpha$.
However, since the constraint function (\ref{summary:constraint_function_for_external_perspective}) is nonlinear in $H_\tx{C}$, the above equation is not a first-order differential equation with respect to $\alpha$.
Under certain reasonable assumptions, it can be approximated by the Schr\"odinger-type equation,
\eqn{
( \ri \pd_\alpha - \hat{ K} ) \fixket{ \psi_\tx{ph}^\tx{ex} (\alpha)} \simeq 0 ~.
\label{summary:Schrodinger_eq_with_nonHermitian_Hamiltonian}}
However, now it comes with the operator $\hat{ K}$, which is non-Hermitian in general; therefore, the internal-time evolution is nonunitary in the sense defined below Eq.(\ref{eq:PW_conditional-probability}).

\

It is straightforward to extend this quantum theory of the local observer to incorporate dynamical matter fields.
In this case, the external part of the kinematical Hilbert space contains $\cl{H}_\tx{m}$ for the matter-field d.o.f.: $\cl{H}_\tx{EX} = \cl{H}_\tx{T} \otimes \cl{H}_\tx{X} \otimes \cl{H}_\tx{m}$. 
Nevertheless, as shown in Appendix \ref{app:observers_with_dynamical_matter_field}, it maintains the same formal structure when expressed in a variant of the interaction picture.
Then, by an argument similar to that presented above, 
it can be concluded that the internal-time evolution is nonunitary in general.

\subsubsection{Violation of equivalence principle \label{subsubsec:summary-viloation_of_EEP}}
In Sec.\ref{sec:violation_of_EEP}, turning off the matter-field detector, we consider a freely falling observer on a curved background and discuss the validity of the EEP in the quantum theory defined by the physical condition (\ref{summary:physical_state_condition_for_external_perspective}).
As a generalization of the weak equivalence principle (WEP),
the EEP states that the outcomes of experiments performed in a free-falling laboratory do not depend on its velocity and location in spacetime $(T, \mb{X})$, i.e., the local Lorentz invariance (LLI) and the local position invariance (LPI) hold.
In the present model, the elapsed clock time is equal to the elapsed proper time, and the local experiments are described with respect to it.
The $q$'s Hamiltonian $H_\tx{D}$ is supposed to describe the local experiment, and thus may be renamed to $H_\tx{exp}$.
According to the LLI and the LPI, the evolution of $q$ relative to $\alpha$ should not depend on the external dynamics, which is indeed the case with the present model at the classical level.

\ 

However, the quantum dynamics of $q$ relative to $\alpha$ based on the PW formulation (\ref{eq:PW_conditional-probability}) turns out to be dependent on the external information due to the internal-time nonunitarity; namely, the LLI and the LPI do not hold, as discussed in Sec.\ref{subsec:violation_of_LLI/LPI}.
The most striking example of this behavior is that the probability of finding the $n$-th eigenstate of $\hat{H}_\tx{exp}$ conditioned by $\alpha = a$ varies with the clock reading $a$ as
\eqn{
\frac{\ri }{2} \pd_a P_\tx{exp}( n | a) \simeq ~ &  \la [ \hat{K} ]_\tx{AH}  \ra_n (a) \\ 
&  -  P( n | a) \sum_k  \la  [ \hat{K} ]_\tx{AH} \ra_k (a)  ~,
\label{summary:internal-time_derivative_of_conditional_probability}}
where $\la  \hat{O} \ra_n (a)$ is the normalized expectation value in the state $\fixket{ \psi_\tx{ph}^\tx{ex} (\alpha)}$ of $\hat{\Pi}(n) \otimes \hat{O}$ with $\hat{O}$ an operator acting on $\cl{H}_\tx{EX}$ and $\hat{\Pi}(n)$ the projector onto the $n$-th eigenstate of $\hat{H}_\tx{exp}$, and $[ \hat{K} ]_\tx{AH}$ is the anti-Hermitian part of $\hat{K}$ in the approximated evolution equation (\ref{summary:Schrodinger_eq_with_nonHermitian_Hamiltonian}).
The right-hand side (rhs) of Eq.(\ref{summary:internal-time_derivative_of_conditional_probability}) is generally not vanishing for the entanglement between the variable $q$ and the external d.o.f.; therefore, it depends on the external information.

\

It should be emphasized that this dependence originates from their ``indirect'' interaction through the square root in the physical condition (\ref{summary:alpha-representation_of_physical_state_condition_for_external_perspective}) in the free fall case considered here; consequently, it manifests itself as $[ \hat{K} ]_\tx{AH}$ acting only on $\cl{H}_\tx{EX}$, not coupling the external and internal d.o.f. directly.
In light of this, the internal-time nonunitarity is identified as the cause of the violation of the EEP.

\subsubsection{Perspective-dependent quantum theory?}
One may wonder if the quantum theory should be formulated from the internal perspective based on a constraint $C^\tx{in}_+ = 0$ with $C^\tx{in}_+$ defined to be strictly linear in the clock Hamiltonian $H_\tx{C}$ while sharing an intersection with $C^\tx{ex}_+ = 0$.
However, as discussed in Sec.\ref{subsec:qauntum_theory_from_internal_pespective?}, it leads to both technical and conceptual difficulties.
Promoting the constraint function $C^\tx{in}_+$ to the corresponding operator requires an \textit{ad hoc} projector with certain ambiguities.
Even if one manages to justify such a quantization procedure, the next problem inevitably arises; the physical Hilbert space defined by $\hat{C}^\tx{in}_+ \dblket{\psi^\tx{in}_\tx{ph}} = 0$ is inequivalent to the one defined by the condition (\ref{summary:physical_state_condition_for_external_perspective}) from the external perspective; especially, the evolution relative to the external time $T$ is no longer unitary. 
Then, one seems to be forced to the conclusion that the quantum theory depends on the perspective.
This is a version of the ``multiple-choice problem'' of time \cite{Kuchar:1991qf}. 

\

We do not explore this possibility of realizing the internal-time unitarity by solving the constraint for the clock Hamiltonian in the present work.
Instead, we exploit diffeomorphism invariance to get the indication that the evolution relative to the internal clock is unitary.

\subsection{On dynamical spacetime}
Diffeomorphism invariance is realized when not only the local observer but also fields, including the metric itself, are treated dynamically, as in Sec.\ref{sec:observers_with_dynamical_gravity}.
We use the term ``gauge'' transformation in a broad sense to refer to diffeomorphisms as transformations connecting physically indistinguishable configurations and hereafter use these two terms interchangeably.
In Sec.\ref{subsec:classical_theory_w/o_gauge-fixing}, we review the canonical formulation with the $(d+1)$ decomposition wherein the observer's time coordinate $T$ is already eliminated by the worldline reparametrization.
Then, we proceed to the one with a partial gauge fixing as follows.
For simplicity, we assume a closed spatial geometry without boundaries.
\begin{figure}[t]
\begin{center}
\includegraphics[width=8cm]{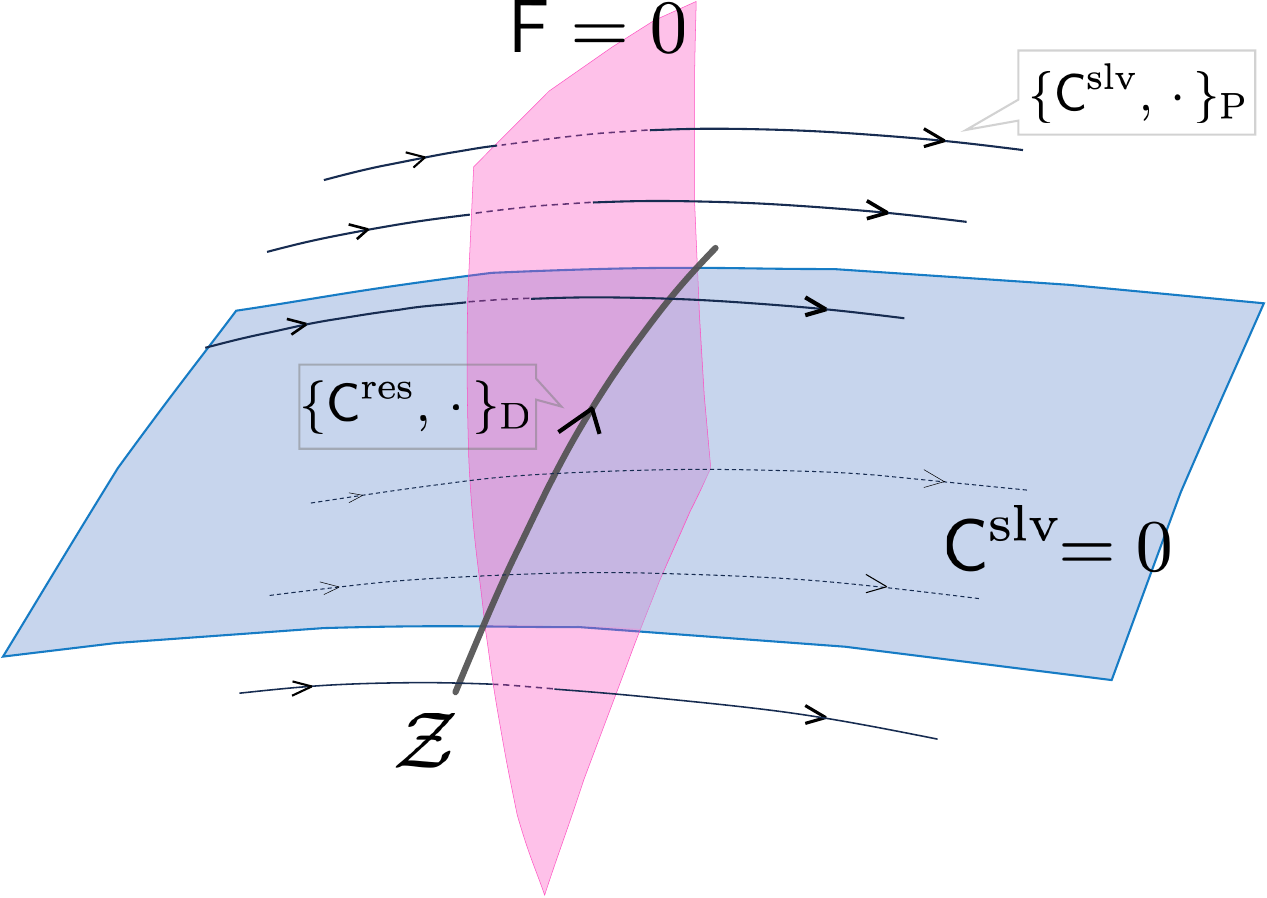}
\caption{The schematic view of the hypersurfaces in the kinematical phase space. The gauge-fixing slice defined by $\ssf{F}=0$ intersects the gauge orbits generated by the elements of $\ssf{C}^\tx{slv}$.
The hypersurface defined by $\ssf{C}^\tx{slv}=0$ and the gauge-fixing slice share the intersections $\cl Z$, referred to as the partially reduced phase space.
The remaining diffeomorphisms are identified with the flow along the gauge orbits in $\cl Z$ generated by the elements of $\ssf{C}^\tx{res}$.
}
\label{fig:partially-reduced_phase_space}
\end{center}
\end{figure}

\subsubsection{Partial reduction of phase space with observer-centric coordinate}
Diffeomorphisms transform the coordinate values of the observer's position.
In other words, we can gauge away these redundant variables, and yet the theory remains consistent through the appropriate reduction of the phase space \`a la Dirac, as explored in Sec.\ref{subsec:observer-centric_coordinate_system}.
Consider a coordinate system in which the observer appears to be at rest: $\mb{X} = \mb{x}_\tx{obs}$ with $\mb{P} = \mb{0}$, which yields $2d$ gauge-fixing conditions, denoted by $\ssf{F}=0$.
For consistency, the diffeomorphisms that act on these variables are forbidden.
There are corresponding $2d$ linearly independent generators
whose set is denoted by
$\ssf{C}^\tx{slv}$
since they are formally ``solved'' in the Dirac procedure that we employ.
There still remain an infinite number of residual constraints $\ssf{ C}^\tx{res}$.
In other words, the set of diffeomorphism generators $\ssf{ C}^\tx{diff}$ is decomposed as
\eqn{
\ssf{ C}^\tx{diff} = \ssf{ C}^\tx{res} \oplus \ssf{ C}^\tx{slv} ~,
\label{summary:res_oplus_slv}}
where the direct sum symbol emphasizes the linear independence of the two sets.

\

The $4 d$ constraint functions comprising those in $\sf F$ and $\ssf{ C}^\tx{slv}$ form a set of second-class constraints $\sf Z$.
Defining the associated Dirac bracket $\{ \cdot , \cdot \}_\tx{D}$ allows us to work on the partially reduced phase space $\cl{ Z}$ on which all the elements in $\sf Z$ vanish; see Fig.\ref{fig:partially-reduced_phase_space}. 
The algebra of the residual constraint functions in $\ssf{ C}^\tx{res} $ is closed under the Dirac bracket:
\eqn{
\{ \ssf{ C}^\tx{res}_r , \ssf{ C}^\tx{res}_{r'} \}_\tx{D}   \weakeq{\tx{res}} 0 ~,
}
where the weak equality $\weakeq{\tx{res}}$ indicates that the equality holds when all the elements of $\ssf{ C}^\tx{res}$ vanish on $\cl Z$.
Therefore, $\ssf{ C}^\tx{res}$ is of the first class, and its elements generate gauge transformations consistently on $\cl{ Z}$.

\subsubsection{Evolution relative to internal clock}
As discussed in Sec.\ref{subsec:physical_evolution_relative_to_internal_clock},
the set $\ssf{ C}^\tx{res}$ of the residual diffeomorphism generators on $\cl Z$ can be decomposed as
\eqn{
\ssf{ C}^\tx{res} = \ssf{ C} \oplus \ssf{ C}^\lozenge ~,
\label{summary:0_oplus_lozenge}}
where $\ssf{ C}$ contains only one element, which we also denote by $\ssf{ C}$, depending on the observer's energy of the same functional form as $\omega$ in Eq.(\ref{summary:constraint_function_for_external_perspective}) but now with $\mb{P} = \mb{0}$ on $\cl{ Z}$, and thus linear in $H_\tx{C}$ as
\eqn{
\ssf{ C} = M + H_\tx{C} + H_\tx{D} + \ssf{ H}_\tx{g+m}[n] ~,
\label{summary:singled_out_from_C^diff}}
where $\ssf{ H}_\tx{g+m}[n]$ is the Hamiltonian of the dynamical fields associated 
with the lapse function $n (\mb{x})$ satisfying $\pd n /\pd \mb{x} |_{\mb{x} = \mb{x}_\tx{obs}}  = \mb{0}$ and normalized as $n(\mb{x_\tx{obs}}) = 1$, whereas the elements of $\ssf{ C}^\lozenge$ have no contributions from the observer's internal d.o.f.
They form a closed subalgebra under the Dirac bracket as
\eqn{
\{ \ssf{ C}^\lozenge_r , \ssf{ C}^\lozenge_{r'} \}_\tx{D} \weakeq{\lozenge} 0 ~,
\label{summary:closed_sub-algebra}}
where $\weakeq{\lozenge}$ indicates that the equality holds when all the elements of $\ssf{ C}^\lozenge$ vanish on $\cl Z$.
In addition, 
\eqn{
\{ \ssf{ C}   , \ssf{ C}^\lozenge_r \}_\tx{D}  \weakeq{\lozenge} 0
\label{summary:invariance_under_sub-diffeo}}
holds; that is, $\ssf{ C} $ is invariant under the ``subgauge'' transformation generated by the elements of $\ssf{ C}^\lozenge$.

\ 

The symplectic structure for the observer's internal variables is preserved on $\cl{ Z}$, meaning that the Dirac brackets involving the internal variables coincide with the Poisson brackets.
Therefore, $\ssf{ C} $ given in Eq.(\ref{summary:singled_out_from_C^diff}) is the only generator that transforms $\alpha$ in the form of a mere shift.
Then, based on quantities $\ssf{ O}_\lozenge$ satisfying $\{ \ssf{ O}_{\lozenge}, \ssf{ C}^\lozenge_r \}_\tx{D}   \weakeq{\lozenge} 0$ to be invariant under the subgauge transformation, we can construct relational Dirac observables as
\eqn{
\ssf{ O}_\tx{ph} (a ; \alpha) &= \sum_{n=0}^\infty \frac{(a - \alpha)^n}{n !} \{   \ssf{ O}_\lozenge  , \ssf{ C}   \}_n   \\
&= \int_\mathbb{R} \rd \eta \left[ \sum_{n=0}^\infty \frac{\eta^n}{n !} \{   \delta (a - \alpha ) \, \ssf{ O}_\lozenge  , \ssf{ C}   \}_n \right] ~
\label{summary:classical_observable_relative_to_internal_time}}
with the internal-time parameter $a$,
where $\{ \cdot , \cdot \}_n$ is the $n$-th nested Dirac bracket defined by $\{ f , g \}_{n+1} = \{ \{ f , g \}_n , g \}_\tx{D}$ with $\{ f , g \}_{0} = f$.
We have written it in the group-averaging form in the second line of Eq.(\ref{summary:classical_observable_relative_to_internal_time}) with the gauge parameter $\eta$ integrated.
This type of quantity satisfies
\eqn{
\{ \ssf{ O}_\tx{ph} (a ; \alpha)  , \ssf{ C}^\tx{res}_r \}_\tx{D}   \weakeq{\lozenge} 0 ~,
\label{summary:weak_physical_condition_with_gravity}}
which is stronger than the condition $\{ \ssf{ O}_\tx{ph}  , \ssf{ C}^\tx{res}_r \}_\tx{D}   \weakeq{\tx{res}} 0$ for $\ssf{ O}_\tx{ph}$ to be physically observable, in that $\ssf{ C} = 0$ is not required.

\ 

It describes the evolution of $\ssf{ O}_\lozenge$ along the Hamiltonian flow and satisfies $\pd_a  \ssf{ O}_\tx{ph} (a ; \alpha) = \{ \ssf{ O}_\tx{ph} (a ; \alpha)  , \ssf{ H}   \}_\tx{D}$ with
\eqn{
\ssf{ H}  := \ssf{ C}  - H_\tx{C} = M + H_\tx{D} + \ssf{ H}_\tx{g+m}[n] ~, 
\label{summary:enrgy_of_universe_relative_to_clock}}
which is nonvanishing on shell and invariant under the residual diffeomorphisms by construction; it can be interpreted as the energy of the rest of the Universe relative to the clock variable. 

\ 
 
Note that this construction of the Dirac observables works regardless of whether the observer is freely falling or not.

Also, if there are multiple observers, we can choose one observer to partially fix the coordinate system and apply the above procedure to obtain the generator $\ssf{ C} $ that is linear in the Hamiltonian of the clock variable carried by that observer.
Applying the procedure with another observer chosen, we get a different $\ssf{ C} $ but with all the desired properties.
Nevertheless, those theories with different gauge fixings should be equivalent.

\subsubsection{Internal-time unitarity \label{summary-subsubsec:QG}}
In Sec.\ref{subsec:unitarity_w.r.t._internal_time}, we assume that there exists an effective quantum field theory consistent with diffeomorphism invariance and
the algebra of the partially gauge-fixed theory is preserved, at least within the regime of validity of that effective theory.
The kinematical Hilbert space is factorized as in the fixed-background case, $\cl{ H}_\tx{kin} = \cl{H}_\tx{IN} \otimes \cl{H}_\tx{EX}$, but now with $\cl{H}_\tx{EX}$ associated with the external d.o.f. consisting of the modes of the matter and metric fields that remain independent after the partial gauge fixing.

\ 

With the decomposition (\ref{summary:0_oplus_lozenge}) in mind, states in the physical Hilbert space $\cl{ H}_\tx{ph}$ are defined to satisfy
\eqn{ 
\hat{\ssf{ C}}  \dblket{\Psi_\tx{ph}} = 0 ~~~ \tx{and} ~~~  \hat{\ssf{C}}^\lozenge_r \dblket{\Psi_\tx{ph}} = 0 ~ 
\label{summary:gravitational_state_condition_with_partial_reduction}}
for all $r$.
These physical conditions are mutually consistent for the closure of the algebra under the commutator corresponding to Eqs.(\ref{summary:closed_sub-algebra}) and (\ref{summary:invariance_under_sub-diffeo}).

\

Projecting the first condition in Eq.(\ref{summary:gravitational_state_condition_with_partial_reduction}) to the clock state $\fixket{\alpha}$ defined above Eq.(\ref{summary:alpha-representation_of_physical_state_condition_for_external_perspective}), we get the Schr\"odinger equation for the state $\fixket{\Psi_\tx{ph} (\alpha)} := \bkk{\alpha}{\Psi_\tx{ph}}$ with respect to the internal reference time $\alpha$, 
\eqn{
(\ri \pd_\alpha  - \hat{\ssf{H}}  ) \fixket{\Psi_\tx{ph} (\alpha)} = 0  ~,
\label{summary:Schrodinger_eq_with_gravity}}
with the Hermitian operator $\hat{\ssf{H}} $ corresponding to the energy of the rest of the Universe (\ref{summary:enrgy_of_universe_relative_to_clock}); hence, the internal-time unitarity follows.
The physical state $\dblket{\Psi_\tx{ph}}$ satisfying the condition in Eq.(\ref{summary:gravitational_state_condition_with_partial_reduction}) is reconstructed by $\dblket{\Psi_\tx{ph}} = \int_\mathbb{R} \rd \alpha \,  \fixket{\alpha} \otimes \fixket{\Psi_\tx{ph} (\alpha)}$.
 
\ 

Unlike Eq.(\ref{summary:Schrodinger_eq_with_nonHermitian_Hamiltonian}) in the fixed-background case, the Schr\"odinger equation (\ref{summary:Schrodinger_eq_with_gravity}) respects the LLI and the LPI.
This is guaranteed not only by the absence of direct coupling between the internal and external d.o.f. in $\sf C$ for the inertial observer but also by the unitarity of the internal-time evolution, 
which is achieved by diffeomorphism invariance that eliminates the unphysical indirect internal-external interaction.

\ 

Local observables are defined such that they leave the physical Hilbert space $\cl{ H}_\tx{ph}$ invariant.
They can be obtained as the quantum version of the group-averaging construction in Eq.(\ref{summary:classical_observable_relative_to_internal_time}) satisfying the stronger condition (\ref{summary:weak_physical_condition_with_gravity}), and solve the Heisenberg equation that describes the evolution with respect to the internal time parameter $a$ generated by the Hamiltonian $\hat{\ssf{H}}$.
Such operators have the crossed-product structure as those obtained in Refs.\cite{Chandrasekaran:2022cip,Witten:2023xze} for a simplified model with a nondynamical observer carrying a clock. 
 
\

In the present formulation, it is the coordinate choice that makes $\ssf{ C} $ linear in $H_\tx{C}$.
Now that the effective quantum field theory consistent with diffeomorphism invariance is assumed to exist,
the unitarity with respect to the internal time must also hold in the formulation without gauge-fixing.
In that case, although the linearity in $H_\tx{C}$ is absent in the resultant WDW equation, the unitarity with respect to the internal clock time as well as the EEP should emerge for diffeomorphism invariance.
It should be noted, however, that the quantization of canonical gravity involves a subtle difficulty intrinsic to the theory itself, irrespective of the observer's presence.
This point is touched upon in Sec.\ref{subsec:challenges}, together with another issue regarding dynamical observers, but a detailed analysis is left for future study.

\section{Relativistic particle with clock and detector \label{sec:relativistic_particle_with_clock_and_detector}}
In this section, we consider the model of a local observer as a relativistic particle\footnote{
With the worldline reparametrization as the diffeomorphism in the $(0+1)$-dimensional spacetime, the Hamiltonian formulation of the relativistic particle on a fixed curved spacetime offers a simplified yet insightful toy model for learning how the multiple-choice problem of time arises \cite{Kuchar:1991qf} in canonical quantum gravity.
The model is also useful for understanding another criticism raised by Kucha\v{r}, which we have not mentioned in Sec.\ref{sec:introduction}, concerning the Klein--Gordon (KG) analogy of the WDW equation.
This issue has been addressed in Ref.\cite{Hoehn:2020epv} by considering a covariant clock.
However, in the present work, we investigate it with the internal d.o.f., aiming to clarify the role of the local observer in a gravitational theory invariant under the diffeomorphism of the target $(d+1)$-dimensional spacetime.
}
that carries the internal d.o.f., namely the clock variable $\alpha$ and the matter-field detector variable $q$ on $(d+1)$-dimensional spacetime with a fixed background metric $g_{\mu \nu}(x)$; see Fig.\ref{fig:observer}.
For the absence of diffeomorphism invariance, the evolution with respect to the external coordinate time has a clear physical meaning, allowing a standard formulation of quantum theory from the external perspective; nevertheless, our focus is on the evolution with respect to the internal clock time.

\ 

In this section, we assume that the detector $q$ is coupled with a nondynamical scalar field $\phi(x)$ for simplicity.
The classical theory is given in Sec.\ref{subsec:classical_theory}.
The quantum theory is formulated from the external perspective in Sec.\ref{subsec:quantization}.
In that formulation, the evolution relative to the internal clock turns out to be nonunitary in Sec.\ref{subsec:evolution_relative_to_the_internal_clock}.
We briefly see that the same conclusion can be drawn with a dynamical scalar field in Appendix \ref{app:observers_with_dynamical_matter_field}.

\subsection{Classical theory \label{subsec:classical_theory}}
The external coordinate values of the observer's location in spacetime are denoted by $X^\mu = (T , \mb{X})$, and the $i$-th component of $\mb{X}$ is denoted by $\tx{X}^i$.
With $\eta$ parametrizing the worldline, the observer's action is given by
\eqn{
S_\tx{obs} = \int    \rd \eta \left[ p \frac{\rd q}{\rd \eta}  +  \Omega  \frac{\rd \alpha}{\rd \eta}   -  \upsilon (M + H_\tx{D} + H_\tx{C}) \right] ~.
\label{eq:observer_action_eta}}
where $\upsilon :=  ( - g_{\mu\nu}(X)  (\rd X^\mu / \rd \eta) (\rd X^\nu / \rd \eta) )^{1/2}$ with the metric evaluated at the observer's location $g_{\mu \nu} (X) = g_{\mu \nu}(T,\mb{X})$, expressed in terms of the lapse $N$, the shift $\bm{\beta}$ and the induced metric $h_{ij}$ on time slices as $g_{00} = -N^2 + h_{ij}\beta^i \beta^j$, $g_{0i} = h_{ij} \beta^j$ and $g_{ij} = h_{ij}$.
The canonical conjugate momenta of the internal degrees of freedom, $q$ and  $\alpha$, are denoted by $p$ and $\Omega$, respectively.
The mass of the relativistic particle itself is denoted by $M$.
The clock Hamiltonian is assumed to be the simplest one,
\eqn{
H_\tx{C} = \Omega ~. 
\label{eq:clock-Hamiltonian}}
To avoid the effective mass of the observer being negative, the clock momentum needs to be bounded from below. We set the lower bound to be zero: 
\eqn{
\Omega \geq 0 ~.
\label{eq:Omega>0}}
The detector interacts with the matter field at the observer's location $\phi (X) = \phi(T,\mb{X})$ through the Hamiltonian
\eqn{ 
H_\tx{D} = H_\tx{D,o}(p, q ) +  H_\tx{D,i} (p, q, \phi (X)) \geq 0 ~,
\label{eq:detector-Hamiltonian}}
whose explicit form is not necessary for the following discussion.
 
\ 

Additionally, we assume that the total Hamiltonian of the internal degrees of freedom remains sufficiently small that the internal energy is subdominant to the effective mass of the observer:
\eqn{  \zeta  := \frac{H_\tx{C} + H_\tx{D} }{M} \ll 1 ~. \label{eq:energy_of_internal_dof_is_subdominant}}
This assumption will be utilized later in Sec.\ref{subsubsec:Nonunitary evolution with respect to internal time}.

\ 

We note that, with the proper-time element along the worldline $\rd \tau = \upsilon \rd \eta$,  the action (\ref{eq:observer_action_eta}) can be expressed as
\eqn{
S_\tx{obs} = \int    \rd \tau \left[ p \frac{\rd q}{\rd \tau}  + \Omega \frac{\rd \alpha}{\rd \tau}   - (M + H_\tx{D} + H_\tx{C}) \right]  ~.
\label{eq:observer_action_tau}}
This expression provides a heuristic, yet intuitive derivation of $\rd \alpha /\rd \tau = 1$, i.e., the elapsed clock time is equal to the elapsed proper time of the observer, 
and thus, the clock time runs slow relative to the external coordinate time as
\eqn{
\rd \alpha  =  \rd T / \gamma   ~,
\label{eq:time_dilation}}
where $\gamma := \upsilon^{-1} (\rd T / \rd \eta) = (- g_{\mu \nu} V^\mu V^\mu)^{-1/2}$ is the Lorentz factor with $V^\mu:= \rd X^\mu /\rd T =( 1 , \rd \mb{X}/\rd T )$.
This type of action is assumed in \cite{Smith:2019imm}, with no matter-field detector in Minkowski spacetime, but with a general form of the clock Hamiltonian.
The discussion in the present work is applicable to cases with a general clock Hamiltonian as long as there is no direct interaction with other d.o.f., as assumed here.

\subsubsection{Constraints \label{subsubsec:constraints}} 
The action is invariant under reparametrization $\eta \to \eta' = f(\eta)$ of the worldline.
Consequently, the system is constrained so that the KG-type constraint function
\eqn{C^\tx{KG}  :=   P^2 + M_\tx{eff}^2
\label{eq:KG-type_constraint_function}}
vanishes, where the Hamiltonians of the d.o.f. contribute to the effective mass of the observer as
\eqn{
M_\tx{eff} := M  + H_\tx{D} + H_\tx{C} 
\label{eq:effective_mass}}
and the canonical conjugate to $X^\mu$ is denoted by $P_\mu = (P_0 , \mb{P})$.
Treating the zeroth component separately, we denote it by $E := - P_0$.
With $n^\mu := (1 , - \bm{\beta}  )/N $ being the unit vector normal to the time slices,
we have
\eqn{
P^2 := P_\mu P_\nu g^{\mu \nu}(X) = - \theta^2 + |\mb{P}|^2_h ~,
}
where 
\eqn{ 
\theta := -P_{\mu} n^\mu(X) ~, ~~~ |\mb{P}|^2_h := h^{ij}(X) \tx{P}_i \tx{P}_i ~.
}
Then, the KG-type constraint function (\ref{eq:KG-type_constraint_function}) is written as
\eqn{C^\tx{KG}  =   - \theta^2 + \omega^2 ~,
\label{eq:KG-type_constraint_function_with_K}}
which tells us that
\eqn{
\omega := ( |\mb{P}|^2_h  + M_\tx{eff}^2 )^{1/2} ~.
\label{eq:relativistic_particle_energy}}
corresponds to the particle's energy $\theta$ measured by the fiducial observer associated with the chosen foliation, whose $(d+1)$ velocity is given by $n^\mu (X)$.

\

This constraint function can be expressed as
\eqn{C^\tx{KG}
=    C^\tx{ex}_- C^\tx{ex}_+  =  C^\tx{in}_- C^\tx{in}_+
\label{eq:factorizations}
}
with
\eqn{
C^\tx{ex}_\pm :=   \mp  \theta   +  \omega  
\label{eq:constraint_function_for_external_perspective}}
and 
\eqn{
C^\tx{in}_\pm :=& M_\tx{eff} \mp (-P^2)^{1/2} ~. 
\label{eq:constraint_function_for_internal_perspective}}
The former is linear in $E$, while the latter is linear in $H_\tx{C}$ instead.
The effective mass $M_\tx{eff}$ is positive by definition, and $\theta$, as the particle's energy, is positive in an acceptable situation.
Therefore, physical configurations are realized on the intersection $\cl{I}$ of the constraint surfaces in the phase space defined by $C^\tx{in}_+ = 0$ and $C^\tx{ex}_+ = 0$,
or equivalently, by $H_\tx{C} = - H^\tx{in}$ and $E= H^\tx{ex}$, where
\eqn{
H^\tx{in} := M + H_\tx{D}   - (-P^2)^{1/2} ~
\label{eq:classical_Hamiltonian_from_internal_perspective}} 
and 
\eqn{
H^\tx{ex} := \omega N(X) - \tx{P}_i \beta^i (X) ~.
\label{eq:classical_Hamiltonian_from_external_perspective}}
 
\ 

The constraint function $C_+^\tx{p}$, where p is either ``ex'' or ``in,'' generates the corresponding gauge transformation, the active diffeomorphism in the present case.
Then, physical quantities $O_\tx{ph}$ are defined to be constant along the gauge orbit on the corresponding constraint surface in the phase space; in other words, $\{ O_\tx{ph} , C_+^\tx{p} \}_\tx{P} \weakeq{\tx{p}} 0$ with the weak equality $\weakeq{\tx{p}}$ indicating the equality when $C_+^\tx{p}=0$.
On the intersection $\cl{ I}_\tx{ph}$, the two constraint functions $C_+^\tx{ex}$ and $C_+^\tx{in}$ define the same set of physical quantities.\footnote{
The Poisson bracket with $C^\tx{KG}$ can be written as
\eqn{
\{ C^\tx{KG} , O \}_\tx{P} &=  \{ C^\tx{ex}_+ , O \}_\tx{P}  C^\tx{ex}_-   + \{  C^\tx{ex}_-   , O \}_\tx{P} C^\tx{ex}_+ \\
&=  \{ C^\tx{in}_+ , O \}_\tx{P} C^\tx{in}_- + \{ C^\tx{in}_- , O \}_\tx{P} C^\tx{in}_+ ~
}
for a given phase space function $O$, and hence, $\{ C^\tx{ex}_+ , O \}_\tx{P}  C^\tx{ex}_- /N^2$ is equal to $\{ C^\tx{in}_+ , O \}_\tx{P} C^\tx{in}_-$ on the intersection $\cl{ I}_\tx{ph}$.
Note that both $C^\tx{ex}_-$ and $C^\tx{in}_-$ are positive on $\cl{ I}_\tx{ph}$.
Therefore, $\{ C^\tx{ex}_+ , O \}_\tx{P} \weakeq{\tx{ex}} 0$ implies $\{ C^\tx{in}_+ , O \}_\tx{P} \weakeq{\tx{in}} 0$, and vice versa.
}

\subsubsection{Relational observables \label{subsubsec:classical_observable_relative_to_external_time}}
Here, let us take the external perspective with $C^\tx{ex}_+$.
It corresponds to taking the viewpoint of an external observer who sets up a coordinate system on spacetime.
The following construction of the physical quantities also works with $C^\tx{in}_+$.
As discussed above, it encodes the same dynamics, at least at the classical level. 

\ 

For a given kinematical ``bare'' function $O$ of the phase-space variables, its finite gauge transform $\cl{G}^\tx{ex}_O (\eta)$ with the gauge parameter $\eta$ described by $\pd_\eta \cl{G}^\tx{ex}_O = \{  \cl{G}^\tx{ex}_O, N  C^\tx{ex}_+ \}_\tx{P}$ with a boundary condition $\cl{G}_O^\tx{ex}(0) = O$ is given as \cite{Dittrich:2004cb,Dittrich:2005kc}
\eqn{
\cl{G}^\tx{ex}_O(\eta) = \sum_{n=0}^\infty \frac{\eta^n}{n !} \{  O , N C^\tx{ex}_+ \}_n ~, 
\label{eq:gauge_flow}}
where $\{ \cdot , \cdot \}_n$ is the nested Poisson bracket defined by $\{ f , g \}_{n+1} = \{ \{ f , g \}_n ,g \}_\tx{P}$ with $\{ f , g \}_0 = f$.
For $O=T$, it reads as $\cl{G}^\tx{ex}_T(\eta) = T+ \eta$.

Introducing an ``evaluation'' time $t$ and solving $\cl{G}_T^\tx{ex}(\eta) = t$ with respect to $\eta$, we get $\eta(t ; T) = t - T$ with $T$ being a ``reference'' time.
Plugging $\eta = \eta(t ; T)$ in Eq.(\ref{eq:gauge_flow}), we obtain a physical ``dressed'' quantity, in the form of the so-called complete observable \cite{Rovelli:1990ph,Rovelli:2001bz}, Dirac relational observable or gauge-invariant extension of a gauge-fixed quantity \cite{henneaux1992quantization},
\eqn{
O_{\tx{ph}}(t ; T ) := \cl{G}_O^\tx{ex}(\eta(t ; T)) ~,
\label{eq:plugging_eta(t;T)}}
whose Poisson bracket with $C^\tx{ex}_+$ vanishes algebraically.\footnote{\label{fn:Liouville}
Equivalently, as a function of $T$, it satisfies 
\eqn{
\pd_T O_{\tx{ph}}(t ; T ) = - \{ O_{\tx{ph}}(t ; T ) , H^\tx{ex}(T) \}_\tx{P} ~,
\label{eq:Liouville}}
which is analogous to the Liouville equation.
We have explicitly shown the $T$ dependence of $H^\tx{ex}$ with its argument.
If we take an ``initial'' phase-space density $\rho_0$ replacing $O$, then $\rho_{\tx{ph}}(T) := \rho_{0, \tx{ph}}(t_0 ; T )$ is the push forward of $\rho_0$ from the initial time $T_0$ to the reference time $T$.
}

\

Although the above construction may seem unfamiliar and somewhat cumbersome, $O_{\tx{ph}}(t ; T ) $ is nothing more than the pull back to the reference time $T$ along the Hamiltonian flow of the observable evaluated at the time $t$, satisfying
\eqn{
\pd_t O_{\tx{ph}}(t ; T )  =& +\{ O_{\tx{ph}}(t ; T ) , H^\tx{ex}_{\tx{ph}}(t ; T ) \}_\tx{P}  + \pd_\tx{xpl} O_{\tx{ph}}(t ; T )  
\label{eq:physical_evolution_relative_to_external_time_on_fixed_background}}
with the boundary condition $O_{\tx{ph}}(t=T ; T ) = O$, where $H^\tx{ex}_{\tx{ph}}(t ; T )$ is the physical Hamiltonian obtained by replacing $O$ with $H^\tx{ex}$ in Eq.(\ref{eq:plugging_eta(t;T)}) and $\pd_\tx{xpl}$ denotes the derivative acting only on the ``explicit'' time dependence coming from the $T$ dependence of the bare quantity $O$ if any, that is, $\pd_\tx{xpl} O_{\tx{ph}}(t ; T ) := \cl{ G}^\tx{ex}_{\pd_T O} (\eta (t;T))$.

\ 

We note that the relational observables can also be written in the group-averaging form as
\eqn{
O_{\tx{ph}}(t ; T ) = \int_\mathbb{R} \rd \eta \left[ \sum_{n=0}^\infty \frac{\eta^n}{n !} \{  \delta (t - T) \, O , N C^\tx{ex}_+ \}_n  \right] ~. 
\label{eq:classical_observable_relative_to_external_time_on_fixed_background}}
Expressed in this form, observables are directly promoted to their corresponding physical operators in quantum theory; see Appendix \ref{appsubsec:physical_state_and_operator}.

\subsection{Quantum theory from the external perspective \label{subsec:quantization}}
Now, the system is quantized with the standard procedure; the phase space variables are promoted to the corresponding operators denoted with hats, and the Poisson bracket is replaced by the commutator $[\cdot , \cdot] \times (-\ri )$.  

\ 

While the two constraint functions $C^\tx{ex}_+$ and $C^\tx{in}_+$ at the classical level result in mutually equivalent theories as noted at the end of Sec.\ref{subsubsec:constraints}, that will be no longer the case at the quantum level as discussed in Sec.\ref{subsec:qauntum_theory_from_internal_pespective?}.
Here, we work with the constraint operator corresponding to $C^\tx{ex}_+$, not only because the external coordinate time is the standard choice on a fixed background, but also because the quantum theory based on $C^\tx{in}_+$ requires an \textit{ad hoc} prescription; see Sec.\ref{subsubsec:possible_formulation_of_internal_perspective}.

\subsubsection{Quantized as constrained system \label{subsubsec:Quantized as a constrained system}} 
Let $\cl{H}_\tx{IN} =  \cl{H}_\tx{C} \otimes  \cl{H}_\tx{D}$ be the Hilbert space associated with the internal d.o.f., where the clock Hilbert space $\cl{H}_\tx{C}  \simeq L^2 (\mathbb{R}_+)$ is spanned by the eigenstates of the clock Hamiltonian $\hat{H}_\tx{C}$ corresponding to the classical one (\ref{eq:clock-Hamiltonian}) and the detector Hilbert space $\cl{H}_\tx{D}$ is spanned by the eigenstates of the free part $\hat{H}_\tx{D,o}$ of the detector Hamiltonian (\ref{eq:detector-Hamiltonian}), which is assumed to have a discrete spectrum such as that of a harmonic oscillator, for definiteness. 
The Hilbert space of the external d.o.f. is naturally endowed with an inner product involving the measure $|g|^{1/2}_{X} := N(X) h^{1/2}(X)$ with $h$ being the determinant of the induced metric $h_{ij}$.
For simplicity, we ``flatten'' the measure by absorbing the factor $|g|^{1/4}_{X}$ into the basis so that the Hilbert space can be factorized as $\cl{H}_\tx{EX} =  \cl{H}_\tx{T} \otimes  \cl{H}_\tx{X}$,
where $\cl{H}_\tx{T} \simeq L^2 (\mathbb{R})$ and $\cl{H}_\tx{X}$ are those for the observer's external time and position coordinates. 
See Appendix \ref{appsubsec:without_flattening} for expressions in terms of the original basis before the flattening.

\ 

The total kinematical Hilbert space is given by $\cl{H}_\tx{kin} = \cl{H}_\tx{IN} \otimes \cl{H}_\tx{EX}$ whose elements are written using the double bra/ket with a subscript ``kin,'' such as $\dblket{\psi_\tx{kin}}$.
The physical Hilbert space $\cl{H}_\tx{ph}^\tx{ex}$ is obtained as the projection of $\cl{H}_\tx{kin}$ consisting of the states $\dblket{\psi_\tx{ph}}$ that are annihilated by the operator corresponding to the classical constraint function (\ref{eq:constraint_function_for_external_perspective});
\eqn{
\hat{C}_+^\tx{ex} \dblket{\psi^\tx{ex}_\tx{ph}} = 0 
\label{eq:physical_state_condition_for_external_perspective}}
The precise definition of the constraint operator is provided in Eq.(\ref{eq:constraint_operator_for_external_perspective}) in Sec.\ref{subsubsec:Internal-External interaction via fields}, while its details are not crucial to our main argument.

As a consequence, the physical Hilbert space $\cl{H}_\tx{ph}^\tx{ex}$ is no longer factorizable:
when a physical state is expanded, for example, as
\eqn{
\dblket{\psi^\tx{ex}_\tx{ph}} =& \int_\mathbb{R}  \rd T  \int  \rd^d \tx{X}   \fixket{T} \otimes \fixket{\mb{X}} \otimes \fixket{\psi_\tx{ph}^\tx{ex}(T ; \mb{X})}  \\
=& \sum_n \int_{\mathbb{R}_+ } \frac{\rd \Omega}{2 \pi} \int_\mathbb{R} \frac{\rd E}{2 \pi} \int \frac{\rd^d \tx{P}}{(2 \pi)^d}  \\ &\hspace{1.5em}  \psi_\tx{ph}^\tx{ex}(\Omega; n ; E ; \mb{P}) \fixket{\Omega} \otimes \fixket{n} \otimes \fixket{E} \otimes \fixket{\mb{P}}  ~,
}
the physical wave functions $\fixket{\psi_\tx{ph}^\tx{ex}(T ; \mb{X})} :=  \fixbra{\mb{X}} \bkk{T}{\psi^\tx{ex}_\tx{ph}}$ $\in \cl{H}_\tx{IN}$ and $\psi^\tx{ex}_\tx{ph}(\Omega; n ; E ; \mb{P}) := \fixbra{\Omega} \fixbra{n}  \fixbra{E} \bkk{\mb{P}}{\psi^\tx{ex}_\tx{ph}}$ are not arbitrary, where $\fixket{T}$ and $\fixket{\mb{X}}:= \bigotimes_{j = 1,\cdots , d} \fixket{\tx{X}^j}$ denote the eigenstates of $\hat{T}$ and $\hat{\mb{X}}$, respectively, and $\fixket{E}$ and $\fixket{\mb{P}}$ are defined analogously as the eigenstates of $\hat{E}$ and $\hat{\mb{P}}$.
For the internal d.o.f., the eigenstates of $\hat{H}_\tx{C} = \hat{\Omega}$ and $\hat{H}_\tx{D,o}$ are denoted by $\fixket{\Omega}$ and $\fixket{n}$, respectively.

\ 

Physical observables $\hat{O}_\tx{ph}$ are defined as operators that commute with the  constraint operator: $[\hat{C}_+^\tx{ex} , \hat{O}_\tx{ph} ] \weakeq{\tx{ex}} 0$,
where the weak equality at the quantum level means that the equality holds on the physical Hilbert space $\cl{H}_\tx{ph}^\tx{ex}$ defined by Eq.(\ref{eq:physical_state_condition_for_external_perspective}),
so that $\cl{H}_\tx{ph}^\tx{ex}$ is invariant under the action of them.
As discussed in Appendix \ref{appsec:Standard_quantum_mechanics_from_group_averaging},
such operators can be obtained from the group-averaging form (\ref{eq:classical_observable_relative_to_external_time_on_fixed_background}) of the classical observables by replacing the phase-space variables with the corresponding operators.
The physical observables so constructed are essentially the Heisenberg operators satisfying the quantum analog of the evolution equation (\ref{eq:physical_evolution_relative_to_external_time_on_fixed_background}).

\subsubsection{Internal-external interaction via fields \label{subsubsec:Internal-External interaction via fields}}
The internal d.o.f. consisting of the clock and the detector couple to the external d.o.f., namely the observer's spacetime coordinate, through the nondynamical field\footnote{
When the matter field is dynamical as in Appendix \ref{app:observers_with_dynamical_matter_field}, it is also a part of the external d.o.f.
}
in a ``direct'' and/or ``indirect'' manner.

\ 

In this work, the direct interaction refers to the coupling between the detector and the external d.o.f. due to the presence of $\hat{H}_\tx{D,i}$ in the effective mass (\ref{eq:effective_mass}),
now promoted to an operator as
\eqn{
\hat{M}_\tx{eff} = M + \hat{H}_\tx{C} + H_\tx{D,o}(\hat{p},\hat{q}) + H_\tx{D,i}(\hat{p},\hat{q}, \phi (\hat{X})) ~, 
\label{eq:effective_mass_operator}}
where
\eqn{
\phi (\hat{X})    = \int_{\mathbb{R}} \rd T \int \rd^d  \tx{X}  \,  \phi (T,\mb{X}) \,  \projector{T} \otimes \projector{\mb{X}} 
\label{eq:field_value_at_observer's_location}}
is the value of the matter field, which is nondynamical for the moment,
evaluated at the observer's spacetime location described by the operator $\hat{X}^\mu = (\hat{T}, \hat{\mb{X}})$, and $\projector{T}$ and $\projector{\mb{X}} := \bigotimes_{j = 1,\cdots , d} \projector{\tx{X}^j}$ are the projection operators onto the eigenstates of $\hat{T}$ and $\hat{\mb{X}}$, respectively.
In the following, $N(\hat{X})$, $\bm{\beta}(\hat{X})$, and $h_{ij}(\hat{X})$ are to be understood analogously.

\ 

The indirect interaction refers to any couplings that persist when the matter-field detector is turned off, namely with $\hat{H}_\tx{D,i} =0$, and comes from the relativistic dependence of the effective energy (\ref{eq:relativistic_particle_energy}) of the particle on its effective mass.
For definiteness, we define the corresponding operator 
\eqn{
\hat{\omega} =   ( |\hat{\mb{P}}|_h^2  + \hat{M}_\tx{eff}^2 )^{1/2} 
\label{eq:relativistic_particle_energy_operator}
}
by the power series expansion around $M$.\footnote{\label{fn:square_root}
The square root of $M + \hat{Z}$ is expanded as
\eqn{
(M^2 + \hat{Z})^{1/2} = - M \sum_{n=0}^{\infty} \frac{(2n -3)!!}{n !}\left(\frac{ - \hat{Z} }{2 M^2} \right)^n ~.
}
For the operator $\hat{\omega}$, we have $\hat{Z} = |\hat{\mb{P}}|_h^2  + ( \hat{M}_\tx{eff}^2 - M^2)$.
}
It gives rise to terms such as $\hat{\zeta} \otimes |\hat{\mb{P}}|_h^2$ and $\hat{\zeta}^2$, where $\hat{\zeta}$ is the operator corresponding to the internal Hamiltonian (\ref{eq:energy_of_internal_dof_is_subdominant}) and
\eqn{
|\hat{\mb{P}}|^2_h :=  h^{-1/4}(\hat{X}) \hat{\tx{P}}_i h^{1/2}(\hat{X}) h^{ij}(\hat{X})  \hat{\tx{P}}_j  h^{-1/4}(\hat{X})
\label{eq:momentum-squared_operator}}
with $h:= \det (h_{ij})$ the determinant of the induced metric.
The operator ordering is fixed so that it gives the $d$-dimensional Laplace–Beltrami operator on the time slices in the $X$ representation where $\hat{X}$ is diagonalized.
 
\ 

On the other hand, the first term in the constraint operator (\ref{eq:physical_state_condition_for_external_perspective}) is purely external:
\eqn{
\hat{\theta} := \hat{N}^{-1/2} (\hat{E} + \{ \hat{P}_i , \beta^i (\hat{X}) \}/2 ) \hat{N}^{-1/2} ~,
\label{eq:theta_with_anticommutator}}
where $\{ \cdot , \cdot \}$ denotes the anticommutator.
With this ordering, it yields a compact form of the differential operator in the $X$ representation, expressed in terms of the spacetime covariant derivative and the unit normal to the time slices.\footnote{
Owing to the definitions given in Eqs.(\ref{eq:momentum-squared_operator}) and (\ref{eq:theta_with_anticommutator}), quantum theories formulated in two different frames are equivalent to each other, as long as the functional forms of the fixed spacetime metric are interconnected via an active foliation-preserving diffeomorphism transformation \cite{Kuchar:1991qf}.
}

\ 

To summarize, the constraint function (\ref{eq:constraint_operator_for_external_perspective}) is written as
\eqn{
\hat{C}_+^\tx{ex} &= - \hat{\theta} + \hat{\omega} \\
&=\hat{N}^{-1/2} ( - \hat{E} + \hat{H}^\tx{ex}  ) \hat{N}^{-1/2} ~,
\label{eq:constraint_operator_for_external_perspective}}
where
\eqn{
\hat{H}^\tx{ex} :=  - \{ \hat{P}_i , \beta^i (\hat{X}) \}/2  + \hat{N}^{1/2} \hat{\omega} \hat{N}^{1/2} ~
\label{eq:Hamiltonian_operator_from_external_perspective}}
is assumed to be self-adjoint. 
The direct and indirect interactions are induced by $\hat{\omega}$ given in Eq.(\ref{eq:relativistic_particle_energy_operator}) and, as will be explained in Secs.\ref{subsec:evolution_relative_to_the_internal_clock} and \ref{subsec:violation_of_LLI/LPI}, they give rise to the nonunitarity of the evolution relative to the internal clock, and consequently, the violation of the EEP.

In the standard picture where the evolution is described with respect to the external coordinate time, the role of the Hamiltonian is played by $\hat{H}^\tx{ex}$ given in Eq.(\ref{eq:Hamiltonian_operator_from_external_perspective}).
Then, the internal-external interactions are described by its second term, in which there is another source of interaction due to the $X$ dependence of the lapse function $N$.
However, we will see that it contributes to the internal-time nonunitarity only through the indirect interaction due to the square root in the definition of $\hat{\omega}$, by our choice of the inner products for the $T$ representation in Eq.(\ref{eq:physical_inner_product}) and for the $\alpha$ representation in Eq.(\ref{eq:inner_product_for_alpha-rep}).

\subsubsection{Internal-external decoupling \label{subsubsec:internal-external_decoupling}}
There are special cases where the square root in Eq.(\ref{eq:relativistic_particle_energy_operator}) does not cause the indirect interaction between the internal and external d.o.f.
Consider an ultrastatic spacetime that accommodates a hypersurface-orthogonal Killing vector of constant norm.
Adopting the rest frame of the corresponding Killing observer in which $N = 1$, $\bm{\beta} = \mb{0}$, and the spatial metric $h_{\tx{st},ij}$ is time independent,
one finds that the internal and external d.o.f. are decoupled since the relation in Eq.(\ref{eq:factorizations}) still holds at the quantum level: the four operators, 
\eqn{
 \hat{C}^\tx{in}_{\pm} := \hat{M}_\tx{eff} \mp (- \hat{P}^2)^{1/2} ~,
\label{eq:constraint_operator_for_internal_perspective_pm}}
\eqn{\hat{C}_-^\tx{ex} :=  + \hat{\theta}  +  \hat{\omega}   ~,
\label{eq:constraint_operator_for_external_perspective_-}}
and $\hat{C}_+^\tx{ex}$ given in Eq.(\ref{eq:constraint_operator_for_external_perspective}), reduce to
\eqn{
\hat{c}^\tx{in}_{\pm} = \hat{M}_\tx{eff,o} \mp (\hat{E}^2 - |\mb{P}|_\tx{st}^2 )^{1/2} ~,
\label{eq:constraint_operator_for_internal_perspective_pm_ultrastatic}}
\eqn{
\hat{c}_{\pm  }^\tx{ex } := \mp \hat{E} + (\hat{M}_\tx{eff,o}^2 + |\mb{P}|_\tx{st}^2  )^{1/2}  ~,
\label{eq:constraint_operator_for_external_perspective_pm_ultrastatic}}
where $\hat{M}_\tx{eff,o} := M + \hat{H}_\tx{C} +  \hat{H}_\tx{D,o}$ is the effective mass with $\hat{H}_\tx{D} = \hat{H}_\tx{D,o}(p,q)$ and the squared norm $|\hat{\mb{P}}|_\tx{st}^2$ is given by Eq.(\ref{eq:momentum-squared_operator}) with $h^{ij} = h_\tx{st}^{ij}$, and thus satisfy  
\eqn{
 \hat{c}_{-  }^\tx{ex } \hat{c}_{+  }^\tx{ex }   = \hat{c}^\tx{in  }_{- } \hat{c}^\tx{in }_{+ }  
}
thanks to $[ \hat{c}_{-  }^\tx{in } , \hat{c}_{+  }^\tx{in } ] = 0$ and $[ \hat{c}_{-  }^\tx{ex } , \hat{c}_{+  }^\tx{ex } ] = 0$.
As a consequence, the physical states defined by $\hat{c}_{+  }^\tx{ex } \dblket{\psi_\tx{ph}^\tx{ex}} = 0$ necessarily satisfy $\hat{c}_{+  }^\tx{in } \dblket{\psi_\tx{ph}^\tx{ex}} = 0$ since $\hat{c}_{-  }^\tx{in} > 0$ on $\cl{H}_\tx{ph}^\tx{ex}$.
The internal and external d.o.f. are decoupled in the sense that, while they are still related via the constraint, there is no cross term between them in $\hat{c}^\tx{in }_{+ }$.

\ 

Alternatively, without any restriction on the background spacetime, one can consider the nonrelativistic expansion up to the leading order of the kinetic contribution,
\eqn{
\hat{\omega} \simeq \hat{M}_\tx{eff} + \frac{|\hat{\mb{P}}|_h^2}{2M} ~,
\label{eq:nonrelativistic_expansion}}
with higher orders in $\hat{\zeta}$ dropped; then the internal d.o.f. decouple from the external d.o.f.
If the momentum's contributions are further neglected in this expression and also in Eq.(\ref{eq:theta_with_anticommutator}), it is essentially equivalent to considering a nondynamical observer that carries the internal d.o.f. on a worldline given by $\mb{x}=\tx{const}$ as in the case of the Unruh-DeWitt (UDW) detector \cite{Unruh:1976db,DeWitt1979}.

\subsubsection{External-time unitarity \label{subsubsec:external-time_unitarity}} 
Projected to an eigenstate $\fixket{T}$ of $\hat{T}$, the physical state condition (\ref{eq:physical_state_condition_for_external_perspective}) is nothing but the Schr\"odinger equation 
\eqn{
(  \ri \pd_T -  \hat{H}^\tx{ex}(T) ) \fixket{\tilde{\psi}^\tx{ex}_\tx{ph} (T)} = 0 ~
\label{eq:Schrodinger_eq_in_T-rep}}
with respect to the reference time $T$,
where 
\eqn{
\fixket{\tilde{\psi}^\tx{ex}_\tx{ph} (T)} := \fixbra{T} \hat{N}^{-1/2} \dblket{\psi^\tx{ex}_\tx{ph} } 
\label{eq:T-representation_of_physical_state_tilde_from_external_perspective}}
is an element of $\cl{H}_{\overline{\tx{T}}} := \cl{H}_\tx{IN} \otimes \cl{H}_\tx{X}$ and $\hat{H}^\tx{ex} (T)$ is the representation of the operator (\ref{eq:Hamiltonian_operator_from_external_perspective}) with respect to the basis $\{ \fixket{T} \}$.
Then, the state $\fixket{\tilde{\psi}^\tx{ex}_\tx{ph} (T)}$ evolves unitarily
with respect to the external reference time $T$ as
\eqn{
 \fixket{\tilde{\psi}^\tx{ex}_\tx{ph} (T)} = \hat{U}^\tx{ex}(T,T_0) \fixket{\tilde{\psi}^\tx{ex}_0}
\label{eq:solution_of_Schrodinger_eq_w.r.t_T}}
with the evolution operator 
\eqn{
\hat{U}^\tx{ex}(T,T') := \cl{T} e^{- \ri  \int^{T}_{T'} \rd \eta \hat{H}^\tx{ex}(\eta) } ~,
\label{eq:unitary_evolution_operator}}
where $\cl{T}$ denotes the time ordering in terms of the external time
and $\fixket{\tilde{\psi}^\tx{ex}_\tx{0}} $ is an arbitrary ``initial'' state at $T=T_0$, normalized as $\fixbraket{\tilde{\psi}^\tx{ex}_\tx{0}}{\tilde{\psi}^\tx{ex}_\tx{0}}  = 1$.
It is when $N=1$ that $\fixket{\psi^\tx{ex}_\tx{ph} (T)} := \bkk{T}{\psi^\tx{ex}_\tx{ph} }$ coincides with the state (\ref{eq:T-representation_of_physical_state_tilde_from_external_perspective}) and solves the Schr\"odinger equation (\ref{eq:Schrodinger_eq_in_T-rep}) or (\ref{summary:Schrodinger_eq_in_T-rep}) in the absence of the shift vector.
 
\

The physical state as an element of $\cl{H}_\tx{ph}^\tx{ex}$ that satisfies Eq.(\ref{eq:physical_state_condition_for_external_perspective}) can be reconstructed as
\eqn{
\dblket{\psi^\tx{ex}_\tx{ph}} &= \hat{N}^{1/2} \int_{\mathbb{R}} \rd T \fixket{T} \otimes  \fixket{\tilde{\psi}^\tx{ex}_\tx{ph} (T)} ~,
\label{eq:reconstruction_for_external_perspective}}
taking the form of the so-called history state.
Then, the unitarity of the evolution with respect to the external time is seen as the $T$-independence of the inner product between two different states $\dblket{\psi^\tx{ex}_\tx{ph}}$ and $\dblket{\chi^\tx{ex}_\tx{ph}}$,
\eqn{
\dblbra{\chi^\tx{ex}_\tx{ph}}  (\hat{\tilde{\Pi}}_\tx{EX}(T) \otimes \hat{I}_\tx{IN})  \dblket{\psi^\tx{ex}_\tx{ph}} = \fixbraket{\tilde{\chi}^\tx{ex}_\tx{0}}{\tilde{\psi}^\tx{ex}_\tx{0}} ~,
\label{eq:physical_inner_product}}
where $\hat{\tilde{\Pi}}_\tx{EX}(T) := \hat{N}^{-1/2} (\projector{T} \otimes \hat{I}_\tx{X})\hat{N}^{-1/2}$.
This inner product corresponds to the physical inner product defined in Eq.(\ref{app:physical_inner_product}) with the group-averaging technique.
We note that it has a more natural expression (\ref{app:inner_product_with_proper_volume_element}) when written in terms of the original basis of $\cl{H}_\tx{EX}$ before the flattening; see Appendix \ref{appsubsec:without_flattening}.  

\ 

If one replaces the identity $\hat{I}_\tx{IN}$ with an operator $\hat{O}_\tx{IN}$ nontrivially acting on $\cl{ H}_\tx{IN}$ in Eq.(\ref{eq:physical_inner_product}) and sandwiches it between identical physical states, then the expectation value of $\hat{O}_\tx{IN}$ is obtained. 
For example, with $\hat{O}_\tx{IN} = \projector{n}$, the projector to a detector eigenstate, one obtains
\eqn{ 
\dblbra{\psi^\tx{ex}_\tx{ph}} ( \hat{\tilde{\Pi}}_\tx{EX}(T) \otimes \projector{n} )   \dblket{\psi^\tx{ex}_\tx{ph}} 
= \| \fixbraket{n}{\tilde{\psi}^\tx{ex}_\tx{ph} ( T )}   \|^2 ~,
\label{eq:P(n|T)}}
which is a PW-type conditional probability given as in Eq.(\ref{eq:PW_conditional-probability}) with the trivial denominator for the unitarity.

\subsection{Evolution relative to internal clock and its nonunitarity
\label{subsec:evolution_relative_to_the_internal_clock}}
Within the quantum theory constructed from the external perspective,
we discuss the evolution of the system relative to the internal clock, following the PW formalism. 

\

Let us consider a physical state that solves the constraint (\ref{eq:physical_state_condition_for_external_perspective}) by taking the projection to a clock state $\fixket{\alpha}$ defined in the following subsection,
\eqn{
\fixket{\psi_\tx{ph}^\tx{ex} (\alpha)} := \bkk{\alpha}{\psi_\tx{ph}^\tx{ex}}  ~,
\label{eq:alpha-representation_of_physical_state_from_external_perspective}}
and discuss how it depends on the value of $\alpha$, i.e., its evolution with respect to the internal reference time.
In the following, it is referred to as the ``$\alpha$ representation.''

\ 

As the inner product, we consider
\eqn{
&\fixbraket{\chi^\tx{ex}_\tx{ph}(\alpha)}{\psi^\tx{ex}_\tx{ph}(\alpha)} = \dblbra{\chi^\tx{ex}_\tx{ph}} ( \hat{ I }_\tx{EX} \otimes \projector{\alpha} ) 
\dblket{\psi^\tx{ex}_\tx{ph}} \\
&\hspace{3em} = \int_\mathbb{R} \rd T \dblbra{\chi^\tx{ex}_\tx{ph}} ( \hat{ \Pi }_\tx{EX}(T) \otimes \projector{\alpha} )   \dblket{\psi^\tx{ex}_\tx{ph}} ~.
\label{eq:inner_product_for_alpha-rep}}
Using $\hat{\Pi}_\tx{EX}(T) :=  \hat{N}^{1/2}  \hat{\tilde{\Pi}}_\tx{EX}(T) \hat{N}^{1/2} =   \projector{T} \otimes \hat{I}_\tx{X}$ on the second line, we have expressed it in an alternative form to be compared with the one for the $T$ representation in Eq.(\ref{eq:physical_inner_product}).
As seen in Appendix \ref{appsubsec:without_flattening},
with the original basis before the flattening, this inner product is expressed by the $(d+1)$-dimensional integration with the proper spacetime volume element.

\subsubsection{Clock states \label{subsubsec:clock_states}}
The clock variable is promoted to an operator satisfying the covariance condition \cite{Holevo:1414149}, namely, transforming covariantly under translations generated by the clock Hamiltonian $\hat{H}_\tx{C} = \hat{\Omega}$:
\eqn{
e^{-\ri  \alpha' \hat{H}_\tx{C} }\hat{\alpha} e^{+\ri  \alpha' \hat{H}_\tx{C} } = \hat{\alpha} - \alpha' \hat{I}_\tx{C} ~,
\label{eq:covarinace_condition}}
which formally satisfies $[\hat{\alpha} ,  \hat{H}_\tx{C}  ] = \ri$.
With a bounded spectrum of the clock Hamiltonian, such an operator is known to be symmetric but not self-adjoint \cite{pauli2012}.
It can be constructed as $\hat{\alpha} =  \int_{\mathbb{R}} \rd \alpha  \fixket{\alpha} \alpha \fixbra{\alpha}$
based on the clock states
\eqn{
\fixket{\alpha} := \int_{\mathbb{R}_+ } \frac{\rd \Omega }{2 \pi} e^{- \ri \alpha \Omega} \fixket{\Omega} ~,
\label{eq:clock_state}}
where $\fixket{\Omega}$ denotes an eigenstate of the clock Hamiltonian with eigenvalue $\Omega$.
In the $\alpha$ representation, the clock Hamiltonian turns into $-\ri \pd_\alpha$ since $\hat{H}_\tx{C} \fixket{\alpha} = + \ri \pd_\alpha \fixket{\alpha}$.

\ 

Although two clock states are not orthogonal, the set of $\fixket{\alpha}$ with $\alpha \in \mathbb{R}$ resolves the identity on $\cl{H}_\tx{C}$ as $\hat{I}_\tx{C} =  \int_{\mathbb{R}} \rd \alpha  \projector{\alpha}$; therefore, the clock states form a positive operator-valued measure (POVM). 
One of the POVM elements corresponding to a clock reading $\alpha$ is denoted by
\eqn{
\hat{\Pi} (\alpha) := \projector{\alpha}  ~.
\label{eq:clock_POVM}}
as if it were a projector.

\

Note that, based on the belief that gravity couples to everything, we do not introduce von Neumann's measurement process from outside.
Then, with the assumption that the clock variable is decoupled from the other internal variable, the value of $\alpha$ is never read off directly. 
Nevertheless, it can serve as a reference time to define the relational dynamics with the PW mechanism.

\subsubsection{Evolution relative to internal clock without internal-external interaction \label{subsubsec:evolution_w.r.t_internal_time_w/o_observer-field_interaction}}
We begin with the trivial case with the observer on an ultrastatic spacetime with the matter-field detector turned off.
As discussed in Sec.\ref{subsubsec:internal-external_decoupling}, the physical states satisfy $\hat{c}_{+  }^\tx{in } \dblket{\psi_\tx{ph}^\tx{ex}} = 0$, which reads, in the $\alpha$ representation, as
\eqn{
(\ri \pd_\alpha -\hat{H}^\tx{in}_\tx{st,o} )  \fixket{\psi_\tx{ph}^\tx{ex} (\alpha)} = 0 ~,
\label{eq:Schrodinger_eq_in_alpha-rep}}
where $\hat{H}^\tx{in}_\tx{st,o} :=  \hat{H}_\tx{D,o} + M -( \hat{E}^2 - |\hat{\mb{P}}|_\tx{st}^2 )^{1/2}   $.
Since the physical states are originally defined by $\hat{c}_{+  }^\tx{ex} \dblket{\psi_\tx{ph}^\tx{ex}} = 0$, the operator $\hat{E}^2 - |\hat{\mb{P}}|_\tx{st}^2$ in the square root is positive on $\fixket{\psi_\tx{ph}^\tx{ex} (\alpha)}$; hence, $\hat{H}^\tx{in}_\tx{st,o}$ is Hermitian and the Schr\"odinger equation (\ref{eq:Schrodinger_eq_in_alpha-rep}) tells us that $\fixket{\psi_\tx{ph}^\tx{ex} (\alpha)}$ evolves unitarily with respect to the internal time.

\

In this case, the quantum state (\ref{eq:T-representation_of_physical_state_tilde_from_external_perspective}) is a certain superposition of simultaneous eigenstates of $\hat{H}_\tx{C}$, $\hat{H}_\tx{D,o}$ and $|\hat{\mb{P}}|_\tx{st}$, denoted by $\fixket{\Omega} \fixket{n} \fixket{\cl{P}}$.
The eigenvalues $\Omega$ and $\epsilon_n$ of the clock and detector Hamiltonians distinguish the effective mass $M_\tx{eff}(\Omega ; n) := M + \Omega + \epsilon_n$, whereas the eigenvalues $\cl P$ of $|\hat{\mb{P}}|_\tx{st}$ determine the spatial behavior of the wave function.
From the expression (\ref{eq:reconstruction_for_external_perspective}), we obtain 
\eqn{
&\fixbra{n}\fixbra{\cl{P}}\fixbraket{T}{\psi_\tx{ph}^\tx{ex} (\alpha)} \\
&= \int_{\mathbb{R}_+} \frac{\rd \Omega}{2 \pi}   e^{\ri \alpha \Omega - \ri \omega(\Omega , n , \cl{P}) (T-T_0) }    \psi_{0}^\tx{ex}(\Omega; n ; \cl{P}) 
\label{eq:wavefunction(T,alpha)}
}
with $\psi_{0}^\tx{ex}(\Omega; n ; \cl{P}) := \fixbra{n}\fixbraket{\cl{P}}{\tilde{\psi}_{0}^\tx{ex}}$ and the eigenvalue of external-time Hamiltonian $\omega(\Omega , n , \cl{P}) := (M_\tx{eff}^2(\Omega ; n) + \cl{ P}^2 )^{1/2}$. 

\ 

To see the correlation between the internal clock time and the external time, one may compute\footnote{
Now that the internal-external interaction is absent, $P(T| \alpha )$, given in Eq.(\ref{eq:P(T|alpha)}) is equivalent to $P(\alpha | T)$ obtained by replacing $\projector{n}$ with $\hat{\Pi} (\alpha)$ in Eq.(\ref{eq:P(n|T)}).
} 
\eqn{
\frac{\dblbra{\psi^\tx{ex}_\tx{ph}} ( \projector{T}   \otimes \hat{\Pi}(\alpha)  ) \dblket{\psi^\tx{ex}_\tx{ph}}}{\dblbra{\psi^\tx{ex}_\tx{ph}}   \hat{\Pi}(\alpha)   \dblket{\psi^\tx{ex}_\tx{ph}}} = \frac{\| \fixbraket{T}{\psi^\tx{ex}_\tx{ph} ( \alpha )}   \|^2}{\| \fixket{\psi^\tx{ex}_\tx{ph} ( \alpha )}   \|^2} ~,
\label{eq:P(T|alpha)}}
which is a conditional probability $P(T | \alpha)$ in the PW sense.
If $\psi_{0}^\tx{ex}(\Omega; n ; \cl{P})$ in Eq.(\ref{eq:wavefunction(T,alpha)}) is sufficiently localized around a certain value $\bar{\Omega}(n ; \cl{P})$ that the expansion of $\omega(\Omega , n , \cl{P})$ in terms of $\Omega - \bar{\Omega}$ is justifiable, then the probability distribution (\ref{eq:P(T|alpha)}) depends on the internal and external times only through the combination $T / \bar{\gamma} - \alpha$ and the relation (\ref{eq:time_dilation}) holds up to the quantum uncertainty\footnote{
The authors of Ref.\cite{Smith:2019imm} discussed a correlation between two different clocks carried by different relativistic particles at the leading order of the relativistic effect as well as the Helstrom-Holevo bound for proper-time estimation \cite{Holevo:1414149}.
}
with the inverse of the Lorentz factor
\eqn{
1/\bar{\gamma}(n , \cl{P}) :=   \left. \frac{\pd \omega(\Omega; n ; \cl{P}) }{\pd \Omega} \right|_{\Omega = \bar{\Omega}(n ; \cl{P})}  = \frac{ M_\tx{eff}(\bar{\Omega} ; n) }{\omega(\bar{\Omega}; n ; \cl{P})}
}
for each eigenstate labeled by $n$ and $\cl{P}$.
We note that, from this internal perspective, the Kennard-Robertson inequality yields the ``external'' time-energy uncertainty relation $\Delta T \Delta E \geq 1/2$.

\subsubsection{Nonunitary evolution relative to internal clock  \label{subsubsec:Nonunitary evolution with respect to internal time}}
Here, we consider the general case where the background spacetime is curved, and the detector is interacting with the matter field that is spacetime dependent.
Then, unlike in the special case in Sec.\ref{subsubsec:evolution_w.r.t_internal_time_w/o_observer-field_interaction}, the constraint operator $\hat{C}^\tx{ex}_+$ is genuinely nonlinear in $\hat{H}_\tx{C}$; thereby, the internal-time nonunitarity is expected from the physical state condition (\ref{eq:physical_state_condition_for_external_perspective}).  
This provides a manifestation of the nonunitarity discussed in Refs.\cite{Gambini:2004pe,Gambini:2006yj,Gambini:2006ph}. However, the nonunitarity obtained below arises from the interaction between the clock and the remaining d.o.f., and thus differs in its origin from the nonunitarity associated with the spreading of the clock's probability distribution primarily discussed in those works.

\

If this is indeed the case,
by approximately recasting the physical condition (\ref{eq:physical_state_condition_for_external_perspective}) into a Schr\"odinger-type first-order differential equation with respect to the clock time,
\eqn{ ( \ri \pd_\alpha - \hat{K} ) \fixket{\psi_\tx{ph}^\tx{ex} (\alpha)} \simeq 0 ~,
\label{eq:Schrodinger_eq_with_nonHermitian_Hamiltonian}}
we can see the nonunitarity as the non-Hermiticity of the operator $\hat{K}$,
contributing to the derivative of the inner product between two states $\fixket{\psi^{\tx{ex}}_{\tx{ph},i}(\alpha)}$ as
\eqn{ \ri \pd_\alpha \fixbraket{\psi^{\tx{ex}}_{\tx{ph},1}(\alpha)}{\psi^\tx{ex}_{\tx{ph},2} (\alpha)} = 2  \fixbra{\psi^{\tx{ex}}_{\tx{ph},1}(\alpha)}  [ \hat{K} ]_\tx{AH} \fixket{\psi^\tx{ex}_{\tx{ph},2} (\alpha)}   ~,}
where $[ \cdot  ]_\tx{AH}$ extracts the anti-Hermitian part of the operator in the square bracket: $[ \hat{K} ]_\tx{AH} = ( \hat{K} - \hat{K}^\dag ) /2$.

\

Let us start with a formal discussion of how the internal-time evolution becomes nonunitary. 
It originates from the fact that the relation in Eq.(\ref{eq:factorizations}) no longer holds at the quantum level due to the noncommutativity of the operators.
For $\hat{P}^2 := -\hat{\theta}^2  +  |\hat{\mb{P}}|_h^2 $, the KG-type operator $\hat{C}^\tx{KG}  =   \hat{M}_\tx{eff}^2 + \hat{P}^2$ can be rewritten into two expressions as
\eqn{
  \hat{C}^\tx{KG}  =&~  \hat{C}^\tx{ex}_{-}  \hat{C}^\tx{ex}_+   - \hat{\Delta}_1   \\
 =&   ~ \hat{C}^\tx{in}_- \hat{C}^\tx{in}_+ -  \hat{\Delta}_2  ~,
\label{eq:factorization_failed}}
where $\hat{C}^\tx{ex}_\pm$ and $\hat{C}^\tx{in}_\pm$ are defined in Eqs.(\ref{eq:constraint_operator_for_external_perspective}), (\ref{eq:constraint_operator_for_external_perspective_-}) and (\ref{eq:constraint_operator_for_internal_perspective_pm}), with the extra contributions
\eqn{
\hat{\Delta}_1 :=  [   \hat{ \theta}      ,  \hat{ \omega }  ]   ~,
\label{eq:Delta_1}}
and
\eqn{\hat{\Delta}_2  :=   [  (- \hat{P}^2 )^{1/2} , \hat{M}_\tx{eff} ] = [  (- \hat{P}^2 )^{1/2} , \hat{H}_\tx{D,i} ]  ~ .
\label{eq:Delta_2}
}
With the inverse of $\hat{C}^\tx{in}_-$ applied from the left, Eq.(\ref{eq:factorization_failed}) reads as
\eqn{
 &  \hat{M}_\tx{eff} - (-  \hat{P}^2 )^{1/2}   \weakeq{\tx{ex}}  -   [\hat{C}^\tx{in}_- ]^{-1}(\hat{\Delta}_1 -  \hat{\Delta}_2 )  ~.
\label{eq:nonlocal_Schrodinger_operator}}
Recall that ``$\weakeq{\tx{ex}}$'' indicates that the equality holds on $\cl{H}_\tx{ph}^\tx{ex}$.
If $\hat{H}_\tx{C}$ in $\hat{\Delta}_1 $, $  \hat{\Delta}_2$ and $\hat{C}^\tx{in}_-$ on the rhs can be somehow neglected or rewritten in terms of other operators that trivially act on $\cl{H}_\tx{C}$, then we achieve the Schr\"odinger-type equation (\ref{eq:Schrodinger_eq_with_nonHermitian_Hamiltonian}) in the $\alpha$ representation with a potentially non-Hermitian operator corresponding to $\hat{K}$.\footnote{
Otherwise, it can be understood as a nonlocal Schr\"odinger equation and formally solved with the Dyson series. The authors of Ref.\cite{Smith:2017pwx} discussed the nonunitarity due to the time-nonlocal Hamiltonian.
}

\

The above discussion, involving the square root of $-\hat{P}^2$ that is not necessarily positive, might seem somewhat questionable.
Thus, we shall obtain the form of Eq.(\ref{eq:Schrodinger_eq_with_nonHermitian_Hamiltonian}) in a brute-force manner, starting from the first line of Eq.(\ref{eq:factorization_failed}), which reads as
\eqn{
\hat{M}_\tx{eff}^2 + (\hat{P}^2 + \hat{\Delta}_1 )   \weakeq{\tx{ex}} 0 ~.
\label{eq:physical_state_condition_for_external_perspective_nonlinear}}
By making use of the assumption (\ref{eq:energy_of_internal_dof_is_subdominant}) that the energy of the internal d.o.f. is subdominant in the effective mass, together with the adiabaticity that the terms with the spacetime derivative of the fields are sufficiently small,
we obtain the desired form as follows.
Let us write Eq.(\ref{eq:physical_state_condition_for_external_perspective_nonlinear}) as $- 2 \hat{\zeta}  \weakeq{\tx{ex}} (  \hat{P}^2  + M^2)/M^2  +  \hat{\Delta}_1   / M^2  + \hat{\zeta}^2  $.
Each $\hat{\zeta} :=(\hat{H}_\tx{C}+ \hat{H}_\tx{D})/M$ appearing on the rhs, acting on the state, can be iteratively rewritten in terms of $(\hat{P}^2   + M^2)/M^2$  and the coefficient operators of the power series expansion of $\hat{\Delta}_1   / M^2 $ with respect to $\hat{\zeta} $,\footnote{\label{fn:expanding_omega}
The operator $\hat{\omega}$ in $\hat{\Delta}_1$ is expanded as $\hat{\omega}= \sum_n \hat{\omega}_n$ with $\hat{\omega}_0 = \hat{\omega}_\tx{rp}:= ( |\hat{\mb{P}}|_h^2 + M^2 )^{1/2}$ and $\hat{\omega}_1 \simeq \hat{\zeta} M^2  \hat{\omega}_\tx{rp}^{-1}$ at the zeroth order of the commutator, where the operator ordering is irrelevant.
Therefore, up to the first order in the commutator, we have $\hat{\Delta}_1 \simeq [\hat{\theta}, \hat{\omega}]  \simeq [\hat{\theta}, \hat{\omega}_\tx{rp}  ] \hat{z} + M^2  [\hat{\theta} ,  \hat{\zeta}  ]   \hat{\omega}_\tx{rp}^{-1} $.
When $\hat{\zeta}$ in $\hat{z}:=( 1 - M^2   \hat{\omega}_\tx{rp}^{-2}  \hat{\zeta} )$ acts on the physical state, it turns into the suppression of $(\hat{P}^2 + M^2)/M^2$ and/or the spacetime derivative; thus, we can set $\hat{z} \simeq 1$.
}
all of which should be small under our assumptions.
By retaining the terms up to the first order in either $(\hat{P}^2 + M^2)/M^2$ or the commutator resulting in the spacetime derivative of the fields, we obtain
\eqn{
&- \left[\hat{\zeta}  +  \frac{  \hat{P}^2   + M^2    }{2M^2}   \right] \dblket{\psi^\tx{ex}_\tx{ph}} \\
& \simeq  \left[  \frac{ [ \hat{  \theta }  ,   \hat{ \omega }_\tx{rp}  ]    }{2  M^2}  + \frac{ [\hat{\theta} , \hat{H}_\tx{D,i} ] }{2  M   \hat{ \omega }_\tx{rp}   } + \frac{ [  \hat{P}^2  , \hat{H}_\tx{D,i} ] }{ 4 M^3}  \right] \dblket{\psi^\tx{ex}_\tx{ph}} ~,
\label{eq:physical_state_condition_for_external_perspective_truncated}}
where $\hat{\omega}_\tx{rp} =  \hat{\omega} |_{\hat{\zeta} \to 0}$ corresponds to the relativistic particle energy without the internal d.o.f.; see footnotes \ref{fn:square_root} and \ref{fn:expanding_omega}.
The first and second terms on the rhs come from $\hat{\Delta}_1$, whereas the third term appears from\footnote{
In fact, $2 M^2 \hat{\zeta}^2 \dblket{\psi^\tx{ex}_\tx{ph}} \simeq  -  [    \hat{P}^2  + M^2  +   \hat{\Delta}_1    ]  \, \hat{\zeta}\,  \dblket{\psi^\tx{ex}_\tx{ph}} \allowbreak + [\hat{P}^2 , \hat{\zeta}] \dblket{\psi^\tx{ex}_\tx{ph}}$, and the extra $\hat{\zeta}$ acting on the state in the first term makes it second and higher order in $(\hat{P}^2 + M^2)/M^2$ and the spacetime derivative, which we neglect in Eq.(\ref{eq:physical_state_condition_for_external_perspective_truncated}).
}
$\hat{\zeta}^2$ and corresponds to the leading-order contribution from the term with $\hat{\Delta}_2$ in Eq.(\ref{eq:physical_state_condition_for_external_perspective_truncated}).
The anti-Hermitian part of $\hat{K}$ in the Schr\"odinger-type equation (\ref{eq:Schrodinger_eq_with_nonHermitian_Hamiltonian}) is identified as
\eqn{
 2 [ \hat{K} ]_\tx{AH} \simeq &~  \frac{[\hat{\theta}  , \hat{\omega}_\tx{rp}]}{M} +\frac{[\hat{\theta}, \hat{H}_\tx{D,i}]}{   \hat{\omega}_\tx{rp} } + \frac{[\hat{P}^2 , \hat{H}_\tx{D,i}]}{2 M^2 } ~.
\label{eq:K_AH}}
One can also check that Eq.(\ref{eq:nonlocal_Schrodinger_operator}) reduces to Eq.(\ref{eq:physical_state_condition_for_external_perspective_truncated}) with the above-mentioned truncations and the square root of $-\hat{P}^2$ understood as its expansion around $M$, given as in Footnote \ref{fn:square_root} with $\hat{Z} = - \hat{P}^2 - M^2$.

\ 

We note that the first term in the rhs of Eq.(\ref{eq:K_AH}) is the contribution of the indirect interaction between the internal and external d.o.f. as they persist when the matter-field detector is turned off. 
In the Newtonian limit where $\Phi := N-1 \ll 1$, $\bm{\beta} = \mb{0}$, $h_{ij} = \delta_{ij}$ and $E$ is dominated by the mass, this part of $ [ \hat{K} ]_\tx{AH}$ becomes $ -  [ \Phi(\hat{X}) ,  \hat{ \omega }_\tx{rp}] /2$ within the present approximation and gives rise to the source of nonunitarity found in Ref.\cite{Paiva:2022mxf} when moving beyond the approximation in Refs.\cite{Castro-Ruiz:2017whw,Smith:2017pwx,Castro-Ruiz:2019nnl}.

\ 

The second and third terms originate from the direct interaction between the detector and the external d.o.f.
However, they should be understood as a consequence of the indirect interaction between the internal and external d.o.f. induced by that direct interaction since, if $\hat{C}^\tx{ex}_+$ were linear in $\hat{\zeta}$ as in Eq.(\ref{eq:nonrelativistic_expansion}), the internal-time nonunitarity would not arise.

\section{Violation of equivalence principle on background spacetime \label{sec:violation_of_EEP}}
The equivalence of inertial and gravitational masses (WEP) implies that the physics in a homogeneous gravitational field is indistinguishable from that in a uniformly accelerating reference frame.
This principle was generalized by Einstein into the EEP, which extends the independence of the outcomes of local nongravitational experiments to both the velocity of the freely falling frame (LLI) and its spacetime location (LPI).

\ 

In this section, we focus on the theory with an inertial observer with the matter-field detector turned off: $H_\tx{D,i} =0$, and let this switched-off detector variable $q$ represent all the d.o.f. involved in the local experiments with the Hamiltonian $H_\tx{exp}$ renamed from $H_\tx{D,o}$ for clarity. 
Then, its evolution is described relative to the clock whose variation corresponds to the elapsed proper time of the observer, $\rd \alpha /\rd \tau = 1$, as already mentioned above Eq.(\ref{eq:time_dilation}).
At the classical level, the fact that the effective mass $M_\tx{eff}$ has no explicit dependence on the external d.o.f. implies that the present model described by the action (\ref{eq:observer_action_eta}) respects the EEP.

\ 

In what follows, we discuss the LLI and the LPI at the quantum level\footnote{
The universality of free fall, as an interpretation of the WEP, is ambiguous at the quantum level because the particle's wave function evolution depends on its mass even against the Minkowski background.
However, if the WEP is strictly interpreted as the equivalence between inertial and gravitational masses, its quantum-mechanical extension can be formulated in the Newtonian limit \cite{Zych:2015fka}.
} within the theory formulated from the external perspective, where the constraint equation (\ref{eq:physical_state_condition_for_external_perspective}) defines the physical states.
In Sec.\ref{subsec:violation_of_LLI/LPI}, we see that they are violated due to the internal-time nonunitarity.
Then, we briefly mention a formulation of quantum theory from the internal perspective based on the internal-time unitarity, in Sec.\ref{subsec:qauntum_theory_from_internal_pespective?}, which raises the multiple-choice problem of time on a fixed background spacetime.

\ 

We denote the Hilbert space spanned by the eigenbasis $\{ \fixket{n} \}$ of the operator $\hat{H}_\tx{exp}$ by $\cl{ H}_\tx{exp}$ rather than $\cl{ H}_\tx{D}$, and the eigenvalue associated with $\fixket{n}$ by $\epsilon_n$.

\subsection{Quantum LLI/LPI violation  \label{subsec:violation_of_LLI/LPI} }
Given the fact that the clock time corresponds to the proper time and that the experimenters in the freely falling laboratory have no direct access to the external d.o.f., the outcomes of the local experiment can be computed from the reduced density matrix obtained by tracing out the external d.o.f.,
\eqn{
\hat{\rho}_\tx{exp} (\alpha) := \frac{\Tr_\tx{EX} [ \fixket{\psi^{\tx{ex}}_{\tx{ph}}(\alpha)} \fixbra{\psi^{\tx{ex}}_{\tx{ph}}(\alpha)} ]}{\fixbraket{\psi^{\tx{ex}}_{\tx{ph}}(\alpha)}{\psi^{\tx{ex}}_{\tx{ph}}(\alpha)}} ~, 
\label{eq:reduced_density_operator}} 
where $\Tr_\tx{EX}$ denotes the trace over $\cl{H}_\tx{EX}$. 

\

Any difference in $\fixket{\psi_\tx{ph}^\tx{ex} (\alpha ; n)} :=  \fixbraket{n}{\psi^{\tx{ex}}_{\tx{ph}}(\alpha)} \in \cl{H}_\tx{EX}$ associated with the external d.o.f. is now interpreted as a difference in the internal state described by the matrix elements $\rho_\tx{exp}(\alpha ; n,n') := \fixbra{n}  \hat{\rho}_\tx{exp} (\alpha) \fixket{n'}$.
Hence, it is impossible to discuss whether the external configuration affects the results of the experiment carried out in the laboratory based merely on the expectation values of observables at a single, chosen internal time. 

\ 

Therefore, let us consider the internal-time evolution of the state.
By taking the derivative with respect to $\alpha$, we find
\eqn{
& \ri \pd_\alpha \rho_\tx{exp}(\alpha ; n,n') -  (\epsilon_n - \epsilon_{n'}) \rho_\tx{exp}(\alpha ; n,n') \\
& \simeq  2 \la    [\hat{K}_\tx{o} ]_\tx{AH} \ra_{n,n'} (\alpha) \\
&\hspace{2em} - 2 \rho_\tx{exp}(\alpha ; n,n')     \sum_{k}  \la   [\hat{K}_\tx{o} ]_\tx{AH}  \ra_k (\alpha) ~,
\label{eq:internal-time_derivative_of_reduced_density_matrix}}
where
\eqn{
[\hat{K}_\tx{o} ]_\tx{AH} \simeq  [ \hat{\theta}  ,  \hat{ \omega}_\tx{rp}]/M
\label{eq:K_o,AH}} is defined as the anti-Hermitian part (\ref{eq:K_AH}) with $\hat{H}_\tx{D,i}=0$; then, for an operator $\hat{O}$ such as $[\hat{K}_\tx{o} ]_\tx{AH}$ acting only on $\cl{H}_\tx{EX}$,
\eqn{
\la  \hat{O}  \ra_{n,n'} (\alpha) :=   \frac{   \fixbra{\psi^{\tx{ex}}_{\tx{ph}}(\alpha ; n')} \hat{O} \fixket{\psi^{\tx{ex}}_{\tx{ph}}(\alpha ; n)} }{  \fixbraket{\psi^{\tx{ex}}_{\tx{ph}}(\alpha)}{\psi^{\tx{ex}}_{\tx{ph}}(\alpha)} }
}
and $\la  \hat{O}  \ra_{n} (\alpha) := \la  \hat{O}  \ra_{n,n} (\alpha)$.
It shows that the LLI and the LPI are violated, since the physical law governing the local experiment on a general curved background is different from the Liouville equation $\ri \pd_\alpha \hat{\rho}_\tx{exp} = [\hat{H}_\tx{exp}, \hat{\rho}_\tx{exp}]$ on the Minkowski background due to the nonunitarity of the internal-time evolution.
If one considers the diagonal components, the equation (\ref{eq:internal-time_derivative_of_reduced_density_matrix}) tells us the evolution of the conditional probability $P_\tx{exp}(n | \alpha) := \rho_\tx{exp}(\alpha ; n,n)$; see Eq.(\ref{summary:internal-time_derivative_of_conditional_probability}).

\ 

We note that, when $q$ is disentangled with the external d.o.f., that is, the state $\fixket{\psi^\tx{ex}_\tx{ph} (\alpha)} = \bkk{\alpha}{\psi^\tx{ex}_\tx{ph}}$ is a product state on $\cl{ H}_\tx{exp} \otimes \cl{ H}_\tx{EX}$,
two terms in the rhs of Eq.(\ref{eq:internal-time_derivative_of_reduced_density_matrix}) cancel out.
Holding at a certain time, it is maintained at any other time since $[ \hat{K} ]_\tx{AH}$ acts only on $\cl{H}_\tx{EX}$, not to generate the entanglement between $q$ and the external d.o.f., reflecting the fact that it originates from the indirect interaction as emphasized below Eq.(\ref{eq:K_AH}).
For such a special physical state, if any, the violation of the LLI/LPI does not appear.

\subsection{Perspective (in)dependence of quantum theories on fixed background \label{subsec:qauntum_theory_from_internal_pespective?}}
One might argue that, if the nonunitarity with respect to the internal time is the underlying cause and it originates from the nonlinear dependence of the constraint operator $\hat{C}_+^\tx{ex}$ on the clock Hamiltonian,\footnote{
One could obtain the constraint function that is linear both in $E$ and $H_\tx{C}$ by completing the square of the KG-type constraint introducing the spin d.o.f. \`a la Dirac. We do not explore this possibility in this work, focusing on the importance of diffeomorphism invariance.
}
an alternative formulation based on the constraint function $C^\tx{in}_+$ given in Eq.(\ref{eq:constraint_function_for_internal_perspective}) should be considered.
The corresponding operator defined in Eq.(\ref{eq:constraint_operator_for_internal_perspective_pm}) could then be used to define the physical state.

\subsubsection{Possible formulation from internal perspective \label{subsubsec:possible_formulation_of_internal_perspective}}
While there is no problem with $C^\tx{in}_+$ at the classical level leading to the dynamics equivalent to that obtained from $C^\tx{ex}_+$,
some care is required in general at the quantum level because (i) it has the square root of $- P^2 $ that may take negative values corresponding to spacelike momenta of the observer and (ii) the constraint itself does not forbid the observer with negative energy. 
Then, the interaction with the fields that vary over spacetime can, in principle, drive the observer into states with spacelike momentum and/or negative energy, referred to as ``unphysical'' for simplicity.\footnote{
The authors of Ref.\cite{Oliveira:2025iha} considered the case with an observer minimally coupled to the background electromagnetic field on Minkowski spacetime and found that quantum theory can be safely formulated from the internal perspective with manifest internal-time unitarity.
We note that, in that setup, quantum theories from the internal and external perspectives can be inequivalent depending on the choice of the gauge field, in the sense of Eq.(\ref{eq:inequivalence}).
It would be interesting to see if the situation changes when the gauge field is dynamical.
}

\
 
There seem to be arbitrarily many possibilities based on the constraint function in Eq.(\ref{eq:constraint_function_for_internal_perspective}) to formulate a theory with the internal-time unitarity while keeping such unphysical states away.
Here, let us proceed with one of the most naive prescriptions, introducing 
\eqn{
\hat{\Pi} := \Theta ( - \hat{P}^2_+)  ~, 
\label{eq:projector}}
which projects out eigenstates with negative eigenvalues of the operator $-\hat{P}^2_+ := \Theta (\hat{E}) (-\hat{P}^2 ) \Theta (\hat{E})$,
and then
\eqn{(- \hat{P}^2 )^{1/2}_\Pi := \{   (- \hat{P}^2_+ ) \hat{\Pi} \}^{1/2} ~, 
\label{eq:projected_P^2}}
where the square root is understood in terms of the spectral decomposition.
With this operator, the constraint operator corresponding to $C_+^\tx{in}$ is defined by
\eqn{
\hat{C}_+^\tx{in} = \hat{H}_\tx{C}  + \hat{H}^\tx{in}_\tx{o} 
\label{eq:constraint_operator_for_internal_perspective}}
with $\hat{H}^\tx{in}_\tx{o} := \hat{H}_\tx{exp} + M - (- \hat{P}^2  )^{1/2}_\Pi $.
The physical states defined by
\eqn{
\hat{C}_+^\tx{in} \dblket{\psi^\tx{in}_\tx{ph}}  = 0 
\label{eq:physical_state_condition_for_internal_perspective}}
form the physical Hilbert space $\cl{H}_\tx{ph}^\tx{in}$.
This condition turns into the Schr\"odinger equation in $\alpha$ representation; therefore, the evolution relative to the internal clock becomes unitary.
 
\ 

It should be noted that generalizing this formulation to the case with the matter-field detector turned on gives rise to another source of ambiguity.
Since $\hat{H}_\tx{D,i}$ does not commute with $\hat{E}$ and $\hat{P}^2$ in general, considering $\hat{P}^2_+$ in Eq.(\ref{eq:projected_P^2}) is not enough; one might replace $\hat{H}_\tx{D,i}$ with $\hat{\Pi}  \hat{H}_\tx{D,i}  \hat{\Pi} $ to keep the unphysical states away.

\subsubsection{Perspective (in)dependence \label{subsubsec:Perspective (in)dependence}}
Suppose that one finds a convincing prescription to define a quantum theory with the internal-time unitarity, referred to as a theory formulated from the internal perspective.
Such a theory must be inequivalent to the one formulated from the external perspective with $\hat{C}_+^\tx{ex}$, which is based on the external-time unitarity and leads to the internal-time nonunitarity as seen in Sec.\ref{subsec:evolution_relative_to_the_internal_clock}.
This inequivalence directly follows from Eq.(\ref{eq:factorization_failed}); the inequality
\eqn{
\hat{C}^\tx{ex}_{-}  \hat{C}^\tx{ex}_+    \ne  \hat{C}^\tx{in}_- \hat{C}^\tx{in}_+    
\label{eq:inequivalence}}
leads to the conclusion that the physical states $\dblket{\psi_\tx{ph}^\tx{ex}}$ defined to be annihilated by $\hat{C}^\tx{ex}_+$ are, in general, not annihilated by $\hat{C}^\tx{in}_+ $, and vice versa.
In other words, the two physical Hilbert spaces $\cl{H}_\tx{ph}^\tx{ex}$ and $\cl{H}_\tx{ph}^\tx{in}$ are different from each other, while both are obtained as projections of the kinematical Hilbert space $\cl{H}_\tx{kin}$.
This is a version of the multiple-choice problem of time \cite{Kuchar:1991qf} that the quantum theory seemingly depends on the perspective.
It makes a contrast with the classical case where the two constraint surfaces defined by $C^\tx{in}_+ = 0$ and $C^\tx{ex}_+ = 0$ in the phase space have the intersection $\cl{I}$, as discussed in Sec.\ref{subsubsec:constraints}.

\

On the other hand, the approach based on the QRF resolves this issue by introducing a physical quantum entity whose location is identified as the origin of the external coordinate system\footnote{
It is typically considered the center-of-mass coordinate of a ``laboratory'' in the literature, whereas what we refer to as the local observer may be called a particle carrying internal d.o.f., or simply, a detector.
}
and considering a superposition of coordinate transformations to realize the quantum local inertial frame associated with the freely falling observer.
In Ref.\cite{Giacomini:2017zju}, the quantum WEP is formulated as the indistinguishability between a superposition of uniform gravitational fields and a superposition of accelerations in flat spacetime, and then its generalization to the quantum EEP on general curved backgrounds is proposed in Ref.\cite{Giacomini:2020ahk}. Furthermore, based on the spacetime QRF approach \cite{Giacomini:2021gei}, the violation of the quantum EEP is formulated in Ref.\cite{Cepollaro:2021ccc} extending the model of Ref.\cite{Zych:2015fka}, and it is shown that the violation of the EEP for QRFs breaks the equivalence between the external and internal perspectives, with which one of our findings in this section is consistent in the sense that the inequivalence (\ref{eq:inequivalence}) between those two perspectives seems unavoidable in the present formulation without the physical object to define the external coordinate system, violating the LLI/LPI at the quantum level as discussed in Sec.\ref{subsec:violation_of_LLI/LPI}.

\

In what follows, we pursue a possibility with dynamical gravity, partially fixing the gauge at the classical level to realize the frame associated with the local observer, but without introducing any extra dynamical object.
When such a theory is extended to the quantum regime with the assumption that quantum gravity is ultimately consistent with diffeomorphism invariance, it should be closely related to the QRF approach.

\section{Observers with dynamical gravity \label{sec:observers_with_dynamical_gravity}}
In this section, we argue that diffeomorphism invariance that typically appears in theories with dynamical gravity may play a crucial role in achieving the unitarity with respect to the internal time and the compatibility between the unitarities with respect to different local observers.

\ 

As seen in Sec.\ref{subsec:evolution_relative_to_the_internal_clock}, in the quantum theory constructed from the external perspective on a fixed background, the evolution relative to the internal clock is not unitary.
Then, to realize the internal unitarity, the constraint is required to be recast into the form linear in the clock Hamiltonian, such as (\ref{eq:constraint_operator_for_internal_perspective}) with the \textit{ad hoc} projector introduced, as briefly discussed in Sec.\ref{subsec:qauntum_theory_from_internal_pespective?}.

\ 

First, we briefly review the canonical formulation of classical gravity with a dynamical observer in Sec.\ref{subsec:classical_theory_w/o_gauge-fixing}.
Then, we introduce a particular type of partial gauge fixing where the observer appears to be at rest in Sec.\ref{subsec:observer-centric_coordinate_system}.
It is argued in Sec.\ref{subsec:physical_evolution_relative_to_internal_clock} that a single constraint function can be distinguished to involve the Hamiltonian of the internal d.o.f. linearly dependent on it, and associated relational observables are consistent with the rest of the diffeomorphism constraints. 
Based on that, it is discussed in Sec.\ref{subsec:unitarity_w.r.t._internal_time} that the evolution relative to the internal clock is expected to be unitary when the system is quantized, temporarily disregarding some issues that will be discussed in Sec.\ref{subsec:challenges}.

\subsection{Classical theory without gauge fixing \label{subsec:classical_theory_w/o_gauge-fixing}}
Now, let us consider the case with a dynamical spacetime described by an action $ \ssf{ S} =  \ssf{ S}_\tx{ g} + \ssf{ S}_\tx{m} +  \ssf{ S}_\tx{obs} $ with diffeomorphism invariance.
The observer part $\sf{S}_\tx{obs}$ is given as in Eq.(\ref{eq:observer_action_eta}) but now with the dynamical metric $g_{\mu \nu}(x)$ and the matter field $\upphi (x)$ evaluated at the observer's location $x^\mu = X^\mu$. 
The detailed expressions of the gravitational part $\sf{S}_\tx{g}$ and the matter part $\sf{S}_\tx{m}$ are not essential for the discussion below.
Following the Arnowitt–Deser–Misner formalism \cite{Arnowitt:1959ah}, the spacetime metric is written, as in Sec.\ref{subsec:classical_theory}, in terms of the lapse $N$, shift $\beta^i$, and induced metric $\ssf{ h}_{ij}$ on a reference time slice, among which only $\ssf{ h}_{ij}$ turns out to be dynamical. 
For simplicity, we consider a spatially closed manifold.

\

The dynamical fields are functions of the spacetime coordinate $x^\mu = (t , \mb{x})$.
Therefore, the action of the full system has two independent time coordinates: the external one $t$ in $\sf{S}_\tx{g} + \sf{S}_\tx{m}$ and the internal one $\eta$ parametrizing the worldline in $\sf{S}_\tx{obs}$. 
Then, to get a form suitable for the Hamiltonian formalism \cite{Rovelli:1990ph}, we fix the gauge associated with the reparametrization of the worldline by imposing $T(\eta) =t$ and regard $t$ as a single time parameter in the action.
Let us take a reference time slice with, say, $t =  T(0)$ on which the phase space variables are introduced.
Those associated with the internal d.o.f. are the same as in the previous discussion, while those associated with the external d.o.f. now consist of the observer's spatial position $\mb{X}$, the metric $\ssf{ h}_{ij}(\mb{x})$ induced on the slice, the matter field $\upphi (\mb{x})$, and their conjugate momenta.

\

The total Hamiltonian is given as a function of those phase space variables and the nondynamical lapse $N$ and shift $\bm{\beta}$ by
\eqn{
\ssf{ H}^\tx{ex}  = \ssf{ C}^\tx{diff}[N , \bm{\beta}] ~, \label{eq:total_Hamiltonian_with_dynamical_gravity}
}
where
\eqn{
\ssf{ C}^\tx{diff}[v , \mb{v}]  = \ssf{ C}_\perp [v] + \ssf{ C}_\parallel [\mb{v}]
\label{eq:generator_of_diffeomorphism}}
is the generator of diffeomorphism transformations of the dynamical variables, composed of the following two distinct terms;
the so-called Hamiltonian constraint
\eqn{\ssf{ C}_\perp [v] := \ssf{ H}_\tx{g+m}[v] + \ssf{ H}_\tx{obs} v(\mb{X}) 
\label{eq:generator_of_temporal_diffeo}}
generates a position-dependent translation of the reference time slice by a given scalar field $v(\mb{x})$ in the normal direction, whereas the so-called momentum constraint
\eqn{
\ssf{ C}_\parallel [\mb{v}] := - \ssf{ P}_\tx{g+m}[\mb{v}]  - \mb{P} {\cdot} \mb{v} (\mb{X}) ~, \label{eq:generator_of_spatial_diffeo}
}
generates a spatial diffeomorphism on the time slice with a given vector field $\mb{v}(\mb{x})$.
The first terms, $\ssf{ H}_\tx{g+m}$ and $\ssf{ P}_\tx{g+m}$, are contributions from $\sf{S}_\tx{g} + \sf{S}_\tx{m}$.
The second terms are from $\sf{S}_\tx{obs}$, cf. Eq.(\ref{eq:classical_Hamiltonian_from_external_perspective}). 
The part in the Hamiltonian constraint has a similar form as the one in Eq.(\ref{eq:relativistic_particle_energy}) but with $\ssf{ h}^{ij}$, the inverse of the dynamical induced metric on the slice; 
\eqn{
\ssf{ H}_\tx{obs} = (  |\mb{P}|_\ssf{h}^2 + \ssf{ M}_\tx{eff}^2  )^{1/2} ~, ~~~
|\mb{P}|_\ssf{h}^2 :=   \tx{P}_i \tx{P}_j \ssf{ h}^{ij}(\mb{X})    ~,
\label{eq:observer_Hamiltonian_with_dynamical_gravity}}
where $\ssf{ M}_\tx{eff}$ is the effective mass, 
\eqn{
 \ssf{ M}_\tx{eff}  :=  M    +    H_\tx{D}(p, q, \upphi (\mb{X}) ) + H_\tx{C}  ~,  ~ \label{eq:effective_mass_with_dynamical_matter}
}
which depends on the dynamical matter field $\upphi$ through the interaction part $H_\tx{D,i}$ of the detector Hamiltonian.
As well known, these generators satisfy $\{ \ssf{ C}_\parallel [\mb{v}] , \ssf{ C}_\parallel [\tilde{\mb{v}}] \}_\tx{P} =   \ssf{ C}_\parallel [ \cl{L}_{\mb{v}} \tilde{\mb{v}} ]$,
$\{ \ssf{ C}_\parallel [\mb{v}], \ssf{ C}_\perp [v] \}_\tx{P} = \ssf{ C}_\perp [\cl{L}_{\mb{v}} v]$
and $\{ \ssf{ C}_\perp [v_1], \ssf{ C}_\perp [v_2] \}_\tx{P} = \ssf{ C}_\parallel [\mb{v}_{12}]$, where $\cl{L}_{\mb{v}}$ denotes the Lie derivative along $\mb{v}$, and $\tx{v}_{12}^i := v_1 \ssf{ h}^{ij} \pd_j v_2 - v_2 \ssf{ h}^{ij} \pd_j v_1$ is a metric-dependent vector field. 
Therefore, the diffeomorphism generators given as in Eq.(\ref{eq:generator_of_diffeomorphism}) form a closed algebra under the Poisson bracket, i.e., $\{ \ssf{ C}^\tx{diff}[v_1 , \mb{v}_1] ,\ssf{ C}^\tx{diff}[v_2 , \mb{v}_2]   \}_\tx{P} \weakeq{} 0$, where the weak equality indicates that the equality holds on the constraint surface on which $\ssf{ C}^\tx{diff}[v , \mb{v}]  = 0$ for any $v$ and $\mb{v}$; hence, no other constraint is required for consistency.
 
\

The flow parameter associated with the total Hamiltonian (\ref{eq:total_Hamiltonian_with_dynamical_gravity}) is interpreted as the external coordinate time.
However, since this flow is itself a diffeomorphism under which physical quantities do not evolve, this leads to the problem of time in gravitational theories.
Therefore, we need a clock variable, such as $\alpha$ carried by the observer, to describe the physical evolution of the rest of the system in a relational manner.

We remark that the problem of time represents only a facet of the more comprehensive problem of how to construct local observables $\ssf{ O}_\tx{ph}$ satisfying $\{ \ssf{ O}_\tx{ph} , \ssf{ C}^\tx{diff}[v , \mb{v}]  \}_\tx{P} \weakeq{} 0$.
Roughly speaking, the temporal part of the problem is resolved by the presence of the internal clock variable,\footnote{
The temporal location of the observer alone cannot do the job, unlike in the theory on a fixed background discussed in Sec.\ref{subsec:classical_theory} or Appendix \ref{app:observers_with_dynamical_matter_field}.
This is because the system is ``doubly gauged'' in the time direction; namely, the action is invariant under not only (i) the reparametrization on the worldline, which is timelike, generated by the constraint discussed in Sec.\ref{subsubsec:constraints} but also (ii) the local reparametrization of the external coordinate time that corresponds to the deformation of the time slice generated by the Hamiltonian constraint (\ref{eq:generator_of_temporal_diffeo}).
Correspondingly, in the current approach, after the external-time coordinate $T$ of the observer's location has been gauged away by (i) with the condition $\rd T /\rd \eta = 1$, the total Hamiltonian (\ref{eq:total_Hamiltonian_with_dynamical_gravity}) is constrained to vanish for the invariance under (ii) together with spatial diffeomorphism invariance, unlike the fixed background case where the total Hamiltonian (\ref{eq:classical_Hamiltonian_from_external_perspective}) or (\ref{app:total_Hamiltonian_with_dynamical_matter}) is not constrained to vanish.
} whereas the spatial part can be handled by the position of the local observer as a physical reference point on the time slice \cite{Rovelli:1990ph,Rovelli:1990pi}. 
In contrast to employing such a physical entity, gauge-invariant observables can also be constructed by anchoring them to geometric boundaries.
One standard approach fixes a boundary at asymptotic infinity where diffeomorphisms are supposed to vanish and shoots geodesics inward to specify bulk points \cite{Heemskerk:2012np,Donnelly:2015hta,Donnelly:2016rvo}.
Otherwise, one can consider edge modes on subregion boundaries and employ them as reference frames \cite{Carrozza:2022xut,Kabel:2023jve,Giesel:2024xtb} to define relational observables; see Ref.\cite{Goeller:2022rsx} for a general framework and references therein.

\ 

The Hamiltonian $H_\tx{C}$ appears in the Hamiltonian constraints (\ref{eq:generator_of_temporal_diffeo}).
As in the previous case on a fixed background, it is nonlinear in $H_\tx{C}$;
naively, this seems to give rise to the nonunitarity of the evolution relative to the internal clock again.
However, now we have an infinite number of constraints for diffeomorphism invariance. Solving some of these constraints changes the situation, as seen in the following.

\subsection{Observer-centric coordinate system \label{subsec:observer-centric_coordinate_system}}
Let us consider a partial gauge fixing in such a way that the coordinate value of the observer's position is constant: $\mb{X} = \mb{x}_\tx{obs}$.
For consistency, the observer's momentum should vanish: $\mb{P}=\mb{0}$, and then the observer's energy in the generators (\ref{eq:generator_of_temporal_diffeo}) reduces to the effective mass: $\ssf{ H}_\tx{obs}  = \ssf{ M}_\tx{eff}$, which is linear in $H_\tx{C}$. Therefore, one naively expects that the quantum theory based on it could be unitary with respect to the clock time. 
In this subsection, we shall discuss this possibility more carefully.

\ 

For simplicity, we suppose that the scalar function $v (\mb{x})$ on the reference time slice is expanded in a complete basis $\{ n^{(l)}(\mb{x}) \}$, and similarly, the vector field $\mb{v}(\mb{x})$ is expanded in $\{ \mb{b}_{(m)}(\mb{x}) \}$, that is,  
\eqn{
v(\mb{x}) = \sum_{l=0}^\infty  v_{(l)} n^{(l)}(\mb{x}) ~, ~~~ \mb{v} (\mb{x}) = \sum_{m=1}^\infty  \tx{v}_{(m)}  \mb{b}_{(m)}(\mb{x}) 
\label{eq:bases}}
with the expansion coefficients $v_{(l)}$ and $\tx{v}_{(m)}$.
It is merely for later convenience that the summation for $l$ starts from $0$, whereas that for $m$ starts from $1$.
Then, the diffeomorphism generators (\ref{eq:generator_of_diffeomorphism}) can be written as
\eqn{
&\ssf{ C}^\tx{diff}[ v(\mb{x}) , \mb{v}(\mb{x}) ] \\
 &= \sum_{l=0}^\infty  v_{(l)}  \ssf{ C}_\perp [ n^{(l)} ] + \sum_{m=1}^\infty  \tx{v}_{(m)}  \ssf{ C}_\parallel [  \mb{b}_{(m)} ] \\
&= \sum_{n}  {\scr V}_{(n)} \ssf{ C}^\tx{diff}_n ~.
\label{eq:generator_of_diffeomorphism_linear_combination}}
On the last line, we have unified the Hamiltonian and momentum constraints into a single summation, defining a basis of the infinite-dimensional space of diffeomorphism constraint functions, 
\eqn{
\ssf{ C}^\tx{diff} = \{ \ssf{ C}_\perp [ n^{(l)} ] \}_{l \geq 0} \cup \{ \ssf{ C}_\parallel [  \mb{b}_{(m)} ] \}_{m \geq 1} ~,
}
with an infinite number of coefficients ${\scr V}_{(n)}$.

\subsubsection{Partial gauge fixing of fully constrained system \label{subsubsec:partial_gauge-fixing}}
Let us start with a general formulation with a set of gauge-fixing functions $\ssf{ F} = \{ \ssf{ F}_k \}_{1 \leq k \leq D}$ without specifying their explicit forms here.
The number of elements $D$ is finite; hence, the gauge fixing of the system remains only partial.

\ 

Suppose that the elements in $\ssf{ F}$ are mutually consistent and not redundant.
For each element to play the role of a gauge-fixing function, it should hold that the slice of the phase space defined by $\ssf{ F} = 0$ and the constraint surface $\cl C$ defined by $\ssf{ C}^\tx{diff} = 0$ share an intersection, denoted by $\cl{C}|_{\ssf{ F}=0}$, whose dimension is $D$ lower than that of $\cl C$.
This intersection $\cl{C}|_{\ssf{ F}=0}$ still has the gauge redundancy and may be referred to as the residual constraint surface.
Since the gauge orbits are generated by the elements of $\ssf{ C}^\tx{diff}$, there must be $D$ linear combinations of them, denoted by $\ssf{ C}^\tx{slv} = \{ \ssf{ C}^\tx{slv}_K \}_{1 \leq K \leq D}$, such that the matrix
\eqn{
\tx{A}_{Kk} := \{ \ssf{ C}^\tx{slv}_K , \ssf{ F}_{k} \}_\tx{P}  
\label{eq:Poisson_bracket_between_F_and_C^slv}}
is invertible in the neighborhood of $\cl{C}|_{\ssf{ F}=0}$,
whereas the rest of the diffeomorphism constraint functions, denoted by $\ssf{ C}^\tx{res} = \{ \ssf{ C}^\tx{res}_r \}$, satisfy 
\eqn{
\{ \ssf{ C}^\tx{res}_r , \ssf{ F}_{k} \}_\tx{P}|_{\ssf{ F}=0}  \weakeq{} 0 ~,
\label{eq:Poisson_bracket_between_F_and_C^res}}
that is, the gauge orbits generated by them stay on the residual constraint surface $\cl{C}|_{\ssf{ F}=0}$ if they start there.
As emphasized in Eq.(\ref{summary:res_oplus_slv}), those two sets, $\ssf{ C}^\tx{slv}$ and $\ssf{ C}^\tx{res}$, are linearly independent of each other.

\

In order to achieve a consistent formulation, one could solve $\ssf{ Z} = 0$, where
\eqn{
\ssf{ Z} := \ssf{ F} \cup \ssf{ C}^\tx{slv} ~,
\label{eq:2nd-class_constraints}
}
to reduce the number of independent d.o.f. by $2 D$ and working on a partially reduced phase space $\cl{Z}$ parametrized by the remaining d.o.f.; see Fig.\ref{fig:partially-reduced_phase_space}.
Then, the elements of $\ssf{ C}^\tx{res}$ must form a closed algebra on $\cl{Z}$ in order to generate consistent gauge transformations on the residual constraint surface $\cl{C}|_{\ssf{ F}=0}$ that coincides with $\cl{Z}|_{\ssf{ C}^\tx{res} = 0}$. 
However, in general, explicitly solving $\ssf{ Z} = 0$ is challenging; to circumvent this difficulty, Dirac's procedure outlined below provides an efficient approach.

\

Since the system is eventually constrained on $\cl{C}|_{\ssf{ F}=0}$, one can focus on its neighborhood.
We assume that the set of $2 D$ constraint functions $\ssf{ Z}$ is second class, namely, 
\eqn{
\Upsilon_{\kappa \kappa'} := \{ \ssf{ Z}_{\kappa} , \ssf{ Z}_{\kappa'} \}_\tx{P} |_{\ssf{ Z} = 0}
\label{eq:2nd-class_constraint_matrix}}
is invertible, at least, in the limit where $\ssf{ C}^\tx{res} \to 0$.
If not, the choice of the gauge-fixing functions in $\sf F$ needs to be revised. 
With the inverse of this constraint matrix, the Dirac bracket is defined as
\eqn{
\{ \, \cdot  \,  , \,  \cdot \,   \}_\tx{D} = \{ \,  \cdot \,   , \,  \cdot  \,  \}_\tx{P} -  \sum_{\kappa , \kappa'}  \{ \,  \cdot  \,  ,  \ssf{ Z}_{\kappa} \}_\tx{P} (\Upsilon^{-1})_{\kappa \kappa'} \{  \ssf{ Z}_{\kappa'} , \,  \cdot \,  \}_\tx{P} ~.
\label{eq:Dirac_bracket_0}}
By definition, the elements of $\ssf{ Z}$ can always be set to zero in this bracket.
Regarding the elements of $\ssf{ C}^\tx{res}$,
we find that they form the first-class constraints, i.e., the algebra is closed under the Dirac bracket as
\eqn{
\{ \ssf{ C}^\tx{res}_r ,  \ssf{ C}^\tx{res}_{r'} \}_\tx{D}   \weakeq{\tx{res}} 0 ~,
\label{eq:algebra_of_C^res}}
where the weak equality $\weakeq{\tx{res}}$ means that the equality is achieved with $\ssf{ C}^\tx{res} = 0$ on $\cl Z$.
In other words, they generate the gauge orbits in the residual constraint surface $\cl{C}|_{\ssf{ F}=0}$.
This follows from the definition of the elements of $\ssf{ C}^\tx{res}$ to have vanishing Poisson brackets with the elements of $\sf F$, in conjunction with the fact that the algebra of the original diffeomorphism constraints is closed under the Poisson bracket.

\subsubsection{Application to observer-centric gauge  \label{subsubsec:implementation}}
Now we identify $D= 2 d$ gauge-fixing functions in $\sf F$ as 
\eqn{
\ssf{ F}_i := \tx{X}^i - \tx{x}^i_\tx{obs} ~, ~~~ \ssf{ F}_{d+i} := \tx{P}_i
}
with $i = 1,\cdots ,d$.
In order to make the decomposition of $\ssf{ C}^\tx{diff}$ into $\ssf{ C}^\tx{slv}$ and $\ssf{ C}^\tx{res}$,
it is convenient to consider bases $\{ n^{(l)}(\mb{x}) \}$ and $\{ \mb{b}_{(m)}(\mb{x}) \}$ in Eq.(\ref{eq:bases}) such that 
\eqn{
v(\mb{x}_\tx{obs}) = v_{(0)} n^{(0)}(\mb{x}_\tx{obs}) ~,
\label{eq:n^(0)}}
\eqn{ 
 \tx{v}^j (\mb{x}_\tx{obs}) = \sum_{J =1}^d  \tx{v}_{(J)}  [\mathbb{ V} ]_{~J}^{j} ~
}
and
\eqn{
\pd_i v(\mb{x}_\tx{obs}) = \sum_{I = 1}^{d} v_{(I)}   [ \mathbb{ S}^{-1}]_{~i}^I  ~,
}
where
\eqn{
[\mathbb{ V} ]_{~I}^{j} := \tx{b}^j_{(I)}(\mb{x}_\tx{obs}) ~, ~~~ [ \mathbb{ S}^{-1}]_{~i}^I :=  \pd_i n^{(I)}(\mb{x}_\tx{obs}) ~, 
\label{eq:V_and_S}}
which are invertible by definition, otherwise the bases $\{ n^{(l)}(\mb{x}) \}$ and $\{ \mb{b}_{(m)}(\mb{x}) \}$ would be incomplete.
Put another way, at the observer's position $\mb{x} = \mb{x}_\tx{obs}$, none of the $n^{(l)} (\mb{x})$ with $l \ne 0$ contribute to the value of the scalar field, none of the $\mb{b}_{(m)} (\mb{x})$ with $m > d$ contribute to the value of the vector field, and none of the $n^{(l)} (\mb{x})$ with $l = 0$ or $n > d$ contribute to the first derivative of the scalar field.
Then, we identify the $D =2d$ elements of $\ssf{ C}^\tx{slv}$, satisfying the condition mentioned in relation with Eq.(\ref{eq:Poisson_bracket_between_F_and_C^slv}), as
\eqn{
\ssf{ C}^\tx{slv} = \{ \ssf{ C}_\perp[n^{(1)}] , \cdots , \ssf{ C}_\perp[n^{(d)}] , \ssf{ C}_\parallel[\mb{b}_{(1)}] , \cdots , \ssf{ C}_\parallel[\mb{b}_{(d)}] \} ~.
\label{eq:definition_of_C^slv}} 

For the generators of residual diffeomorphisms, let us define $\{ \mb{u}_{(q)}(\mb{x}) \}_{q \geq d+1}$ and $\{ u^{(p)}(\mb{x}) \}_{p =0, p\geq d+1}$\footnote{
With a lapse function and a shift vector constructed from these bases, for $U^\mu (x)$ such that $U^\mu U_\mu = -1$ and $U^i=0$ at $\mb{x}=\mb{x}_\tx{obs}$, we find that, on the observer's worldline,
\eqn{U^\nu \nabla_\nu U_\mu  = \delta_\mu^i\pd_i \ln u^{(0)}  = - \sum_{I=1}^d \mf{D}_I  [ \mathbb{ S}^{-1}]_{~i}^I \delta_\mu^i ~,} reproducing the proper acceleration due to the coupling with the matter field.
}
by $\mb{u}_{(q)}(\mb{x}) = \mb{b}_{(q)}(\mb{x})$, $u^{(p)}(\mb{x}) = n^{(p)}(\mb{x})$ for $p \geq d+1$, and
\eqn{
u^{(0)}(\mb{x}) = \frac{n^{(0)}(\mb{x})}{n^{(0)}(\mb{x}_\tx{obs})} -  \sum_{I=1}^d \mf{D}_I n^{(I)}(\mb{x})
\label{eq:u^(0)}}
with
\eqn{
{\mf{D}_I :=   [  \mathbb{ S} ]_{~I}^i \left. \frac{\pd \ln \ssf{ M}_\tx{eff} (\mb{X})}{\pd \tx{X}^i}\right|_{\mb{X} = \mb{x}_\tx{obs}}  }   ~.  
}
Then, the elements of
\eqn{
\ssf{ C}^\tx{res} =  \{ \ssf{ C}_\perp [ u^{(p)} ] \}_{p=0, p \geq d+1} \cup \{ \ssf{ C}_\parallel [  \mb{u}_{(q)} ] \}_{q \geq d+1}
\label{eq:definition_of_C^res}}
can be checked to satisfy the condition (\ref{eq:Poisson_bracket_between_F_and_C^res}).

\

Now, let us label $2 D = 4 d$ elements in $\sf Z$ defined by Eq.(\ref{eq:2nd-class_constraints}) as
\eqn{
\ssf{ Z}^{1}_I := \ssf{ C}_\perp [n^{(I)}] ~,~~~ \ssf{ Z}^{2}_I := \ssf{ C}_\parallel [ \mb{b}_{(I)}] ~,
}
\eqn{
\ssf{ Z}^{3}_i := \tx{X}^i  - \tx{x}^i_\tx{obs}  ~, ~~~ \ssf{ Z}^{4}_i := \tx{P}_i ~ }
with the indices $i,I = 1,\cdots,d$.
Then, the $4d \times 4d$ constraint matrix (\ref{eq:2nd-class_constraint_matrix}) is treated as a $4 \times 4$ block matrix of $d \times d$ blocks, where the $4 \times 4$ indices are written as superscripts and the $d \times d$ indices as subscripts.
We find
\eqn{
\Upsilon^{11}_{IJ} = \{ \ssf{ Z}^{1}_I , \ssf{ Z}^{1}_J \}_\tx{P}|_{\ssf{ Z}=0} = \{ \ssf{ Z}^{1}_I , \ssf{ Z}^{1}_J \}_\tx{P}   = \ssf{ C}_\parallel [ \mb{w} ] 
}
with the vector field $\tx{w}^i := n^{(I)} \ssf{ h}^{ij} \pd_j n^{(J)} - n^{(J)} \ssf{ h}^{ij} \pd_j n^{(I)}$ that vanishes at $\mb{x}=\mb{x}_\tx{obs}$,
\eqn{
\Upsilon^{12}_{IJ} = \{ \ssf{ Z}^{1}_I , \ssf{ Z}^{2}_J \}_\tx{P}  |_{\ssf{ Z}=0} = - C_\perp [ \cl{L}_{\mb{b}_{(J)}} n^{(I)}] |_{\ssf{ Z}=0} ~,
} 
\eqn{
\Upsilon^{22}_{IJ} = \{ \ssf{ Z}^{2}_I , \ssf{ Z}^{2}_J \}_\tx{P}  |_{\ssf{ Z}=0} =  C_\parallel [ \cl{L}_{\mb{b}_{(I)}} \mb{b}_{(J)}]  |_{\ssf{ Z}=0} ~, 
}
\eqn{
\Upsilon^{14}_{Ij} = \{ \ssf{ Z}^{1}_I , \ssf{ Z}^{4}_j \}_\tx{P}  |_{\ssf{ Z}=0} 
=    \ssf{ M}_\tx{eff} [  \mathbb{ S}^{-1}]^I_{~j} ~,  
}
\eqn{
\Upsilon^{23}_{Ij} =  \{ \ssf{ Z}^{2}_I ,  \ssf{ Z}^{3}_j  \}_\tx{P} |_{\ssf{ Z}=0} = + [ \mathbb{ V}^\tx{t} ]_{I}^{~j}  ~,
}
where the superscript ``$\tx{t}$'' denotes the transpose, and
\eqn{
\Upsilon^{34}_{ij} =  \{ \ssf{ Z}^{3}_i , \ssf{ Z}^{4}_j \}_\tx{P}|_{\ssf{ Z}=0} =  \{ \ssf{ Z}^{3}_i , \ssf{ Z}^{4}_j \}_\tx{P}   =  [\, \mathbb{I} \, ]^i_{~j}  
}
with $[ \, \mathbb{I} \, ]^i_{~j} := \delta^i_j$.
The other combinations of the $4 \times 4$ indices give the $d \times d$ zero matrix $\mathbb{O}$.

\

To express the inverse of $\Upsilon$ in a compact form, let us write it as a $2\times 2$ block matrix as 
\eqn{
 \Upsilon = \pmtx{ \tx{C} &  \tx{D}^{-1}  \\  - (\tx{D}^{-1})^\tx{t} & \tx{E} } ~
 }
 with
 \eqn{
\tx{C} := \pmtx{ \Upsilon^{11}  & \Upsilon^{12} \\ \Upsilon^{21}  & \Upsilon^{22} } ~,
}
\eqn{
\tx{E} :=&~ \pmtx{  \Upsilon^{33} & \Upsilon^{34} \\ \Upsilon^{43}  & \Upsilon^{44} } = \pmtx{ \mathbb{O} & \mathbb{I} \\ - \mathbb{I}  & \mathbb{O} } ~,
}
and
\eqn{
\tx{D}  :=&~ \pmtx{ \Upsilon^{13} & \Upsilon^{14} \\ \Upsilon^{23}  &  \Upsilon^{24} }^{-1} = \pmtx{ \mathbb{O} & (\mathbb{V}^{-1})^\tx{t} \\   \mathbb{S}/ \ssf{ M}_\tx{eff}   & \mathbb{O} } ~, 
}
which is nothing but the inverse of the $2 d \times 2 d$ matrix (\ref{eq:Poisson_bracket_between_F_and_C^slv}) evaluated on the partially reduced phase space $\cl{Z}$.
Then, the inverse is obtained as
\eqn{
\Upsilon^{-1} = \Xi^\tx{t} \pmtx{ \tx{F}^{-1} &  \tx{F}^{-1}     \\  \tx{F}^{-1} &      \tx{F}^{-1} - \tx{G}   } \Xi  ~,
\label{eq:Upsilon_inverse}}
where $\tx{F} := \tx{C} + \tx{G}^{-1}$,
\eqn{
 \tx{G} :=  \tx{D}^\tx{t} \tx{E} \tx{D}    = \pmtx{  \mathbb{ O}   &  - (\mathbb{ V}^{-1} \mathbb{ S})^\tx{t}    \\   \mathbb{ V}^{-1} \mathbb{ S}    &  \mathbb{ O} }  / \ssf{ M}_\tx{eff} ~,
\label{eq:G}}
and
\eqn{
\Xi := \pmtx{ \tx{I} & \tx{O} \\ \tx{O} & -(\tx{ED})^{-1}   } 
}
with $\tx{I}$ and $\tx{O}$ being the $2 d \times 2 d $ identity and zero matrices, respectively.
Since the matrix $\tx{C}$ vanishes in the limit of $\ssf{ C}^\tx{res} \to 0$, we expand the inverse of $\tx{F}$ as $\tx{F}^{-1} =   \sum_{n = 0}^\infty  ( - \tx{G} \tx{C}   )^n \tx{G}$ and obtain
\eqn{
\Upsilon^{-1} = \Upsilon^{-1}_{0}   + \sum_{n=1}^\infty  \pmtx{  \tx{G}   & \tx{O} \\  \tx{O}  &   \tx{D}  }    \pmtx{  \tx{C}_{n}   & \tx{C}_n \\  \tx{C}_n  &   \tx{C}_n  }    \pmtx{  \tx{G}   & \tx{O} \\  \tx{O}  &   \tx{D}   }^{\!\tx{t}}   ~,
}
where $\tx{C}_n  := ( - \tx{C} \tx{G}  )^{n} (-\tx{G})^{-1}$ and  
\eqn{
\Upsilon^{-1}_{0}  :=  \pmtx{    \tx{G}  &  - \tx{D}^\tx{t}  \\ \tx{D}   &  \tx{O}}  ~.
\label{Upsilon_inverse_0}}

\

Then, the Dirac bracket (\ref{eq:Dirac_bracket_0}) is now given as
\eqn{
\{ \, \cdot \,  , \,  \cdot \,  \}_\tx{D} =&\, \{ \, \cdot \,  , \, \cdot \,  \}_\tx{P} \\
&- \, \sum_{a,a'=1}^4 \sum_{\iota,\iota' =1}^d  \{ \, \cdot \,  ,  \ssf{ Z}^{a}_{\iota} \}_\tx{P} (\Upsilon^{-1})^{a a'}_{\iota \iota'} \{  \ssf{ Z}^{a'}_{\iota'} , \, \cdot \,  \}_\tx{P} ~.
\label{eq:Dirac_bracket}}
We note that the symplectic structure for the observer's internal variables is preserved on $\cl{ Z}$, that is,
the Dirac brackets involving the internal d.o.f. are the same as their Poisson brackets.
This is ensured by the fact that $n^{(I)}$ with $1 \leq I \leq d$ are chosen to vanish at the observer's position $\mb{x} =\mb{x}_\tx{obs}$.

\subsection{Physical evolution relative to internal clock \label{subsec:physical_evolution_relative_to_internal_clock}}
On the partially reduced phase space associated with the observer-centric coordinate, physical quantities $\ssf{ O}_\tx{ph}$ are defined via
\eqn{
\{ \ssf{ O}_\tx{ph} ,  \ssf{ C}^\tx{res}_r \}_\tx{D}   \weakeq{\tx{res}}  0 
\label{eq:weak_physical_condition_with_gravity}}
with the Dirac bracket given in Eq.(\ref{eq:Dirac_bracket}).
Those that relationally satisfy this condition with the internal clock reveal how the rest of the Universe evolves.

\subsubsection{Diffeomorphisms on partially reduced phase space \label{subsubsec:residual_diffeos}}
Under the use of the Dirac bracket, we can set the elements of $\ssf{ Z}$ to zero; especially $\ssf{ C}_\perp [n^{(I)}] = 0$ with $1\leq I \leq d$ as well as $\mb{P} = \mb{0}$.
Therefore, $\ssf{ C}_\perp [u^{(0)}]$ with $u^{(0)}$ defined in Eq.(\ref{eq:u^(0)}), which is the only element of $\ssf{ C}^\tx{res}$ involving the Hamiltonian of the internal d.o.f., reduces to $\ssf{ C}$ shown in Eq.(\ref{summary:singled_out_from_C^diff}),
\eqn{
\ssf{ C}  := &\, \ssf{ C}_\perp [u^{(0)}] |_{\ssf{ Z}= 0} = \ssf{ C}_\perp [n] |_{\ssf{ Z}= 0} \\
=& \, M + H_\tx{C} + H_\tx{D}  + \ssf{ H}_\tx{g+m}[n]  ~,
\label{eq:singled_out_from_C^diff}}
where $n(\mb{x}) := n^{(0)}(\mb{x})/ n^{(0)}(\mb{x}_\tx{obs})$.

\

Denoting the rest of the elements of $\ssf{ C}^\tx{res}$ by $\ssf{ C}^\lozenge = \{ \ssf{ C}^\lozenge_r \}$, now we have the decomposition of $\ssf{ C}^\tx{res}$ as in Eq.(\ref{summary:0_oplus_lozenge}).
Owing to the fact that the values of $u^{(p > d)}$, $\pd_i u^{(p > d)}$ and $\mb{u}_{(q > d)}$ all vanish at the observer's position, we find $\{ \ssf{ C}^\lozenge_r ,  \ssf{ C}^\lozenge_{r'} \}_\tx{P}|_{\ssf{ C}^\lozenge =0 }   = 0$ and $\{ \ssf{ C}^\lozenge_r ,  \ssf{ Z}_\kappa \}_\tx{P}|_{\ssf{ Z}=0 , \ssf{ C}^\lozenge \!=0 } =  0$, from which it follows that
\eqn{
\{ \ssf{ C}^\lozenge_r ,  \ssf{ C}^\lozenge_{r'} \}_\tx{D}    \weakeq{\lozenge}  0 ~
\label{eq:subalgebra_on_Z}}
with the weak equality $\weakeq{\lozenge}$ meaning that the equality holds on a submanifold $\cl{ Z}_\lozenge$ in $\cl Z$ where $\ssf{ C}^\lozenge = 0$ while $\sf C$ does not necessarily vanish; therefore, they generate a consistent set of subgauge transformations on $\cl{ Z}_\lozenge$.
For the expression of the Dirac bracket with $\sf C$ kept finite; see Appendix \ref{app:Dirac_bracket_with_C}. 

\

In addition, since $\{ \ssf{ C} ,  \ssf{ C}^\lozenge_r \}_\tx{P}|_{\ssf{ C}^\lozenge \!=0 }  =  0$, we find 
\eqn{
\{ \ssf{ C} ,  \ssf{ C}^\lozenge_r \}_\tx{D}    \weakeq{\lozenge}  0 ~,
\label{eq:C_is_invaraint_under_subgauge_transformation}}
that is, $\ssf{ C}$ is invariant under the subgauge transformations.
This also implies that
\eqn{
\cl{G}_{\ssf{ C}^\lozenge_r}(\eta)  \weakeq{\lozenge}  0 ~,
\label{eq:gauge_flow_with_gravity}}
where
\eqn{
\cl{G}_\ssf{ O}(\eta) := \sum_{n=0}^\infty \frac{\eta^n }{n ! } \{  \ssf{ O} , \ssf{ C}  \}_n
}
with $\{ \cdot , \cdot \}_n$ being the $n$-th nested Dirac bracket defined similarly to the nested Poisson bracket in Eq.(\ref{eq:gauge_flow}).

\subsubsection{Relational observables \label{subsubsec:gravitational_relational_observable}}
As mentioned below Eq.(\ref{eq:Dirac_bracket}), the symplectic structure for the observer's internal variables is preserved on $\cl{ Z}$. Especially, $\alpha$ and $H_\tx{C}$ have a nonvanishing bracket only with each other, $\{ \alpha  , H_\tx{C} \}_\tx{D} = 1$ and thus, $\ssf{ C} $ given in Eq.(\ref{eq:singled_out_from_C^diff}) is the only generator in $\ssf{ C}^\tx{res}$ that transforms $\alpha$, and furthermore, 
\eqn{
\{ \alpha , \ssf{ C} \}_\tx{D} = 1 ~.
\label{eq:C_generates_mere_shift}}
Therefore, it can play the role analogous to that of $\tilde{C}^\tx{ex}_+ = N C^\tx{ex}_+$ defining the observables evolving with respect to the external-time parameter $t$ in Sec.\ref{subsubsec:classical_observable_relative_to_external_time}. 

\ 

In order to proceed in parallel with the discussion in that section, we consider a physical condition stronger than in Eq.(\ref{eq:weak_physical_condition_with_gravity}), 
\eqn{
\{ \ssf{ O}_\tx{ph} ,  \ssf{ C}^\tx{res}_r \}_\tx{D}    \weakeq{\lozenge}  0 ~,
}
that is, the equality holds without $\ssf{ C} = 0$, corresponding to the fact that $C_+^\tx{ex} = 0$ is not required for the Poisson bracket between the relational observable given as in Eq.(\ref{eq:classical_observable_relative_to_external_time_on_fixed_background}) and the constraint function $C_+^\tx{ex}$ in the fixed-background case. 

\ 

Considering quantities $\ssf{ O}_\lozenge$ that are invariant under the subgauge transformations satisfying\footnote{\label{fn:O^lozenge}
Recall that the set $\ssf{ C}^\lozenge$ is obtained by dropping the element with $u^{(0)}$ in Eq.(\ref{eq:definition_of_C^res}).
Therefore, it is sufficient for $\ssf{ O}_\lozenge$ to be (i) evaluated at the observer's position, (ii) scalar under the spatial diffeomorphisms, and (iii) not yielding the second- and higher-order derivatives of $u$ from their Poisson brackets with $\ssf{ C}_\perp [u]$.
Functions of the internal variables obviously satisfy these conditions.
Concerning field variables, if one assumes the Einstein-Hilbert with the minimally coupled scalar field $\upphi$, a trivial example of quantities satisfying the above conditions is $\upphi (\mb{x}_\tx{obs})$. A nontrivial example of geometric quantities is $[^{(d)}\!\ssf{ R} - \ssf{ K}^{ij} \ssf{ K}_{ij} + \ssf{ K}^2]_{\mb{x} = \mb{x}_\tx{obs}}$, constructed solely from the local data on a time slice, i.e., its Ricci scalar $^{(d)}\!\ssf{ R}$ and extrinsic curvature $\ssf{ K}_{ij}$. To see that this combination satisfies (iii), it may be enough to note that it is essentially the gravitational part of the Hamiltonian density.
}
\eqn{
\{ \ssf{ O}_\lozenge ,  \ssf{ C}^\lozenge_r \}_\tx{D}   \weakeq{\lozenge}  0 ~,
\label{eq:O^lozenge_is_invaraint_under_subgauge_transformation}}
we can construct the relational observables as
\eqn{
\ssf{ O}_\tx{ph} (a ; \alpha) :=& \, \cl{G}_{\ssf{ O}_\lozenge}( a -\alpha ) \\
=& \int_\mathbb{R} \rd \eta \left[ \sum_{n=0}^\infty \frac{\eta^n}{n !} \{   \delta (a - \alpha ) \, \ssf{ O}_\lozenge  , \ssf{ C}   \}_n  \right]  ~,
\label{eq:classical_observable_relative_to_internal_time}}
where $a$ is an internal-time parameter. 
For the group averaging with the generator $\sf C$, the Dirac brackets with $\sf C$ vanish algebraically.
On the other hand, their Dirac brackets with the elements of $\ssf{ C}^\lozenge$ vanish on $\cl{ Z}_\lozenge$ due to the weak equalities in Eqs.(\ref{eq:O^lozenge_is_invaraint_under_subgauge_transformation}) and (\ref{eq:C_is_invaraint_under_subgauge_transformation}).
Correspondingly, the set of physical quantities, constructed as above from all possible $\ssf{ O}_\lozenge$ and different values of $a$, forms a closed algebra under the Dirac bracket on $\cl{ Z}_\lozenge$.

\ 

Provided that $\ssf{ O}_\lozenge$ does not depend on $\alpha$, those relational observables satisfy
\eqn{
\pd_a  \ssf{ O}_\tx{ph} (a ; \alpha) = \{ \ssf{ O}_\tx{ph} (a ; \alpha)  , \ssf{ H}   \}_\tx{D}
}
with the Hamiltonian 
\eqn{
\ssf{ H}  := \ssf{ C}  - H_\tx{C} = M + H_\tx{D} + \ssf{ H}_\tx{g+m}[n] ~,
\label{eq:enrgy_of_universe_relative_to_clock}}
which is interpreted as the energy of the Universe evolving relative to the clock variable. 
Cf. the internal-time Hamiltonian (\ref{eq:classical_Hamiltonian_from_internal_perspective}) in the fixed background case.
Then, if one treats only the matter field dynamically, it becomes nonlinearly dependent on the matter Hamiltonian $\ssf{ H}_\tx{m}$; see Eq.(\ref{app:classical_Hamiltonian_from_internal_perspective_with_dynamical_matter}).

\ 

We note that this construction of the Dirac observables works regardless of whether the observer is freely falling or not.
With the matter-field detector turned on, in the definition of $u^{(0)}(\mb{x})$ given in Eq.(\ref{eq:u^(0)}), we have nonvanishing ${\mf D}_I$ whose nonlinear dependence on $H_\tx{C}$ seemingly complicates the analysis. 
However, ${\mf D}_I$ comes with $\ssf{ C}_\perp [u^{(I)}] \in \ssf{ Z}$ in $\ssf{ C}_\perp [u^{(0)}] \in \ssf{ C}^\tx{res}$ and such a term can be set to zero in the Dirac bracket; accordingly, we get $\sf C$ as in Eq.(\ref{eq:singled_out_from_C^diff}) that satisfies Eq.(\ref{eq:C_generates_mere_shift}) and generates the mere shift of $\alpha$.

\subsubsection{Multiple local observers \label{subsubsec:multiple_local_observers}}
In the presence of multiple observers with their own clocks, 
we can apply the same discussion to each observer, employing the partial gauge fixing such that a chosen observer appears to be at rest.

\ 

Suppose that there are Alice and Bob.
We adopt the Bob-centric coordinate system in which Bob's position $\mb{X}_\tx{B}$ is set to $\mb{x}_\tx{B}$, which plays the role of $\mb{x}_\tx{obs}$ in Eqs.(\ref{eq:n^(0)})--(\ref{eq:V_and_S}) defining the bases $\{ n^{(l)}(\mb{x}) \}$ and $\{ \mb{b}_{(m)}(\mb{x}) \}$.
Then, in turn, $\{ u^{(p)}(\mb{x}) \}_{p =0, p\geq d+1}$ and $\{ \mb{u}_{(q)}(\mb{x}) \}_{q \geq d+1}$ for the residual diffeomorphisms are defined in a similar manner but with Bob's effective mass $\ssf{ M}_\tx{eff,B}$ in Eq.(\ref{eq:u^(0)}).
The set of corresponding generators, denoted by $\ssf{ C}^\tx{res}_\tx{B}$ and decomposed into $\sf C_\tx{B}$ and $\ssf{ C}^\lozenge_\tx{B}$, satisfies the properties Eqs.(\ref{eq:subalgebra_on_Z}) and (\ref{eq:C_is_invaraint_under_subgauge_transformation}).
We note that their elements have the contribution from Alice; especially,
\eqn{
\ssf{ C}_\tx{B} = \ssf{ M}_\tx{B,eff} + \ssf{ H}_\tx{A} n(\mb{X}_\tx{A}) + \ssf{ H}_\tx{g+m}[n] ~,
}
cf. Eq.(\ref{eq:singled_out_from_C^diff}).  
Here, $\ssf{ H}_\tx{A}$ is Alice's Hamiltonian defined in the same manner as in Eq.(\ref{eq:observer_Hamiltonian_with_dynamical_gravity}) with Alice's position $\mb{X}_\tx{A}$ and momentum $\mb{P}_\tx{A}$.
With Bob's clock variable $\alpha_\tx{B}$ shifted by $\ssf{ C}_\tx{B}$, the relational observables that evolve relative to Bob's internal time are constructed similarly to in Sec.\ref{subsubsec:gravitational_relational_observable}.   

\ 

When we adopt the Alice-centric coordinate system, a similar discussion holds with ``A'' and ``B'' interchanged.
If Alice and Bob are assumed never to occupy the same spacetime point, it seems to also be possible to adopt a coordinate system where both Alice and Bob appear to be at rest.
However, it requires reconsidering the choice of expansion bases and is therefore not discussed in this paper.

\subsection{Quantum theory \label{subsec:unitarity_w.r.t._internal_time}} 
While formulating a quantum theory of gravity is a formidable task, we here proceed under the assumption that a quantum framework can be constructed in such a way that diffeomorphism invariance remains intact and it is effectively equivalent to the canonical quantum theory obtained from the above-discussed partially gauge-fixed theory by promoting the algebra under the Dirac bracket on the partially reduced phase space $\cl{ Z}$ to the corresponding algebra of operators under the commutator.

\subsubsection{Physical states}
As mentioned at the beginning of Sec.\ref{subsubsec:gravitational_relational_observable}, the symplectic structure for the observer's internal variables is preserved on the partially reduced phase space; correspondingly, the kinematical Hilbert space is factorized as in the fixed-background case, $\cl{ H}_\tx{kin} = \cl{H}_\tx{IN} \otimes \cl{H}_\tx{EX}$, where $\cl{H}_\tx{IN} = \cl{ H}_\tx{C} \otimes \cl{ H}_\tx{D}$ is associated with the internal d.o.f. but, on the other hand, $\cl{H}_\tx{EX}$ is associated with the external d.o.f. consisting of the modes of the matter and metric fields that remain independent after the partial gauge fixing with which the observer's location is gauged away. 

\ 

With the decomposition of $\ssf{ C}^\tx{res}$ discussed in Sec.\ref{subsubsec:residual_diffeos} in mind, the corresponding constraint operators define physical states $\dblket{\Psi_\tx{ph}}$ by
\eqn{ 
\hat{\ssf{C}}  \dblket{\Psi_\tx{ph}} = 0  \label{eq:gravitational_state_condition_0}}
and
\eqn{ 
\hat{\ssf{C}}^\lozenge_r \dblket{\Psi_\tx{ph}} = 0 ~ \label{eq:gravitational_state_condition_lozenge}
}
for all $r$. 
These physical conditions are mutually consistent due to the closure of the algebra, corresponding to that in Eqs.(\ref{eq:subalgebra_on_Z}) and (\ref{eq:C_is_invaraint_under_subgauge_transformation}), under the commutator replacing the Dirac bracket, which is now expressed as
\eqn{
[  \hat{\ssf{ C}}^\lozenge_r , \hat{\ssf{ C}}^\lozenge_{r'} ] \weakeq{\lozenge} 0 ~,~~~
[  \hat{\ssf{ C}}   , \hat{\ssf{ C}}^\lozenge_r  ]  \weakeq{\lozenge} 0 ~,
\label{eq:closed_algebra_at_quantum_level}}
where the weak equality $\weakeq{\lozenge}$ at the quantum level means that the equality holds on $\dblket{\Psi_\lozenge} \in \cl{ H}_\tx{kin}$ such that $\hat{\ssf{ C}}^\lozenge_r \dblket{\Psi_\lozenge}  = 0$.
Note that the operators $\hat{\ssf{ C}}^\lozenge = \{ \hat{\ssf{ C}}^\lozenge_r \}$ act trivially on $\cl{ H}_\tx{IN} $ since the internal Hamiltonian $\hat{H}_\tx{C} + \hat{H}_\tx{D}$ appears only in $\hat{\ssf{ C}} $; see Eq.(\ref{eq:singled_out_from_C^diff}).
It may be convenient to express them as $\hat{\ssf{ C}}^\lozenge = \hat{I}_\tx{IN} \otimes \hat{\ssf{ C}}^\lozenge_\tx{EX}$ with the identity operator on $\cl{H}_\tx{IN}$ and the constraint operators acting on $\cl{H}_\tx{EX}$.
Then, the subspace of $\cl{ H}_\tx{kin}$ formed by the states $\dblket{\Psi_\lozenge}$ can be written as $\cl{ H}_\lozenge :=\cl{ H}_\tx{IN}  \otimes \cl{ H}_{\tx{EX},\lozenge}$ with $\cl{ H}_{\tx{EX},\lozenge}$ being the subspace of $\cl{H}_\tx{EX}$ to be annihilated by the operators $\hat{\ssf{ C}}^\lozenge_\tx{EX}$.

\subsubsection{Unitarity with respect to internal time}
Projecting the first condition (\ref{eq:gravitational_state_condition_0}) to the clock state $\fixket{\alpha}$ defined in Sec.\ref{subsubsec:clock_states}, we get the Schr\"odinger equation for  $\fixket{\Psi_\tx{ph} (\alpha)} := \bkk{\alpha}{\Psi_\tx{ph}}$ with respect to the internal reference time $\alpha$, 
\eqn{
(\ri \pd_\alpha  - \hat{\ssf{ H}}  ) \fixket{\Psi_\tx{ph} (\alpha)} = 0  ~,
\label{eq:Schrodinger_eq_with_gravity}}
with the Hermitian operator $\hat{\ssf{ H}} $ corresponding to the energy of the rest of the Universe (\ref{eq:enrgy_of_universe_relative_to_clock}).
It is solved by 
\eqn{\fixket{\Psi_\tx{ph} (\alpha)} =  \hat{\ssf{ U}} (\alpha ; \alpha_0) \fixket{\Psi_{\lozenge,0} }
}
with the unitary evolution operator
\eqn{\hat{\ssf{U}} (\alpha ; \alpha_0 ) := e^{-\ri (\alpha - \alpha_0) \hat{\ssf{ H}}  } }
and the ``initial'' state $\fixket{\Psi_{\lozenge,0}} \in \cl{ H}_\tx{D} \otimes \cl{ H}_{\tx{EX},\lozenge}$ defined with a given fixed value $\alpha_0$ and normalized as $\fixbraket{\Psi_{\lozenge,0}}{\Psi_{\lozenge,0}} = 1$.

\ 

As in Eq.(\ref{eq:reconstruction_for_external_perspective}), the physical states $\dblket{\Psi_\tx{ph}}$ are reconstructed by $\dblket{\Psi_\tx{ph}} = \int_\mathbb{R} \rd \alpha \,  \fixket{\alpha} \otimes \fixket{\Psi_\tx{ph} (\alpha)}$ and satisfy the second condition (\ref{eq:gravitational_state_condition_lozenge}) as well, since $\hat{\ssf{ U}} (\alpha ; \alpha_0)^\dag \hat{\ssf{C}}_r^\lozenge \hat{\ssf{U}} (\alpha ; \alpha_0)$ is nothing but the quantum equivalent of $\cl{ G}_{\ssf{C}_r^\lozenge} (\eta)$ given in Eq.(\ref{eq:gauge_flow_with_gravity}) with $\eta = \alpha - \alpha_0$ and supposed to annihilate the state $\fixket{\Psi_{\lozenge,0}}$.

\subsubsection{EEP at quantum level}
In the present case, the fact that $\sf H$ has no interaction term between the external and internal d.o.f. when the matter-field detector is turned off directly implies that the outcomes of local experiments are independent of the external dynamics; more explicitly, the evolution of the reduced density matrix obtained by tracing out the external d.o.f. as in Eq.(\ref{eq:reduced_density_operator}),
\eqn{
\hat{\uprho}_\tx{exp} (\alpha) :=&  \Tr_\tx{EX} [ \fixket{\Psi_{\tx{ph}}(\alpha)} \fixbra{\Psi_{\tx{ph}}(\alpha)} ] ~, 
} 
is governed by the Liouville equation $\ri \pd_\alpha \hat{\uprho}_\tx{exp} = [\hat{H}_\tx{exp}, \hat{\uprho}_\tx{exp}]$ with the Hamiltonian $\hat{H}_\tx{exp} = \hat{H}_\tx{D,o}$ of the switched-off detector variable that represents the d.o.f. involved in the experiments.

\ 

Let us recall that, in the fixed-background case, the EEP is violated due to the nonunitarity of the internal-time evolution encapsulated by $[\hat{K}_\tx{o}]_\tx{AH}$ given in Eq.(\ref{eq:K_o,AH}), which is the anti-Hermitian part of the operator in the Schr\"odinger-type equation (\ref{eq:Schrodinger_eq_with_nonHermitian_Hamiltonian}) and originates from the indirect interaction between the internal and external d.o.f. through the square root in Eq.(\ref{eq:relativistic_particle_energy_operator}).
In this sense, the EEP at the quantum level is realized with diffeomorphism invariance that can eliminate such an unphysical indirect interaction.

\subsubsection{Relational observables} 
Physical operators $\hat{\ssf{O}}_\tx{ph}$ are defined by
\eqn{
[\hat{\ssf{C}} , \hat{\ssf{O}}_\tx{ph} ] \weakeq{\tx{res}} 0 ~, ~~~ [\hat{\ssf{C}}^\lozenge_r , \hat{\ssf{O}}_\tx{ph} ] \weakeq{\tx{res}} 0 ~,
\label{eq:physical_operator_condition_with_residual_diffeo}}
where the weak equality $\weakeq{\tx{res}}$ at the quantum level means that the equality holds on the physical Hilbert space $\cl{H}_\tx{ph}$ formed by the states satisfying the conditions (\ref{eq:gravitational_state_condition_0}) and (\ref{eq:gravitational_state_condition_lozenge}),
so that $\cl{H}_\tx{ph}$ is invariant under the action of those operators.

\

The relational Dirac observables given in the group-averaging form of Eq.(\ref{eq:classical_observable_relative_to_internal_time}) are promoted to the corresponding physical operators as
\eqn{
\hat{\ssf{O}}_\tx{ph} (a) =&\, \int_\mathbb{R} \rd \eta  e^{+ \ri \eta \hat{\ssf{C}}   } \left(  \projector{a} \otimes \hat{\ssf{O}}_\lozenge  \right)  e^{- \ri \eta \hat{\ssf{C}}  } \\
=&\, \int_\mathbb{R} \rd \alpha  \,  \projector{\alpha}  \otimes \hat{\ssf{U}}(a; \alpha)^\dag \hat{\ssf{O}}_\lozenge \hat{\ssf{U}}(a ; \alpha)   ~,
\label{eq:quantum_observable_relative_to_internal_time}}
where the bare operators $\hat{\ssf{O}}_\lozenge $, acting on $\cl{ H}_\tx{D} \otimes \cl{ H}_\tx{EX}$, are required to satisfy
\eqn{
[ \hat{\ssf{C}}^\lozenge_r  ,  \hat{\ssf{O}}_\lozenge   ]  \weakeq{\lozenge} 0 
\label{eq:O^lozenge}}
with $\weakeq{\lozenge}$ meaning that the equality holds on $\cl{ H}_\lozenge$,
see footnote \ref{fn:O^lozenge} for examples.
The operators so constructed algebraically commute with $\hat{\ssf{C}}$ and weakly commute with $\hat{\ssf{C}}^\lozenge_r$ due to the closure of the algebra
(\ref{eq:closed_algebra_at_quantum_level}) to satisfy the physical conditions (\ref{eq:physical_operator_condition_with_residual_diffeo}).
Those observables solve the Heisenberg equation $\ri \pd_a \hat{\ssf{O}}_\tx{ph} (a) = - [\hat{\ssf{H}} , \hat{\ssf{O}}_\tx{ph} (a) ]$ that describes the evolution with respect to the internal-time parameter $a$ generated by the Hamiltonian $\hat{\ssf{H}}$. 
{The structure is essentially the same as, or even simpler than, that in the fixed-background case discussed in Appendix \ref{appsec:Standard_quantum_mechanics_from_group_averaging}.}

\

We note that, if one ignored the majority of diffeomorphisms with a nondynamical observer staying at $\mb{x} = \mb{x}_\tx{obs}$ and considered only the one parameter family of transformations generated by $\ssf{ C}$, then the physical condition would be given by the same form of the first one (\ref{eq:gravitational_state_condition_0}) that leads to the Schr\"odinger equation (\ref{eq:Schrodinger_eq_with_gravity}) and physical observables would be given by the same form as in Eq.(\ref{eq:quantum_observable_relative_to_internal_time}) but with no additional condition (\ref{eq:O^lozenge}) on the bare operators, as in the fixed background case.  
With the matter-field detector turned off, it coincides with the model discussed in Refs.\cite{Chandrasekaran:2022cip,Witten:2023xze}.

\subsection{Challenges with dynamical gravity  \label{subsec:challenges}}
In the preceding discussion, we have proceeded by temporarily setting aside certain technical difficulties.
Here, we highlight the following two issues for future investigation: the first concerns the dynamical local observer, whereas the second is intrinsic to the theory of quantum gravity itself, irrespective of the observer's presence.
 
\ 

First, the dynamical local observer necessarily curves the spacetime and, modeled as a point particle, generates a singularity.
In addition, the point particle can have sensitivity to infinitely short scales, and then, such an effective description may seem inconsistent.
To avoid it, one could consider smearing the position of the observer.
However, in the Hamiltonian formalism with the $(d+1)$ decomposition, it is not clear how to implement such smearing while respecting full diffeomorphism invariance.

Alternatively, the observer could be modeled as an extended object described by a Nambu-Goto-type action with internal d.o.f.,
\eqn{
S_\tx{obs} =&\, \int_{\mathbb R} \rd \eta \int_{B^d} \rd^d \tx{y} \, |\mf g|^{1/2} [ {\scr L}_\tx{C} - \scr{M} ] \\
=&\, \int_{\mathbb R} \rd \eta \int_{B^d} \rd^d \tx{y} [\Omega \pd_\eta \alpha   -  |\mf g|^{1/2} ( {\scr M} + {\scr H}_\tx{C}  ) ] 
}
supplemented with an appropriate boundary term, where ${\scr L}_\tx{C}$ is a Lagrangian density of $\alpha (\eta , \mb{y})$, the clock scalar field on the worldvolume $\Sigma := {\mathbb R} \times B^d$ endowed with the metric $\mf{g}_\tx{ab} := g_{\mu \nu}(X) \pd_\tx{a} X^\mu \pd_\tx{b} X^\nu$ induced by the map $X$ from $\Sigma$ to the spacetime manifold, and $\mf{g} := \det \mf{g}_\tx{ab}$.
On the second line, we have used the clock's conjugate field $\Omega := |\mf g|^{1/2} \pd {\scr L}_\tx{C} / \pd (\pd_\eta \alpha)$.
Similarly to the action (\ref{eq:observer_action_eta}) of the point particle, the clock Hamiltonian density ${\scr H}_\tx{C} := |\mf g|^{-1/2} \Omega \pd_\eta \alpha  - {\scr L}_\tx{C}$ contributes to the effective mass density whose constant part is given by $\scr M$.
Generalization to the case with a matter-field detector is immediate. 
This type of model is formally distinguished from those with a reference field defined globally in spacetime \cite{Kuchar:1990vy,Brown:1994py,Kabel:2024lzr,Husain:2011tk,Hoehn:2023axh,Chen:2026kui} and warrants an independent investigation, which will be presented elsewhere, while the EEP falls outside the scope of consideration due to the possible presence of tidal forces.
 
\

Second, there is a difficulty of canonical quantum gravity that appears regardless of whether the observer is present or not.
It is the so-called factor-ordering problem \cite{Anderson:1959zzc,anderson1962,Schwinger:1963zzb,Komar:1979vd}: due to the noncommutativity of the operators, the operators corresponding to the generators in Eqs.(\ref{eq:generator_of_temporal_diffeo}) and (\ref{eq:generator_of_spatial_diffeo}) do not close under the commutator, and hence, diffeomorphism invariance appears to be spoiled.
It is argued in Ref.\cite{DeWitt:1967yk} that, since the ordering ambiguities come with $\delta (0)$, they should be ignored, as is done with dimensional regularization; however, the definitive treatment of this issue remains an open question \cite{Tsamis:1987wf,Friedman:1988sf}.
It may be naturally circumvented by recent developments in the relational path integral \cite{Falls:2025tid,Francois:2026qjv,Aguilar-Gutierrez:2026svf}.

We note that, as made clear by the author of \cite{Kuchar:1991qf} who regarded this difficulty as the ``problem of functional evolution,'' it should be separated from the multiple-choice problem of time; see Sec.\ref{subsubsec:Perspective (in)dependence}.

\section{Conclusion and outlook \label{sec:conclusion_and_outlook}}
The passage of time can be understood as a correlation between a variable that serves as a clock and the rest of the Universe.
There appears to be no \textit{a priori} choice for the clock; however, one natural possibility is that such a variable is internal to a spatially localized observer.
In this paper, we discussed whether the evolution relative to such a local observer's clock is unitary and how it is related to the EEP.

Our findings highlight the significance of diffeomorphism invariance, as summarized in Sec.\ref{sec:summary/overview}.
On a fixed background, the internal-time unitarity does not hold due to the artifactual indirect coupling between the internal and external d.o.f. through the square root in the relativistic form (\ref{eq:relativistic_particle_energy_operator}) of particle energy; it gives rise to the anti-Hermitian part of the effective Hamiltonian (\ref{eq:K_AH}), and consequently, the EEP is violated in the sense that the internal-time evolution of the reduced density matrix described by Eq.(\ref{eq:internal-time_derivative_of_reduced_density_matrix}) depends on the quantum state of the external d.o.f.
On the other hand, with dynamical gravity, such an indirect coupling is physically absent, as manifested by the linear dependence on the internal Hamiltonian $H_\tx{C}+ H_\tx{D}$ of the constraint function (\ref{eq:singled_out_from_C^diff}),
shown by adopting the observer-centric coordinate system and working on the associated partially reduced phase space.
Assuming that there exists a quantum equivalent as an effective field theory with diffeomorphism invariance kept intact, we argued that the evolution relative to the internal clock is unitary, and the EEP is preserved at the quantum level.

\

It appears that our formulation with the partial gauge fixing is closely related to the QRF approach, both based on the language of constrained Hamiltonian systems.
As emphasized in Refs.\cite{Vanrietvelde:2018pgb,Vanrietvelde:2018dit},
choosing an internal perspective is equivalent to choosing a gauge, albeit for global reparametrizations.
While there are proposals \cite{Hardy:2018kbp,Hardy:2019cef,Giacomini:2020ahk,Giacomini:2021aof,Giacomini:2021gei,delaHamette:2021iwx,Giacomini:2022hco,Chen:2026kui}, it seems fair to say that a complete understanding of the quantum general coordinate transformation in terms of the QRF transformation is still lacking, and further investigation is required in future work to fill this gap.
In the context of the present work, our primary interest will be whether the ``perspective-neutral'' theory permits the unitarity with respect to the local internal clock as well as the EEP.
In that process, however, the challenges mentioned in Sec.\ref{subsec:challenges} will need to be properly addressed or circumvented.

As a specific application, an appropriate global QRF transformation can recast the dynamics of an object under a superposition of mass configurations into that of the object in a spatial superposition under a definite mass configuration \cite{delaHamette:2022cka,Foo:2023vbr}, reinterpreting the gravitationally induced entanglement setup \cite{Bose:2017nin,Marletto:2017kzi} that involves two massive objects both spatially superposed. 
If the local QRF transformation or the quantum general coordinate transformation is established, then such a setup can be described without the spatial superposition of the objects, relying solely on the superposition of gravitational field configurations, consistently with the fact that what matters is the physical distance between the two objects \cite{Christodoulou:2018cmk}.
In the current formulation, it corresponds to adopting coordinate systems in which two objects appear to be at rest, mentioned at the end of Sec.\ref{subsubsec:multiple_local_observers} to be left for future work. 

\

Local observers carrying internal d.o.f., modeled as UDW-like particle detectors, are employed across a wide spectrum of theoretical and phenomenological scenarios.
Originally introduced to study the thermodynamic properties of causal horizons, such as the Rindler, black hole \cite{Unruh:1976db}, and de Sitter \cite{Gibbons:1977mu} horizons, this operational framework is used to analyze modern protocols, ranging from environment-induced decoherence \cite{Unruh:1989dd,Hu:1993qa,Alsing:2003es} to entanglement/correlation harvesting \cite{Valentini:1991eah,Reznik:2002fz,Pozas-Kerstjens:2015gta}. 
In the context of relativistic quantum measurement theory, the long-standing paradox concerning causality \cite{Sorkin:1993gg} is circumvented by considering a spatially extended detector composed of fields \cite{Fewster:2018qbm,Bostelmann:2020unl}.
Furthermore, the utility of the UDW formalism extends into empirical domains, offering effective models for laboratory instrumentation. In quantum optics, it naturally maps onto light-matter interactions \cite{Martin-Martinez:2012ysv,Lopp:2020qwx}. 
In gravitational-wave physics based on the theory of a massless spin-2 field on a fixed background, the geodesic separation between two freely falling particles can be regarded as the internal detector variable \cite{Parikh:2020nrd,Kanno:2020usf}.
Furthermore, Refs.\cite{Barbado:2020snx,Foo:2020xqn,Foo:2020jmi,Foo:2021exb,Foo:2022dnz} consider the superposition of accelerations/curvatures or topologies, and Refs.\cite{Danielson:2022tdw,Danielson:2022sga,Wilson-Gerow:2024ljx,Biggs:2024dgp,Danielson:2024yru} discuss the horizon-induced decoherence.

As in the present work, one might associate those detectors in the various setups with the observers themselves who describe their local experiments with respect to their own local clocks.\footnote{
If the observer measures its own clock, the nonunitarity arises \cite{Paiva:2021wbv} unless certain classes of interactions are assumed
\cite{Kuypers:2024gqr,Rijavec:2025vti}.
Such a stance is not adopted in this paper; see the last paragraph of Sec.\ref{subsubsec:clock_states}.
}
When the observer's position is assumed to be nondynamical or strictly nonrelativistic, as mentioned below Eq.(\ref{eq:nonrelativistic_expansion}), the internal and external perspectives are trivially equivalent; thus, results obtained from those particle detector models remain true when described with respect to the internal clock.
However, if the observers are dynamical, then the two perspectives are inequivalent (\ref{eq:inequivalence}) on a general fixed background spacetime and, to realize the internal-time unitarity, a fully diffeomorphism-invariant model with dynamical spacetime discussed in Sec.\ref{sec:observers_with_dynamical_gravity} would be one of the most natural formulations, in which it may be worth revisiting the above-mentioned setups to see whether the evolution of detector alters or not, especially in the case of a noninertial observer since, aside from direct gravitational wave detections, our primary tools of cosmological observations still remain nongravitational fields, such as electromagnetism.

\ 

In recent discussions on the type reduction of the von Neumann algebra of observables,
introducing an observer, or equivalently a QRF for time \cite{DeVuyst:2024khu,DeVuyst:2024fxc}, is crucial.
If it is associated with the anti-de Sitter boundaries of a two-sided black hole, the type $\text{III}_1$ algebra that emerges in the dual theory \cite{Leutheusser:2021qhd,Leutheusser:2021frk} reduces to the type $\tx{II}_\infty$ \cite{Witten:2021unn,Chandrasekaran:2022eqq}.
If an observer with the clock variable is placed at the origin of the static patch of the de Sitter spacetime, one obtains the $\tx{II}_1$ algebra \cite{Chandrasekaran:2022cip}.
As emphasized in Ref.\cite{Fewster:2024pur}, this mechanism requires a time-translation isometry within a spacetime subregion that coincides with the modular flow.
In the present formulation with the dynamical observer, we obtained a similar form of the observables with the crossed-product structure (\ref{eq:quantum_observable_relative_to_internal_time}) with the additional restriction (\ref{eq:O^lozenge}).
However, since no decomposition of fluctuations from the background with the desired features is assumed, the type-reduction argument cannot be directly applied here.
Rather, our construction may contribute to further developments in the direction of defining an algebra of local observables in a background-independent manner \cite{Witten:2023xze}.

\ 

While the Hamiltonian formalism that we employed in this paper provides an intuitive grasp of physical quantities, the path-integral formulation is indispensable for the broader scope of quantum theory.
The importance of incorporating a local observer into the gravitational path integral is demonstrated in Refs.\cite{Abdalla:2025gzn,Harlow:2025pvj}.
It resolves nonperturbative Hilbert space puzzles, such as preventing the Hilbert space of a closed universe from collapsing to one dimension and extending the validity of effective field theory for infalling observers to exponentially long times.
We hope to revisit the unitarity of the evolution from a local observer's perspective based on the path-integral formalism.

\ 

Finally, we note that throughout this paper, we assumed the existence of an effective local d.o.f. that is decoupled from nongravitational fields and serves as the local clock. At a more fundamental level, however, since physical clocks are ultimately composed of matter fields, such a decoupling may be only effective. Residual interactions between the clock and the surrounding fields could therefore provide an additional, potentially fundamental source of nonunitarity in evolution relative to the internal clock. Investigating whether such interaction-induced nonunitarity is fundamental and, if so, exploring its phenomenological consequences is an interesting direction for future work.

\begin{acknowledgments}
 The author thanks S. Aoki, Y. Hidaka and S. Iso for valuable discussions on the present and related topics. 
 This work is supported in part by the JSPS Grant-in-Aid for Scientific Research No.~JP22H00129.
\end{acknowledgments}

\appendix

\section{Standard quantum mechanics from group averaging \label{appsec:Standard_quantum_mechanics_from_group_averaging}}
As seen in Secs.\ref{sec:relativistic_particle_with_clock_and_detector} and \ref{sec:observers_with_dynamical_gravity}, the standard Hamilton dynamics emerge in fully constrained systems.
Taking the quantum theory from the external perspective as an example, we discuss how the physical quantities are constructed from the so-called group-averaging technique \cite{Higuchi:1991tm,Higuchi:1991tp,Ashtekar:1995zh,Marolf:1995cn,Marolf:2000iq} to implement a certain constraint.
Now with the nontrivial lapse function $\hat{N} := N(\hat{X})$, it can be seen as a generalization of the standard formulation commonly found in the literature; see \cite{Hoehn:2019fsy} for example.
We also discuss how the concept of reference time fits in the Schr\"odinger/Heisenberg picture.

\subsection{Physical state, operator and inner product \label{appsubsec:physical_state_and_operator}}
By sandwiching the constraint operator (\ref{eq:constraint_operator_for_external_perspective}) with $\hat{N}^{1/2}$, we have
\eqn{
\hat{\tilde{C}}^\tx{ex}_+ := \hat{N}^{1/2} \hat{C}^\tx{ex}_+ \hat{N}^{1/2} = -\hat{E} + \hat{H}^\tx{ex} 
}
and can write the physical states that satisfy the condition (\ref{eq:physical_state_condition_for_external_perspective}) as
\eqn{
\dblket{\psi_\tx{ph}^\tx{ex}} = \hat{\delta}_\tx{ph} \dblket{\psi_\tx{kin}} ~,
\label{app:physical_state}}
where
\eqn{
 \hat{\delta}_\tx{ph} := \hat{N}^{1/2} 2 \pi  \delta (\hat{\tilde{C}}^\tx{ex}_+) \hat{N}^{1/2} ~
\label{app:physical_state_projection}}
is the ``projector'' onto the physical Hilbert space.
In fact, as obtained in Ref.\cite{Castro-Ruiz:2019nnl} using the Trotter-Suzuki formula,
\eqn{
2\pi \delta (\hat{\tilde{C}}^\tx{ex}_+) =&\, \int_\mathbb{R} \rd \eta \, e^{- \ri \eta \hat{\tilde{C}}^\tx{ex}_+ }  \\
=&\,  \int_\mathbb{R}\rd T \rd T' \, \fixket{T}\hat{U}^\tx{ex}(T,T') \fixbra{T'} ~
\label{app:projector_fourier_transformed}}
with $\hat{U}^\tx{ex}(T,T') := \fixbra{T} 2\pi \delta (\hat{\tilde{C}}^\tx{ex}_+) \fixket{T'}$ corresponding to the unitary evolution operator in Eq.(\ref{eq:unitary_evolution_operator}), and thus 
we find 
\eqn{
\dblket{\psi_\tx{ph}^\tx{ex}} =&\,  \hat{N}^{1/2} \int_\mathbb{R}\rd T \rd T' \, \fixket{T}\hat{U}^\tx{ex}(T,T') \fixbra{T'} \hat{N}^{1/2} \dblket{\psi_\tx{kin}} \\
=&\,  \hat{N}^{1/2} \int_\mathbb{R}\rd T  \, \fixket{T} \otimes \hat{U}^\tx{ex}(T,T_0)  \fixket{\tilde{\psi}^\tx{ex}_0}~,
}
reproducing the expression (\ref{eq:reconstruction_for_external_perspective}) with the corresponding $T$ representation $\fixket{\tilde{\psi}^\tx{ex}_\tx{ph} (T)} = \fixbra{T}  \hat{N}^{-1/2}\dblket{\psi^\tx{ex}_\tx{ph}}$ that satisfies the Schr\"odinger equation (\ref{eq:Schrodinger_eq_in_T-rep}), in which
\eqn{
\fixket{\tilde{\psi}^\tx{ex}_0}:= \int_\mathbb{R}  \rd T' \,  \hat{U}^\tx{ex}(T_0,T') \fixbra{T'} \hat{N}^{1/2} \dblket{\psi_\tx{kin}}
\label{app:initial-state}}
plays the role of the initial state.
Here, the introduction of the initial time $T_0$ and the state at that time $\fixket{\tilde{\psi}^\tx{ex}_0} = \fixket{\tilde{\psi}^\tx{ex}_\tx{ph} (T_0)}$ are merely to get a familiar form of the state in the Schr\"odinger picture (\ref{eq:solution_of_Schrodinger_eq_w.r.t_T}),
\eqn{
\fixket{\tilde{\psi}^\tx{ex}_\tx{ph} (T)} = \hat{U}^\tx{ex}(T,T_0) \fixket{\tilde{\psi}^\tx{ex}_0} ~.
\label{app:solution_of_Schrodinger_eq_w.r.t_T}
}

\ 

The physical inner product between two states $\dblket{\psi_\tx{ph}^\tx{ex}}$ and $\dblket{\chi_\tx{ph}^\tx{ex}}$ is defined as
\eqn{
\dblbraket{\chi_\tx{ph}^\tx{ex}}{\psi_\tx{ph}^\tx{ex}}_\tx{ph} := &\, \dblbra{\chi_\tx{kin}}   \hat{\delta}_\tx{ph}    \dblket{\psi_\tx{kin}} \\
=&\, \fixbraket{\tilde{\chi}_\tx{ph}^\tx{ex} (T)}{\tilde{\psi}_\tx{ph}^\tx{ex} (T)} = \fixbraket{\tilde{\chi}_0^\tx{ex}}{\tilde{\psi}_0^\tx{ex}}  ~.
\label{app:physical_inner_product}}
With the expression in Eq.(\ref{app:projector_fourier_transformed}), one can verify the second line. 
By definition, it has no external-time dependence and corresponds to the one defined in Eq.(\ref{eq:physical_inner_product}),
\eqn{
\dblbraket{\chi_\tx{ph}^\tx{ex}}{\psi_\tx{ph}^\tx{ex}}_\tx{ph} = \dblbra{\chi^\tx{ex}_\tx{ph}}  (\hat{\tilde{\Pi}}_\tx{EX}(T) \otimes \hat{I}_\tx{IN})  \dblket{\psi^\tx{ex}_\tx{ph}}   
\label{app:physical_inner_product_with_projector}
}
with the projector $\hat{\tilde{\Pi}}_\tx{EX}(T) := \hat{N}^{-1/2} (\projector{T} \otimes \hat{I}_\tx{X})\hat{N}^{-1/2}$.

\ 

To construct physical operators, we start with quantizing the expression (\ref{eq:classical_observable_relative_to_external_time_on_fixed_background}).
With the nested Poisson bracket replaced by the nested commutator with $(-\ri)^n$ and the delta function understood in terms of spectral decomposition, i.e., $\delta (t- \hat{T}) = \int_\mathbb{R} \rd T \fixket{T} \delta (t-T) \fixbra{T} = \projector{t}$, we obtain
\eqn{
\hat{\tilde{O}}_\tx{ph}  (t) = &\, \int_\mathbb{R} \rd \eta  e^{+ \ri \eta \hat{\tilde{C}}^\tx{ex}_+ } \left(  \projector{t}  \otimes \hat{O} \right)  e^{- \ri \eta \hat{\tilde{C}}^\tx{ex}_+ }  \\
=&\, \int_\mathbb{R} \rd T \, \projector{T}\otimes \hat{\tilde{O}}_\tx{ph} (t;T)   ~,
\label{app:quantum_observable_relative_to_external_time_on_fixed_background_group_averaging}
}
where
\eqn{\hat{\tilde{O}}_\tx{ph} (t;T) := \hat{U}^\tx{ex}(T,t) \hat{O} \hat{U}^\tx{ex}(t,T) ~.
\label{app:Heisenberg_operator_tilde}}
One can also consider bare functions of operators explicitly depending on $\hat{T}$; then the operator inside the parentheses in Eq.(\ref{app:quantum_observable_relative_to_external_time_on_fixed_background_group_averaging}) is replaced by $\hat{O}(\hat{T}) (\projector{t}\otimes \hat{I}_{\overline{\tx{T}}}) = \projector{t}  \otimes \hat{O}(t)$, and we have $\hat{O}(t)$ instead of $\hat{O}$ in Eq.(\ref{app:Heisenberg_operator_tilde}).
They algebraically commute with $\hat{\tilde{C}}^\tx{ex}_+$ and satisfy the Heisenberg equation with respect to $t$, corresponding to the classical equation (\ref{eq:physical_evolution_relative_to_external_time_on_fixed_background}), as
\eqn{
\pd_t \hat{\tilde{O}}_{\tx{ph}}(t  )  =& +\{ \hat{\tilde{O}}_{\tx{ph}}(t  ) , \hat{\tilde{H}}^\tx{ex}_{\tx{ph}}(t ; T ) \}_\tx{P}  + \pd_\tx{xpl} \hat{\tilde{O}}_{\tx{ph}}(t )  ~,
\label{app:Heisenberg_eq}}
where $\hat{\tilde{H}}^\tx{ex}_{\tx{ph}}(t ; T )$ and $\pd_\tx{xpl} \hat{\tilde{O}}_{\tx{ph}}(t )$ are given by Eq.(\ref{app:Heisenberg_operator_tilde}) with $\hat{O}$ replaced by $\hat{H}^\tx{ex}(t)$ and $\pd_t \hat{O}(t)$, respectively.
Then, operators 
\eqn{
\hat{O}_\tx{ph}  (t) := \hat{N}^{+1/2} \hat{\tilde{O}}_\tx{ph} (t)   \hat{N}^{-1/2}
\label{app:physical_operator}}
weakly commute with the constraint operator in the sense that the commutator vanishes on physical state kets,
\eqn{
[\hat{O}_\tx{ph}  (t) , \hat{C}^\tx{ex}_+  ] \dblket{\psi_\tx{ph}^\tx{ex}} = 0 ~,
}
whereas their Hermitian conjugates commute with the constraint operator on physical state bras,
\eqn{
\dblbra{\chi_\tx{ph}^\tx{ex}} [\hat{O}_\tx{ph}^\dag  (t) , \hat{C}^\tx{ex}_+  ]  = 0 ~.
}

\subsection{Schr\"odinger/Heisenberg picture}
Let us consider a physical inner product with an inserted physical operator,
\eqn{
\dblbra{\chi_\tx{ph}^\tx{ex}} \hat{O}_\tx{ph}  (t) \dblket{\psi_\tx{ph}^\tx{ex}}_\tx{ph} := &\, \dblbra{\chi_\tx{kin}} \hat{O}_\tx{ph}  (t)\hat{\delta}_\tx{ph} \dblket{\psi_\tx{kin}} \\
= &\, \dblbra{\chi_\tx{kin}}  \hat{\delta}_\tx{ph} \hat{O}_\tx{ph}^\dag  (t) \dblket{\psi_\tx{kin}} ~.
\label{app:expectation_value}}
Using Eq.(\ref{app:projector_fourier_transformed}) along with the definitions (\ref{app:solution_of_Schrodinger_eq_w.r.t_T}) and (\ref{app:quantum_observable_relative_to_external_time_on_fixed_background_group_averaging}), one can check that
\eqn{
&\dblbra{\chi_\tx{ph}^\tx{ex}} \hat{O}_\tx{ph}  (t) \dblket{\psi_\tx{ph}^\tx{ex}}_\tx{ph} \\
&=\, \dblbra{\chi_\tx{ph}^\tx{ex}} \hat{N}^{-1/2} \left( \projector{t} \otimes \hat{O} \right) \hat{N}^{-1/2} \dblket{\tilde{\psi}_\tx{ph}^\tx{ex}}  ~,
}
which can be understood as the expression (\ref{app:physical_inner_product_with_projector}) with the identity operator on $\cl{H}_{\overline{\tx{T}}} := \cl{H}_\tx{IN} \otimes \cl{H}_\tx{X}$ replaced by $\hat{O}$.
Then, it is straightforward to see that
\eqn{
\dblbra{\chi_\tx{ph}^\tx{ex}} \hat{O}_\tx{ph}  (t) \dblket{\psi_\tx{ph}^\tx{ex}}_\tx{ph}
=&\, \fixbra{\tilde{\chi}_\tx{ph}^\tx{ex} (t)} \hat{O} \fixket{\tilde{\psi}_\tx{ph}^\tx{ex} (t)} \\
=&\, \fixbra{\tilde{\chi}_\tx{0}^\tx{ex}}  \hat{\tilde{O}}_\tx{ph} (t;T_0) \fixket{\tilde{\psi}_\tx{0}^\tx{ex} } \\
=&\, \fixbra{\tilde{\chi}_\tx{ph}^\tx{ex} (T)}  \hat{\tilde{O}}_\tx{ph} (t;T) \fixket{\tilde{\psi}_\tx{ph}^\tx{ex} (T)} ~,
\label{app:Schrodinger/Heisenberg}}
where the first and second lines give the corresponding expressions in the Schr\"odinger and Heisenberg pictures, respectively.

\ 

\begin{figure}[t]
\begin{center}
\includegraphics[width=8cm]{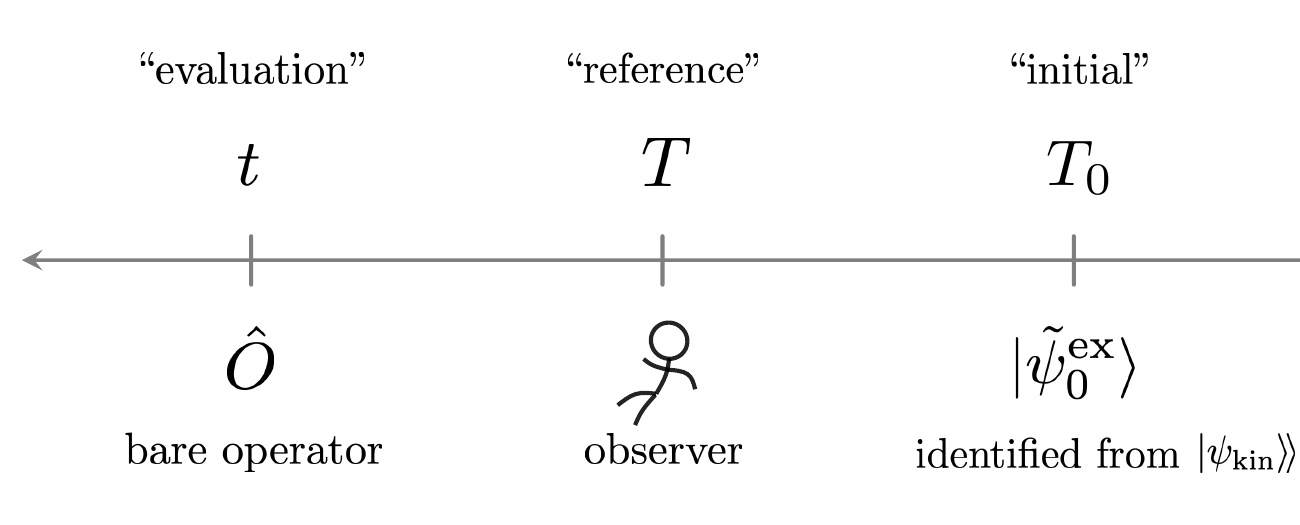}
\caption{The roles of the three distinct times in the formulation.
At the initial time $T_0$, the state $\fixket{\tilde{\psi}^\tx{ex}_0}$ is obtained from the kinematical state $\dblket{\psi_\tx{kin}}$ as in Eq.(\ref{app:initial-state}).
The reference time $T$ is interpreted as the observer's temporal location. 
A physical measurement is intended to occur at the evaluation time $t$ to which the time-independent bare function of operators $\hat{O}$ is assigned to be pulled back along the dynamical flow to the reference time.
For practical computational purposes in evaluating single-time expectations, one can safely evaluate these three times at the same instant ($t = T =T_0$) without loss of generality, as the quantity in Eq.(\ref{app:Schrodinger/Heisenberg}) is only dependent on $t$.
When evaluating multitime expectations, one has the corresponding number of time parameters.
}
\label{fig:time}
\end{center}
\end{figure}

Then, although the last equality is a trivial rewriting, it suggests an interpretation of these equivalent pictures interfaced by the reference time as follows. 
The reference time $T$, specified via the projection to $\fixket{T}$, can be interpreted as the temporal location of the observer; see Fig.\ref{fig:time}.
Suppose that the initial time $T_0$ introduced in Eq.(\ref{app:initial-state}) is fixed.
If the reference time $T$ is varied while the relative interval $t-T$ is kept constant (typically set to zero), we recover the Schr\"odinger picture, which yields the ``present'' values of observables at $T$.
On the other hand, if the evaluation time $t$ is varied while the reference interval $T-T_0$ is kept constant (typically set to zero), we obtain the Heisenberg picture, which predicts the ``future (or past)'' values of observables at $t \ne T$.

\subsection{Without flattening  \label{appsubsec:without_flattening}}
The eigenstates $\fixket{X}$ of $\hat{X}^\mu = (\hat{T} , \hat{\mb{X}})$ before the flattening are related to those after the flattening, $\fixket{T} \otimes \fixket{\mb{X}}$, as $ \fixket{X} := |g|^{-1/4}_X \fixket{T} \otimes \fixket{\mb{X}}$ with $|g|_X = N(X) h^{1/2}(X)$, and hence, the identity operator on $\cl{ H}_\tx{EX}$ is written as
\eqn{
\hat{I}_\tx{EX} &= \int \rd T \rd^d \tx{X} \projector{T} \otimes \projector{\mb{X}}  \\ 
&= \int \rd^{d+1} X |g|^{1/2}_X \projector{X}  ~.
}
The conjugate momentum is represented as $\fixbra{X}P_\mu \fixket{Y} = |g|^{-1/4}_X  (- \ri ) (\pd /\pd X^\mu) \fixbraket{X}{Y} |g|^{+1/4}_Y$ with respect to the basis $\fixket{X}$, where $\fixbraket{X}{Y} = \delta_g^{(d+1)}(X,Y)$ is the scalar Dirac delta function.

\ 

The projector (\ref{app:physical_state_projection}) is now expressed as 
\eqn{ 
\hat{\delta}_\tx{ph} =& \int \rd^{d+1} X |g|_X^{1/2} \int \rd^{d+1} X'  |g|_{X'}^{1/2}   \\ 
& \hspace{5em} \fixket{X}  \fixbra{X'} \otimes \hat{\cl{W}} (X , X') 
}
with the Wightman-type ``vacuum propagator'' 
\eqn{
\hat{\cl{W}} (X , X') :=& \fixbra{X} \hat{\delta}_\tx{ph} \fixket{X'} \\
=&  h_X^{-1/4}  \fixbra{\mb{X}}\hat{U}^\tx{ex} (T , T') \fixket{\mb{X}'} h_{X'}^{-1/4} ~,
}
where $h_X := h (X)$, that satisfies 
\eqn{
\int \rd^{d+1} X'' |g|_{X''}^{1/2} \fixbra{X} \hat{C}^\tx{ex}_+ \fixket{X''} \hat{\cl{W}} (X'' , X') = 0 ~ 
}
as well as
\eqn{
\int \rd^d \tx{X}'' h_{X''}^{1/2}   \hat{\cl{W}} (X , X'') \hat{\cl{W}} (X'' , X') = \hat{\cl{W}} (X , X') 
}
without the integration of $T''$.
For later convenience, let us also introduce
\eqn{
\hat{\cl{U}} (X , X') :=& \fixbra{X}  2 \pi  \delta (\hat{\tilde{C}}^\tx{ex}_+) \fixket{X'} \\
=& N^{-1/2}(X) \hat{\cl{W}} (X , X') N^{-1/2}(X') 
}
which is the equivalent of $\fixbra{\mb{X}}\hat{U}^\tx{ex} (T , T') \fixket{\mb{X}'}$ in the case with the flattened measure.

\ 

The physical states defined by Eq.(\ref{app:physical_state}) can be written as
\eqn{
\dblket{\psi_\tx{ph}^\tx{ex}} =\,   \int \rd^{d+1} X |g|^{1/2}_X \, \fixket{X} \otimes \bkk{X }{\psi_\tx{ph}^\tx{ex}} ~,
}
\eqn{
\bkk{X }{\psi_\tx{ph}^\tx{ex}} = &\int \rd^{d+1} X' |g|^{1/2}_{X'}  \hat{\cl{W}}(X,X') \bkk{X'}{\psi_\tx{kin}} \\
=&  \int \rd^d \tx{X}_0 h^{1/2}_{X_0}  \hat{\cl{W}}(X,X_0)  \fixket{\psi^\tx{ex}_0 (\mb{X}_0)} ~,
}
where
\eqn{
\fixket{\psi^\tx{ex}_0 (\mb{X}_0)} := \bkk{X_0 }{\psi_\tx{ph}^\tx{ex}} = h_{X_0}^{-1/4} \fixbraket{\mb{X}_0}{\tilde{\psi}^\tx{ex}_0}
}
is the initial state given at $T_0$, which is the temporal component of $X_0^\mu$.
In the last expression, we have used the state (\ref{app:initial-state}) defined with the basis after the flattening.
One might also consider 
\eqn{
&\bkk{X}{\tilde{\psi}_\tx{ph}^\tx{ex}} : =\fixbra{X } \hat{N}^{-1/2} \dblket{\psi_\tx{ph}^\tx{ex}} \\
&= \int \rd^{d+1} X' |g|^{1/2}_{X'}  \hat{\cl{U}}(X,X') \fixbra{X } \hat{N}^{1/2} \dblket{\psi_\tx{kin} } \\
&=  \int \rd^d \tx{X}_0 |g|^{1/2}_{X_0}  \hat{\cl{U}}(X,X_0)  \fixket{\tilde{\psi}^\tx{ex}_0 (\mb{X}_0)} ~,
}
which corresponds to the state (\ref{app:solution_of_Schrodinger_eq_w.r.t_T}),
where
$\fixket{\tilde{\psi}^\tx{ex}_0 (\mb{X}_0)} := \bkk{X_0}{\tilde{\psi}_\tx{ph}^\tx{ex}}$.

\ 

The physical inner product is defined as in Eq.(\ref{app:physical_inner_product}).
One can write it in the form of Eq.(\ref{app:physical_inner_product_with_projector}) using the projector $\hat{\tilde{\Pi}}_\tx{EX}(T) = \int \rd^d \tx{X} \,  h^{1/2}_X \projector{ X }$ now written in terms of the original basis before the flattening.
While the value of $T = X^0$ is specified by the projector, the inner product itself does not depend on it;
\eqn{
\dblbraket{\chi_\tx{ph}^\tx{ex}}{\psi_\tx{ph}^\tx{ex}}_\tx{ph} &= \dblbra{\chi^\tx{ex}_\tx{ph}}  (\hat{\tilde{\Pi}}_\tx{EX}(T) \otimes \hat{I}_\tx{IN})  \dblket{\psi^\tx{ex}_\tx{ph}}  \\
& = \int \rd^d \tx{X} h^{1/2}_X  \bbk{X }{\chi_\tx{ph}^\tx{ex}} \bkk{X }{\psi_\tx{ph}^\tx{ex}} \\
&= \int \rd^d \tx{X}_0 h^{1/2}_{X_0}  \fixbraket{\chi^\tx{ex}_0 (\mb{X}_0)}{\psi^\tx{ex}_0 (\mb{X}_0)} ~.
\label{app:inner_product_with_proper_volume_element}}
We note that it comes with the proper volume element on the $d$-dimensional time slice.
Therefore, when we consider the norm of physical states using this inner product, $\| \bkk{X}{ \psi^{\tx{ex}}_\tx{ph}} \|^2 := \bbk{X }{\psi_\tx{ph}^\tx{ex}} \bkk{X }{\psi_\tx{ph}^\tx{ex}}$ can be interpreted as the scalar probability distribution function of $\mb{X}$ on the time slice specified by a given value of $T = X^0$.

\

In the definition of the physical operators, we have $\projector{t} = \hat{\Pi}_\tx{EX}(t)$, where $\hat{\Pi}_\tx{EX}(t) := \int \rd^d \tx{X} \,  |g|^{1/2}_X \projector{ X } = \hat{N}^{1/2} \hat{\tilde{\Pi}}_\tx{EX}(t)\hat{N}^{1/2}$.
With bare operators $\hat{O}$ commuting with $\hat{T}$ and hence with $\hat{\Pi}_\tx{EX}(t)$, the operators (\ref{app:quantum_observable_relative_to_external_time_on_fixed_background_group_averaging}) that algebraically commute with $\hat{\tilde{C}}^\tx{ex}_+$ can be expressed as
\eqn{
\hat{\tilde{O}}_\tx{ph}  (t) = &\, \int_\mathbb{R} \rd \eta  e^{+ \ri \eta \hat{\tilde{C}}^\tx{ex}_+ } \left(  \hat{\Pi}_\tx{EX}(t) \, \hat{O} \right) e^{- \ri \eta \hat{\tilde{C}}^\tx{ex}_+ }  \\
=&\, \int \rd^{d+1} X |g|_X^{1/2} \int \rd^{d+1} Y |g|_{Y}^{1/2}  \\
& \hspace{1em} \delta (X^0 - Y^0) \, \fixket{X} \fixbra{Y} \otimes   \hat{\tilde{O}}_\tx{ph} (t;X,Y)   ~,
\label{app:quantum_observable_in_terms_of_original_basis}
}
where
\eqn{
\hat{\tilde{O}}_\tx{ph} (t;X,Y) \! :=& \!\int \rd^{d+1} Z |g|_Z^{1/2} \int \rd^{d+1} Z' |g|_{Z'}^{1/2} \\
&\hspace{1em}\hat{\cl{U}}(X , Z)  \fixbra{Z}  \hat{\Pi}_\tx{EX}(t) \hat{O}   \fixket{Z'} \hat{\cl{U}}(Z' , Y) ~.
}
 
\ 

Again, one can see that the expectation values of physical operators (\ref{app:physical_operator}) defined with the inner product (\ref{app:expectation_value}) are given in the form equivalent to Eq.(\ref{app:expectation_value}) as
\eqn{
&\dblbra{\chi_\tx{ph}^\tx{ex}} \hat{O}_\tx{ph}  (t) \dblket{\psi_\tx{ph}^\tx{ex}}_\tx{ph} \\
&= \dblbra{\chi_\tx{ph}^\tx{ex}} \hat{N}^{-1/2} \left(  \hat{\Pi}_\tx{EX}(t) \, \hat{O} \right) \hat{N}^{-1/2} \dblket{ \psi_\tx{ph}^\tx{ex}}  ~,
}
which can be rewritten as
\eqn{
&\dblbra{\chi_\tx{ph}^\tx{ex}} \hat{O}_\tx{ph}  (t) \dblket{\psi_\tx{ph}^\tx{ex}}_\tx{ph} \\
&\hspace{1em}=\int \rd^{d+1} X |g|_X^{1/2} \int \rd^{d+1} Y |g|_{Y}^{1/2} \\
&\hspace{4em} \bbk{X}{\tilde{\chi}_\tx{ph}^\tx{ex}}  \fixbra{X}  \hat{\Pi}_\tx{EX}(t) \hat{O}   \fixket{Y}   \bkk{Y}{\tilde{\psi}_\tx{ph}^\tx{ex}}  \\
&\hspace{1em}=\int \rd^d \tx{X}_0 |g|_{X_0}^{1/2} \int \rd^d \tx{Y}_0 |g|_{Y_0}^{1/2} \\
&\hspace{4em} \fixbra{\tilde{\chi}^\tx{ex}_0 (\mb{X}_0)}    \hat{\tilde{O}}_\tx{ph} (t;X_0,Y_0)  \fixket{\tilde{\psi}^\tx{ex}_0 (\mb{Y}_0)} ~.
}
The first expression corresponds to that in the Schr\"odinger picture, while the second corresponds to that in the Heisenberg picture, provided that the initial times for the bra and ket states are set to coincide, $X_0^0 = Y_0^0$.
Note that the value of the above quantity is independent of the choice of these initial times.

\section{Observers with dynamical matter field \label{app:observers_with_dynamical_matter_field}}
In this section, extending the discussion in Sec.\ref{sec:relativistic_particle_with_clock_and_detector} by promoting the matter field to be dynamical, we see that the theory has the same formal structure when expressed in a variant of the interaction picture with the unitary transformation given in Eq.(\ref{app:Schrodinger->Interaction}).

\

The full theory is described by an action  $\ssf{ S} =   \ssf{ S}_\tx{m}  +  \ssf{ S}_\tx{obs}$
where
$\ssf{ S}_\tx{m}$ is the action functional of the dynamical matter field $\upphi (t, \mb{x})$ in a $(d+1)$-dimensional spacetime with a background metric $g_{\mu \nu}(t,\mb{x})$, and its detailed expression is not important for the present purpose.
The observer's action $S_\tx{obs}$ now depends on the dynamical field through the interaction part of the detector Hamiltonian (\ref{eq:detector-Hamiltonian}).

\subsection{Classical theory} 
As in Sec.\ref{subsec:classical_theory_w/o_gauge-fixing}, we fix the gauge associated with the reparametrization of the worldline by imposing $T(\eta) =t$ and regarding $t$ as a single time parameter in the action.
Let us take a reference time slice at $t = T_\tx{r} :=  T(0)$ on which the phase space variables are introduced.
Those associated with the internal d.o.f. are the same as in the previous discussion, while those associated with the external d.o.f. now consist of the observer's spatial position $\mb{X}:=\mb{X}(0)$, the matter field $\upphi (\mb{x}) := \upphi ( T_\tx{r} , \mb{x})$, and their conjugate momenta.

\ 

The Hamiltonian is obtained as a function of these variables and written as   
\eqn{
\ssf{ H}^\tx{ex}(t)  =   \ssf{ H}_\tx{m} (t) + \ssf{ H}_\tx{obs}(t) ~ 
\label{app:total_Hamiltonian_with_dynamical_matter}}
with the argument $t$ to show its explicit time dependence through the nondynamical metric field $g_{\mu \nu}(t,\mb{x})$,
where the matter-field part $\ssf{ H}_\tx{m} (t)$ comes from $\ssf{ S}_\tx{m}$ and the observer part
\eqn{
\ssf{ H}_\tx{obs}(t) := \upomega (t) \, N(t,\mb{X})   - \tx{P}_i \beta^i(t,\mb{X})   \label{eq:observer_Hamiltonian_with_dynamical_matter}
}
comes from $\ssf{ S}_\tx{obs}$ being of a similar form as in Eq.(\ref{eq:classical_Hamiltonian_from_external_perspective})
with
\eqn{
\upomega (t) := (  h^{ij}(t,\mb{X}) \tx{P}_i \tx{P}_i + \ssf{ M}_\tx{eff}^2 )^{1/2}
}
depending on $t$ only through the nondynamical induced metric;
the effective mass
\eqn{
 \ssf{ M}_\tx{eff}  :=  M    +    H_\tx{D}(p, q, \upphi (\mb{X}) ) + H_\tx{C}   ~ \label{app:effective_mass_with_dynamical_matter}
}
does not have such an explicit time dependence because the matter field in the interaction part $H_\tx{D,i}$ in the detector Hamiltonian is now dynamical and $\upphi (\mb{x})$ is defined on the reference time slice.
The evolution of observables with respect to the parameter $t$ can be computed from its Poisson bracket with the total Hamiltonian (\ref{app:total_Hamiltonian_with_dynamical_matter}), as in Eq.(\ref{eq:physical_evolution_relative_to_external_time_on_fixed_background}).

\ 

The formulation as a constrained system, analogous to that given in Sec.\ref{subsubsec:constraints}, can be obtained by restoring $T$ as a dynamical variable with its canonical conjugate $\sf E$ and extending the Poisson bracket so that $\{ \ssf{ E} , T \}_\tx{P} = 1$.
We impose that 
\eqn{
\tilde{\ssf{C}}^\tx{ex}_+ := - \ssf{ E} + \ssf{ H}^\tx{ex}(T)  = - \ssf{ E} +  \ssf{ H}_\tx{m}  + \ssf{ H}_\tx{obs} 
\label{app:constraint_function_for_external_perspective_with_dynamical_matter}}
vanishes, where $\ssf{ H}_\tx{m} :=  \ssf{ H}_\tx{m}(T)$ and $\ssf{ H}_\tx{obs} :=  \ssf{ H}_\tx{obs}(T)$.
When the metric is time independent, the constraint surface is nothing but the surface of constant total energy set by the value of $\sf E$.

\ 

We note that this constraint can be reproduced by reparametrizing the external-time parameter in $ \ssf{ S}_\tx{m} = \int \rd t \int \rd^d \tx{x} \,\cl{ L}_\tx{m}$ via $t= T(\eta)$ and regarding $\eta$ as a single time parameter in the total action $\ssf{ S}_\tx{obs} + \ssf{ S}_\tx{m}$.
Writing $\ssf{ S}_\tx{m} = \int \rd \eta \int \rd^d \tx{x} \, \cl{ L}^{\!(\eta)}_\tx{m}$, let us consider the matter-field Lagrangian density $\cl{ L}^{\!(\eta)}_\tx{m}$ including the spacetime volume factor in terms of this new external-time coordinate $\eta$.
It can be obtained by making replacements in $\cl{ L}_\tx{m}$ that $N\to N (\rd  T/\rd \eta )$, $\beta^i \to \beta^i (\rd  T/\rd \eta )$ and $\pd_t \upphi \to \pd_\eta \upphi$.
The dynamical field $\upphi$ is understood as a function in the new coordinate system via the redefinition $\upphi (T(\eta),\mb{x}) \to \upphi (\eta,\mb{x})$.
Then, the canonical conjugate $- \ssf{ E}$ to $T$ is given by
\eqn{
- \ssf{ E} = \frac{\ssf{ M}_\tx{eff} g_{0 \nu}(X) }{\upsilon} \frac{\rd X^\nu }{  \rd \eta } - \ssf{ H}_\tx{m}  ~, 
}
where the first term comes from the observer's action while the second term follows from the fact that
\eqn{
\frac{\pd \int \rd^d \tx{x} \, \cl{ L}^{\!(\eta)}_\tx{m} }{\pd (\rd T / \rd \eta)} = \left( \frac{\rd T}{\rd \eta} \right)^{-1} \! \! \int \rd^d \tx{x}  \left[  N \frac{\pd }{\pd N} + \beta^i \frac{\pd }{\pd \beta^i} \right] \cl{ L}^{\!(\eta)}_\tx{m} 
}
corresponds to $- \ssf{ H}^{\!(\eta)}_\tx{m} = \int \rd^d \tx{x} ( \cl{ L}^{\!(\eta)}_\tx{m} - \uppi (\mb{x}) \pd_\eta \upphi (\mb{x}) )$ and, as a functional of the phase space variables $\upphi $ and $\uppi  := \pd \cl{ L}^{\!(\eta)}_\tx{m} /\pd (\pd_\eta \upphi)$, is identical to $-\ssf{ H}_\tx{m}$ introduced in Eq.(\ref{app:constraint_function_for_external_perspective_with_dynamical_matter}).
On the other hand, there is no contribution from the matter-field Lagrangian to the canonical conjugate to $\mb{X}$, i.e., $\tx{P}_i =  \upsilon^{-1} \ssf{ M}_\tx{eff} g_{i \nu}(X)  (\rd X^\nu /  \rd \eta )$.
Therefore, we find the KG-type constraint that $\ssf{ C}^\tx{KG}$ vanishes, where $\ssf{ C}^\tx{KG}$ is defined by replacing $E=-P_0$ in Eq.(\ref{eq:KG-type_constraint_function}) with
\eqn{
\ssf{ E}_\tx{obs} := \ssf{ E}  - \ssf{ H}_\tx{m} ~.
\label{app:observer_energy}}

\

As discussed in relation to Eq.(\ref{eq:factorizations}), such a KG-type constraint function can be expressed in two ways:
\eqn{
\ssf{ C}^\tx{KG} = \ssf{ C}^\tx{ex}_- \ssf{ C}^\tx{ex}_+ = \ssf{ C}^\tx{in}_- \ssf{ C}^\tx{in}_+ ~,
\label{app:factorizations}}
where $\ssf{ C}^\tx{ex}_\pm$ and $\ssf{ C}^\tx{in}_\pm$ are given as in Eqs.(\ref{eq:constraint_function_for_external_perspective}) and (\ref{eq:constraint_function_for_internal_perspective}) with $\ssf{ E}_\tx{obs}$ instead of $E$.
Then, the physical assumption that the energy of the relativistic particle with the internal d.o.f. measured by the fiducial observer $(\ssf{ E}_\tx{obs} + \tx{P}_i \beta^i) / N$ is positive implies that $\tilde{\ssf{C}}^\tx{ex}_+ = \ssf{ C}^\tx{ex}_+ N$, defined in Eq.(\ref{app:constraint_function_for_external_perspective_with_dynamical_matter}), vanishes for the KG-type constraint to be satisfied. 
Regarding $\ssf{ C}^\tx{in}_+$ defined to be linear in $H_\tx{C}$, let us remark that the corresponding internal-time Hamiltonian
\eqn{
\ssf{ H}^\tx{in} := &\, \ssf{ C}^\tx{in}_+ - H_\tx{C} \\
= &\,  M + H_\tx{D}   -    [   (\ssf{ E}_\tx{obs} + \tx{P}_i \beta^i)^2 / N^2  - |\mb{P}|_h^2 ]^{1/2}
\label{app:classical_Hamiltonian_from_internal_perspective_with_dynamical_matter}}
nonlinearly depends on $\ssf{ H}_\tx{m}$ through $\ssf{ E}_\tx{obs}$.

\subsection{Quantum theory \label{subsec:quantum_theory_with_dynamical_fields}}
Now, the phase-space variables are replaced with the corresponding operators.
The kinematical Hilbert space is written as $\cl{H}_\tx{kin} = \cl{H}_\tx{IN} \otimes  \cl{H}_\tx{EX}$, as discussed in Sec.\ref{subsec:quantization}, albeit containing  $\cl{H}_\tx{m}$ for the dynamical matter field in the external part, $\cl{H}_\tx{EX} = \cl{H}_\tx{TX}  \otimes \cl{H}_\tx{m}$.
For simplicity, the Hilbert space of the observer's spacetime location is flattened, and then it can be written as $\cl{H}_\tx{TX} = \cl{H}_\tx{T} \otimes \cl{H}_\tx{X}$.

\

Differences from the theory with the nondynamical field at the classical level are that we have $\ssf{ E}_\tx{obs}$ instead of $E$ and the dynamical field $\upphi (\mb{X})$ in the observer's effective mass (\ref{app:effective_mass_with_dynamical_matter}).
Correspondingly, in the quantized theories, we have
\eqn{
\hat{\ssf{E}}_\tx{obs} = \hat{\ssf{E}} - \hat{\ssf{H}}_\tx{m} 
}
and 
\eqn{
 \hat{\ssf{M} }_\tx{eff}  =  M    +   H_\tx{D} (\hat{p}, \hat{q}, \hat{\upphi } (\hat{\mb{X}}) ) +   \hat{H}_\tx{C}  ~,
\label{app:effective_mass_operator} }
where
\eqn{
\hat{\upphi }(\hat{\mb{X}})  =  \int \rd^d  \tx{X} \,   \projector{\mb{X}}   \otimes  \hat{ \upphi } (\mb{X})  
}
acting on $\cl{H}_\tx{X} \otimes \cl{H}_\tx{m}$.\footnote{
Without the flattening, it is considered an operator on $\cl{H}_\tx{TX} \otimes \cl{H}_\tx{m}$ and we have $\hat{\Pi}_\tx{TX}(\mb{X}) := \int_\mathbb{R} \rd T |g|^{1/2}_X \projector{X}$ acting on $\cl{H}_\tx{TX}$ instead of the projector $\projector{\mb{X}}$, commuting with $\hat{\ssf{E}}$ that generates the shift of $T$.
Similarly, in Eq.(\ref{app:Schrodinger->Interaction}), we have, instead of $\projector{T}$, the projector $\hat{\Pi}_\tx{TX}(T) := \int \rd^d \tx{X} \,  |g|^{1/2}_X \projector{ X }$ commuting with the momentum operator $\hat{\mb{P}}$, which is the same as $\hat{\Pi}_\tx{EX}$ in Eq.(\ref{app:quantum_observable_in_terms_of_original_basis}).
}
Cf. the effective mass (\ref{eq:effective_mass_operator}) in the case with the nondynamical matter field.

\ 

The matter Hamiltonian $\hat{\ssf{H}}_\tx{m}$ is made up of the matter field operator in the Schr\"odinger picture $\hat{\upphi } (\mb{x})$ and its conjugate momentum, as well as the external time operator $\hat{T}$ through the dependence on $g_{\mu \nu} (\hat{T}, \mb{x})$.
The observer Hamiltonian $\hat{\ssf{H}}_\tx{obs}$ has basically the same form as $\hat{H}^\tx{ex}$ given in Eq.(\ref{eq:Hamiltonian_operator_from_external_perspective}), but now with the effective mass (\ref{app:effective_mass_operator}).

\ 

To make it possible to develop the discussion in a manner parallel to the nondynamical-matter case, we take a kind of interaction picture where operators $\hat{\ssf{O}}_\mf{O} $ are related to those in the original (Schr\"odinger) picture $\hat{\ssf{O}}$ as
\eqn{
\hat{\ssf{O}}_\mf{O} := \hat{V}^\dag \hat{\ssf{O}} \hat{V} ~, 
}
where
\eqn{\hat{V} =  \int_\mathbb{R} \rd T  \, \projector{T} \otimes    \hat{V}(T)  ~,  ~~~  \hat{V}(T) :=  \cl{T}e^{- \ri \int_{T_\tx{r}}^T \rd \eta \hat{\ssf{H}}_\tx{m}(\eta) } ~  
\label{app:Schrodinger->Interaction}}
with $\hat{\ssf{H}}_\tx{m} (\eta)$ being self-adjoint acting on $\cl{H}_\tx{m}$ and related to $\hat{\ssf{H}}_\tx{m}$ as $\hat{\ssf{H}}_\tx{m} = \hat{\ssf{H}}_\tx{m}(\hat{T})$.
The operators involving the field operators and $\hat{E}$ are those that are affected by this transformation.
In this picture,
the zero point of the total energy is shifted by the matter field's Hamiltonian.
Then, the observer's energy is associated with
\eqn{
\hat{\ssf{E}}_\tx{obs, \mf{O}} = \hat{\ssf{E}} ~,
}
and the constraint operator corresponding to the classical function (\ref{app:constraint_function_for_external_perspective_with_dynamical_matter}) looks the same as the one with the nondynamical field (\ref{eq:constraint_operator_for_external_perspective}), apart from the effective mass,
\eqn{
 \hat{\ssf{M}}_\tx{eff,\mf{O}}   =  M    +   H_\tx{D} (\hat{p}, \hat{q}, \hat{ \upphi }_\mf{O}(\hat{\mb{X}}) ) +   \hat{\Omega}    \label{eq:M_eff_Interaction}
}
which is now dependent on $\hat{T}$ through the field operator.
However, it coincides with the result yielded simply by making the following replacement in Eq.(\ref{eq:field_value_at_observer's_location}):
\eqn{
\phi (T ,\mb{X}) \to \hat{\upphi }_\mf{O} (T ,\mb{X}) :=  \hat{V}^\dag(T)   \hat{ \upphi }(\mb{X}) \hat{V}(T) ~ \label{eq:replacement}
}
This variant of the interaction picture may be referred to as an ``observer picture,'' and hence, we have labeled the operators with the subscript $\mf{O}$.

\ 

Then, all the expressions in Secs. \ref{sec:relativistic_particle_with_clock_and_detector} and \ref{sec:violation_of_EEP} can be reused after the following replacements,
\eqn{\hat{E} \to \hat{\ssf{E}} ~, ~~~ \hat{M}_\tx{eff} \to \hat{\ssf{M}}_\tx{eff,\mf{O}} ~,
\label{app:E_and_M_redefined}}
and 
\eqn{
\phi (T, \mb{X}) \to \hat{\phi}(T, \mb{X}) := \hat{ \upphi }_\mf{O}(T, \mb{X}) ~. \label{app:matter_field_operator_in_observer_picture}
}
Here, it should be emphasized that, while the nondynamical field $\phi$ had an arbitrary external-time dependence, the operator $\hat{\phi}$ obeys the Heisenberg equation $\ri \pd_T \hat{\phi} = [\hat{\phi}, \hat{H}_\tx{m}(T) ]$ in the absence of interaction with the observer, where $\hat{H}_\tx{m}(T) := \hat{V}^\dag(T)  \hat{\ssf{H}}_\tx{m}(T) \hat{V}(T)$.

\ 

We note that the physical states obtained in this manner are in the interaction picture, denoted as $\dblket{\uppsi^\tx{ex}_\tx{ph}}_\mf{O}$ with the subscript $\mf{O}$, and are related to those in the original picture as
\eqn{
\dblket{\uppsi^\tx{ex}_\tx{ph}}_\mf{O} =  \hat{V}^\dag \dblket{\uppsi^\tx{ex}_\tx{ph}} ~.
}
Since $\hat{V}$, defined in Eq.(\ref{app:Schrodinger->Interaction}), is not only unitary but also trivially acting on $\cl{ H}_\tx{IN}$,
the statements about the internal-time evolution in the interaction picture obtained by following the discussion in Secs. \ref{sec:relativistic_particle_with_clock_and_detector} and \ref{sec:violation_of_EEP} also hold in the original picture.
More explicitly, the reduced density matrix  
\eqn{
\hat{\uprho}_\tx{exp} (\alpha) := \frac{\Tr_\tx{EX} \Tr_\tx{C} [  (\hat{\Pi}(\alpha)  \otimes \hat{I}_{\overline{\tx{C}}} )  \dblket{\uppsi_\tx{ph}^\tx{ex}} \dblbra{\uppsi_\tx{ph}^\tx{ex}} ] }{ \Tr [ (\hat{\Pi}(\alpha)  \otimes \hat{I}_{\overline{\tx{C}}} ) \dblket{\uppsi_\tx{ph}^\tx{ex}} \dblbra{\uppsi_\tx{ph}^\tx{ex}} ] }  ~,
\label{app:reduced_density_operator}} 
corresponding to that in the nondynamical-matter case (\ref{eq:reduced_density_operator}), is identical in both pictures, and hence we can conclude that the EEP is violated due to the nonunitarity of the internal-time evolution, based on the analysis in the interaction picture.

\ 

In addition, to formulate a quantum theory from the internal perspective associated with the constraint function $\ssf{ C}^\tx{in}_+$ in Eq.(\ref{app:factorizations}), we can work with the interaction picture and apply those replacements in Eqs.(\ref{app:E_and_M_redefined}) and (\ref{app:matter_field_operator_in_observer_picture}) to the expressions in Sec.\ref{subsubsec:possible_formulation_of_internal_perspective}.
However, in the original picture, they become highly complicated, suggesting that such a formulation is not a promising direction.

\section{Inverse of constraint matrix \label{app:Dirac_bracket_with_C}}
In Sec.\ref{subsubsec:implementation}, the inverse of the constraint matrix $\Upsilon$ was computed with all the elements in $\ssf{ C}^\tx{res}$ assumed to be infinitesimally small.
However, in order to discuss the consistency of the subgauge transformation generated by the elements of $\ssf{ C}^\lozenge$ in Sec.\ref{subsubsec:residual_diffeos} and the construction of the relational observables in Sec.\ref{subsubsec:gravitational_relational_observable}, we should keep $\sf C$ finite.
Therefore, here we give a perturbative computation of $\Upsilon^{-1}$ in terms of the elements of $\ssf{ C}^\lozenge$, without assuming the smallness of $\sf C$.

\

Now, let us write $\tx F$ in Eq.(\ref{eq:Upsilon_inverse}) as
\eqn{
\tx{F} =  \tx{C}^\lozenge + \tx{H}  ~,
}
where
\eqn{
 \tx{C}^\lozenge := \tx{C}  - \tx{C}|_{\ssf{ C}^\lozenge \!= 0}
}
with
\eqn{
\tx{C}|_{\ssf{ C}^\lozenge \!= 0} =   -\ssf{ C}_\perp [n]  \pmtx{ \mathbb{ O} &  \mathbb{ S}^{-1} \mathbb{ V}  \\ - \mathbb{ V}^\tx{t} (\mathbb{ S}^{-1})^\tx{t}  & \mathbb{ O} } 
}
and
\eqn{
\tx{H} :=&~ \tx{G}^{-1}  + \tx{C}|_{\ssf{ C}^\lozenge \!= 0}  \\ 
=&~   \ssf{ H}_{\tx{g}+\tx{m}}[n]   \pmtx{ \mathbb{ O} & - \mathbb{ S}^{-1} \mathbb{ V}  \\  \mathbb{ V}^\tx{t} (\mathbb{ S}^{-1})^\tx{t}  & \mathbb{ O} }
}
Comparing it with $\tx{G}$ in Eq.(\ref{eq:G}), we find
\eqn{
\tx{H}^{-1}= \ssf{ r} \times \tx{G}  ~, ~~~ \ssf{ r} := \frac{ - \ssf{ M}_\tx{eff}}{ \ssf{ H}_{\tx{g}+\tx{m}}[n]} ~.
}
With the expansion $\tx{F}^{-1} =   \sum_{n = 0}^\infty  ( - \tx{H}^{-1} \tx{C}^\lozenge   )^n \tx{H}^{-1}$ and
\eqn{ \tx{C}_n^\lozenge :=  (- \tx{C}^\lozenge  \tx{G})^n (-\tx{G})^{-1} = - (\tx{C}_n^\lozenge )^\tx{t} ~,}
we find
\eqn{
\Upsilon^{-1} = \Upsilon^{-1}_\lozenge  +  \sum_{n=1}^\infty \ssf{ r}^{n+1}  \pmtx{  \tx{G}   & \tx{O} \\  \tx{O}  &   \tx{D}  } \pmtx{  \tx{C}_{n}^\lozenge  & \tx{C}_n^\lozenge \\  \tx{C}_n^\lozenge &   \tx{C}_n^\lozenge }    \pmtx{  \tx{G}   & \tx{O} \\  \tx{O}  &   \tx{D}  }^{\!\tx{t}} ~,
}
where 
\eqn{
\Upsilon^{-1}_\lozenge  :=&~  \Xi^\tx{t} \pmtx{  \tx{H}^{-1} &  \tx{H}^{-1}     \\  \tx{H}^{-1} & \tx{H}^{-1} - \tx{G}     } \Xi \\
=&~   \pmtx{   \ssf{ r}  \tx{G}  &  -  \ssf{ r}  \tx{D}^\tx{t}  \\  \ssf{ r}  \tx{D}   &  ( \ssf{ r} -1 ) \tx{E} } 
}
is the $\ssf{ C}^\lozenge \to 0$ limit of $\Upsilon^{-1}$.
When $\ssf{ C} = 0$, we have $\ssf{ r} = 1$, and it reduces to $\Upsilon_0^{-1}$ given in Eq.(\ref{Upsilon_inverse_0}).

\bibliography{Ref}

\end{document}